\documentclass[a4paper,11pt]{article}
\pdfoutput=1

\usepackage{jheparxiv} 
\usepackage[T1]{fontenc}
\usepackage[utf8]{inputenc}
\usepackage{graphicx}
\usepackage{amsmath}
\usepackage{amssymb}
\usepackage{dsfont}
\usepackage{latexsym}
\usepackage{mathrsfs}
\usepackage{braket}		%Bra Ket Notation
\usepackage{hyperref}
\usepackage{xcolor}
\usepackage{twistor}
\usepackage{mathtools}
\usepackage{comment}
\usepackage{tikz-cd}
\usepackage{bbm}
\usepackage[normalem]{ulem}
\usepackage[mathscr]{eucal}
\usepackage[version=3]{mhchem}
\usepackage{enumitem}
\newcommand\scalemath[2]{\scalebox{#1}{\mbox{\ensuremath{\displaystyle #2}}}}
\usepackage{BOONDOX-uprscr}

\newcommand{\msf}[1]{\mathsf{#1}}

\newcommand{\eps}{\epsilon}
\newcommand{\veps}{\varepsilon}

\newcommand{\sh}{\mathsf{h}}
\newcommand{\bsh}{\bar{\mathsf{h}}}

\newcommand{\bL}{\mathbf{L}}
\newcommand{\bbL}{\overline{\bL}}

\newcommand{\sa}{\msf{a}}

 \newcommand{\badat}{\begin{alignedat}}
 \newcommand{\eadat}{\end{alignedat}}

\newcommand{\s}{\text{s}}
\newcommand{\at}{\text{a}}
\newcommand{\sJ}{\mathscr{J}}
\newcommand{\sL}{\mathscr{L}}
\newcommand{\sC}{\mathscr{C}}

\newcommand{\belta}{\boldsymbol\delta}

\newcommand{\highlight}[1]{%
  \colorbox{gray!30}{$\displaystyle#1$}}

\def\bw{{\bar w}}
\def\bh{{\bar h}}
\def\bb{{\bar b}}
\def\bc{{\bar c}}
\def\ba{{\bar a}}
\def\bd{{\bar d}}
\def\bz{{\bar z}}
\def\O{{\mathscr O}}

\def\bal{{\bar \alpha}}

\title{Celestial amplitudes in an ambidextrous basis}

\author{
Carmen Jorge-Diaz,$^a$ Sabrina Pasterski$^{b}$ \& Atul Sharma$^c$}

\affiliation[a]{The Mathematical Institute, University of Oxford, \\  Woodstock Road, OX2 6GG, U.K.}

\affiliation[b]{Perimeter Institute for Theoretical Physics,\\ Waterloo, ON N2L 2Y5, Canada}

\affiliation[c]{Center for the Fundamental Laws of Nature \& Black Hole Initiative,\\
Harvard University, Cambridge MA 02138, USA}
 
\emailAdd{carmen.jorgediaz@maths.ox.ac.uk, sabrina@perimeterinstitute.ca,  atulsharma@fas.harvard.edu}

\abstract{We start by constructing a conformally covariant improvement of the celestial light transform which keeps track of the mixing between incoming and outgoing states under finite Lorentz transformations in $\R^{2,2}$. We then compute generic 2, 3 and 4-point celestial amplitudes for massless external states in the ambidextrous basis prepared by composing this SL$(2,\mathbb{R})$ intertwiner with the usual celestial map between momentum and boost eigenstates. The results are non-distributional in the celestial coordinates $(z,\bar z)$ and conformally covariant in all scattering channels. Finally, we focus on the tree level 4-gluon amplitude where we present a streamlined route to the ambidextrous correlator based on Grassmannian formulae and examine its alpha space representation. In the process, we gain insights into the operator dictionary and CFT data of the holographic dual.
}

\begin{document}

\maketitle

\section{Introduction}

Celestial conformal field theories have assumed a central role in the study of holography in asymptotically flat spacetimes~\cite{Pasterski:2021raf}. In the absence of an intrinsic construction of a boundary dual (until recently! \cite{Costello:2022jpg}), the general methodology has been to build a holographic dictionary starting from the kinematics: the generators of asymptotic symmetries in 4D naturally organize into currents in a 2D theory~\cite{Strominger:2017zoo}, and we can choose to prepare scattering states that transform in representations amenable to such a 2D description. Such {\it celestial amplitudes} recast ${\cal S}$-matrix elements as correlators in this codimension-2 CFT by considering external wavepackets that are boost eigenstates~\cite{Pasterski:2016qvg,Pasterski:2017kqt,Pasterski:2017ylz}. However, the existence of intertwiners between such representations of the Lorentz group poses a potential ambiguity in our choice of holographic map. 

The standard dictionary proceeds as follows~\cite{deBoer:2003vf,Cheung:2016iub,Pasterski:2017ylz}. Consider the $(2,2)$ signature amplitude $A(p_i)$ with massless external momenta
\be\label{pi}
p_i^\mu=\frac{\veps_i\omega_i}{\sqrt2}\,(1+z_i\bz_i,z_i+\bz_i,z_i-\bz_i,1-z_i\bz_i)\,.
\ee
By performing a Mellin transform in the energy variables $\omega_i$, one obtains the so-called \emph{celestial amplitude}
\be\label{eq:mellinmap}
\langle \O^{\veps_1}_{h_1,\bh_1}(z_1,\bz_1)\cdots\O^{\veps_n}_{h_n,\bh_n}(z_n,\bz_n) \rangle=\prod_{i=1}^n\int \d\omega_i\, \omega_i^{\Delta_i-1}\; A(p_1,\dots,p_n)
\ee
that transforms as a correlator of quasi-primaries with weights
\be
h_i=\frac{1}{2}(\Delta_i+J_i),~~~\bh_i=\frac{1}{2}(\Delta_i-J_i)\,,
\ee
where $\Delta_i$ is the Rindler energy and $J_i$ is the helicity of the external state analytically continued from $\mathbb{R}^{1,3}$. 

While this celestial CFT (CCFT) correlator has the desired Lorentz covariance, the additional requirement of translation invariance implies that the celestial amplitudes for $n\le4 $ particles are necessarily distributional. 
At higher points, the imprint of translation invariance manifests itself as channel dependent constraints on the allowed positions of the celestial operators and the invariant data the kinematics-stripped amplitude depends on (see~\cite{Mizera:2022sln} for a detailed discussion). However, we can still interpret the collinear limits as encoding an OPE structure~\cite{Fotopoulos:2019tpe,Pate:2019lpp, Fotopoulos:2019vac} within these higher point amplitudes, and know the spectrum required for a complete basis of scattering states~\cite{Pasterski:2017kqt,Donnay:2020guq}.  Namely, we have what would ordinarily be the data of the CFT, but the presence of extra symmetries imposes non-standard constraints that complicate the standard procedure for extracting CFT data from a conformal block decomposition of the four-point function.

Given these exotic features of the celestial CFT, there are two parallel approaches that have met with some success thus far.  On the one hand, we can take our correlators as they are and re-derive properties of the CCFT based on our knowledge of the bulk physics, making sure not to take our standard CFT intuition for granted.   On the other hand, we can try to exploit the flexibility of our dictionary to reorganize CCFT in a basis where the behavior of the correlators looks more familiar. This paper takes the latter route. By composing a light transform on the celestial torus with a Mellin transform in the energy scale, we can go from plane wave scattering in $(2,2)$ signature to a so-called ambidextrous basis~\cite{Sharma:2021gcz}.  This basis choice tames the distributional nature of low-point massless amplitudes and makes them more amenable to standard CFT technology in $(2,2)$ signature.

Our starting point is a modified definition of the light transform adapted to celestial CFT.  This builds off the results of \cite{Sharma:2021gcz}, but remedies certain shortcomings. In particular, the prescription for the light transform used in the earlier celestial literature failed to produce conformally covariant results for all channels. We resolve this here and work out the fully covariant, non-distributional light transformed amplitudes at 2, 3 and 4 points for generic massless processes. Postulating these to be our 2D CFT operators, we can read off the boundary operator algebra for light transformed conformal primaries. The results contain the beta function OPE coefficients familiar from past studies of collinear limits. Focusing on the gluon case we proceed with an alpha space analysis of the conformal block decomposition.

Since examining the CCFT data is an active area of interest, it is worth highlighting relevant precedents in the literature that our investigations have benefited from. Conformal block decompositions in the standard basis have been examined in~\cite{Lam:2017ofc,Nandan:2019jas}, while conformal block decompositions involving light transformed operators (using the previous prescription) have been considered in~\cite{Atanasov:2021cje,Guevara:2021tvr,Hu:2022syq,Banerjee:2022hgc}.  Meanwhile, shadow transforms have been used to tame the 4-point amplitudes in $(1,3)$ signature in~\cite{Fan:2021isc,Fan:2021pbp,Fan:2022vbz}, and have appeared in proposals for the state operator dictionary for 2D out states in radial quantization~\cite{Fan:2021isc,Crawley:2021ivb} as well as earlier investigations of how to make low point correlators more standard~\cite{ss}.

This paper is organized as follows. In order to build up to our construction of an improved ambidextrous basis, we start with a review of the celestial torus in section~\ref{sec:celestial_torus}, a discussion of the ordinary conformal basis in Klein space in section~\ref{sec:Klein_basis}, and the light transforms of these states in section~\ref{sec:LT_torus}. A main theme that we encounter is how to handle the additional $\mathbb{Z}_2$ action of the $(2,2)$ signature Lorentz group on the choice of crossing channels, which would not arise in ordinary Minkowski space.  We then apply our basis transformation to generic low-point massless amplitudes in section \ref{sec:LT_amplitudes}, before focusing on tree-level gluon amplitudes in section \ref{sec:gluon_decomp}.  There we are able to streamline our construction of the light transformed amplitudes using a twistor-inspired presentation of the external wavepackets combined with Grassmannian formulae for momentum space amplitudes. We run our results through the machinery of the alpha space transform of \cite{Hogervorst:2017sfd} and extract CCFT data like its operator spectrum, obtaining results consistent with expectations from past work on celestial OPE \cite{Pate:2019lpp}. We close with some discussions of how to think of the celestial dictionary for this basis in section~\ref{sec:conclusions}. 

We work in four dimensions throughout but include some additional perspectives on the ambidextrous prescription coming from higher dimensions in appendix~\ref{app:higherd}. Appendix \ref{app:fac} is dedicated to the conformal block decompositions of light transformed unitarity cuts (the $(2,2)$ signature analogue of optical theorems), providing a relatively simple toy example displaying the highlights of our methods. Appendix \ref{app:ope} contains some further details on (partial) OPE data obtainable from alpha space expansions of gluon 4-point amplitudes.

%%%%%%%%%%%%%%%%%%%%%%%%%%%%%%%%%%%
%%%%%%%%%%%%%%%%%%%%%%%%%%%%%%%%%%%

\section{The celestial torus}\label{sec:celestial_torus}
 
Let $\R^{2,2}$ denote flat space-time equipped with a metric of ultrahyperbolic signature,
\be\label{22met}
\d s^2 = -(\d X^0)^2+(\d X^1)^2-(\d X^2)^2+(\d X^3)^2\,.
\ee
Following \cite{Atanasov:2021oyu}, we refer to $\R^{2,2}$ as Klein space. In \cite{Atanasov:2021oyu,Mason:2005qu}, it was observed that the asymptotic boundary of Klein space is foliated by tori. In this section, we briefly review the notion of such a celestial torus. We also explain how  it as an analytic continuation of the celestial sphere of Minkowski space $\R^{1,3}$.  In the sections that follow, our momentum space will also be equipped with the same bilinear form as in \eqref{22met}. The discussion of this section will then carry over by replacing null vectors $X^\mu$ by null momenta $p^\mu$. 

The light cone of the origin of $\R^{2,2}$ is given by the quadric
\be\label{lc}
(X^0)^2+(X^2)^2 = (X^1)^2+(X^3)^2\,.
\ee
Cross sections of this light cone are found by quotienting out the dilatations $X^\mu\mapsto t\,X^\mu$, $t>0$. Using such a rescaling, we can normalize $X^\mu$ so that
\be\label{cteq}
(X^0)^2+(X^2)^2 = (X^1)^2+(X^3)^2 = 1\,.
\ee
The resulting geometry is a torus $S^1\times S^1$, colloquially referred to as the \emph{celestial torus} of Klein space. Importantly, it has a single connected component. 

To study its geometry, introduce complex coordinates
\be\label{wdef}
w_1 = X^0 + \im\,X^2\,,\qquad w_2 = X^3+\im\, X^1\,.
\ee
In terms of these, the metric of Klein space reads
\be
\d s^2 = -|\d w_1|^2 + |\d w_2|^2\,.
\ee
The equations in \eqref{cteq} are now given by $|w_1|=|w_2|=1$. So the celestial torus can be coordinatized by angles $\psi,\phi\in S^1$,
\be\label{angles}
w_1 = \e^{\im\psi}\,,\qquad w_2 = \e^{\im\phi}\,,
\ee
defined up to $\psi\sim\psi+2\pi$, $\phi\sim\phi+2\pi$. These provide global coordinates on the torus with induced Lorentzian metric
\be
\d s^2 = -\d\psi^2+\d\phi^2\,.
\ee
Comparing \eqref{angles} to \eqref{wdef}, one can also read off the embedding of the torus in the original coordinates,
\be\label{Xangles}
X^\mu = (\cos\psi,\sin\phi,\sin\psi,\cos\phi)\,.
\ee
Unfortunately, the Lorentz group $\SL(2,\R)\times\SL(2,\R)$ acts linearly on the $X^\mu$ but \emph{non-linearly} on $\psi,\phi$. 

Instead, a more convenient choice of coordinates on the celestial torus is
\be\label{z22}
z = \frac{X^1+X^2}{X^0+X^3}\,,\qquad\bar z = \frac{X^1-X^2}{X^0+X^3}\,.
\ee
These are real and independent, as well as manifestly invariant under projective rescalings of $X^\mu$. Up to conformal rescaling, the induced metric in these coordinates is given by
\be
\d s^2 = -\d z\,\d\bar z\,,
\ee
and we can recognize these coordinates as spanning 
a Poincar\'e patch $\R^{1,1}\subset S^1\times S^1$. The Lorentz group acts on $z,\bz$ via real and independent M\"obius transformations,
\be
z\mapsto \frac{az+b}{cz+d}\,,\qquad\bar z\mapsto \frac{\bar a\bar z+\bar b}{\bar c\bar z+\bar d}
\ee
where $ad-bc=1=\bar a\bar d-\bar b\bar c$. The price to pay for this simplicity is that the $(z,\bar z)$ coordinates are no longer global on the celestial torus. 

Substituting \eqref{Xangles} into \eqref{z22}, we can express $z,\bar z$ in terms of the global coordinates $\psi,\phi$:
\be
z = \tan\left(\frac{\phi+\psi}{2}\right)\,,\qquad\bar z = \tan\left(\frac{\phi-\psi}{2}\right)\,.
\ee
Notice that even though $(\psi,\phi)$ and $(\psi+\pi,\phi+\pi)$ are distinct points on the torus, they are both mapped to the same $z,\bar z$ (since $\tan(y+\pi)=\tan(y)$). As a result, $z,\bar z$ can only be used as coordinates on half the celestial torus. More precisely, they give coordinates on $(S^1\times S^1)/\Z_2$, where $\Z_2$ acts as the diagonal antipodal map $\sigma:(\psi,\phi)\mapsto(\psi+\pi,\phi+\pi)$. At the level of the embedding space coordinates \eqref{Xangles}, the antipodal map acts as
\be
\sigma\,:\, X^\mu\mapsto -X^\mu\,.
\ee
This $\Z_2$ action is obviously a symmetry of the light cone \eqref{lc}. It is an analog of the $\{1,\mathrm{PT}\}$ subgroup of the Klein four-group $\mathrm{O}(1,3)/\SO^+(1,3)$ from $\mathbb{R}^{1,3}$. 

To cover the whole torus, one conventionally cuts it along the hyperplane $X^0+X^3=0$ where the expressions in \eqref{z22} have poles. The resulting subspace is $\Z_2\times\R^{1,1}\subset S^1\times S^1$ and consists of two patches: 
\begin{itemize}
    \item a ``future'' Poincar\'e patch $\R^{1,1}$ on which $X^0+X^3>0$, and
    \item a ``past'' Poincar\'e patch $\R^{1,1}$ on which $X^0+X^3<0$.
\end{itemize}
These get exchanged by the antipodal map $X^\mu\mapsto-X^\mu$. We then use the same expressions \eqref{z22} for $z,\bar z$ as coordinates on either patch.
Since $\sgn(X^0+X^3)$ is not Lorentz invariant in $(2,2)$ signature, the choice of the two Poincar\'e patches is not Lorentz invariant. That is, points in one Poincar\'e patch can be mapped to points in the other by Lorentz transformations (as we will see more explicitly in momentum space in the next section). Nevertheless, once a choice of two Poincar\'e patches on the celestial torus is fixed, one can think of the $X^0+X^3>0$ patch as an analogue of the future celestial sphere of Minkowski space, and the $X^0+X^3<0$ patch as that of the past celestial sphere. This allows us to at least formally define the notion of incoming/outgoing states and their scattering amplitudes \cite{Mason:2005qu}. We can also make this intuition precise by means of a Wick rotation from Minkowski space.

\paragraph{Analytic continuation of the celestial sphere.} Let $X^\mu$ be a point on the light cone of the origin of $\R^{1,3}$. It satisfies
\be
(X^0)^2 = (X^1)^2+(X^2)^2+(X^3)^2\,.
\ee
Quotienting out by positive rescalings of $X^\mu$, one can normalize this so as to satisfy
\be
(X^0)^2 = (X^1)^2+(X^2)^2+(X^3)^2 = 1\,.
\ee
This has two connected components $X^0=\pm1$ that act as the future and past celestial spheres of Minkowski space. Since (orthochronous) Lorentz transformations in $\R^{1,3}$ preserve the sign of $X^0$, they leave the two connected components invariant. That is, unlike the celestial torus, Lorentz transformations of $\R^{1,3}$ that are homotopic to the identity do not map points on the past celestial sphere to points on the future celestial sphere or vice versa.

Since $X^\mu$ is null, we find
\be
X^0+X^3 = X^0\pm\sqrt{(X^0)^2-(X^1)^2-(X^2)^2}\,.
\ee
Therefore, $X^0+X^3\geq0$ on the future celestial sphere and $X^0+X^3\leq0$ on the past, just as with our choice of Poincar\'e patches on the celestial torus. Away from the pole $X^0+X^3=0$, the standard stereographic coordinates on either celestial sphere are given by
\be
z = \frac{X^1+\im\,X^2}{X^0+X^3}\,,\qquad \bar z = \frac{X^1-\im\,X^2}{X^0+X^3}\,,
\ee
where $\bar z$ is now the complex conjugate of $z\in\C$. We can Wick rotate $X^2\mapsto-\im\,X^2$ to turn the Lorentzian metric $\d s^2 = -(\d X^0)^2+(\d X^1)^2+(\d X^2)^2+(\d X^3)^2$ into the $(2,2)$ signature metric \eqref{22met}. This Wick rotation maps the stereographic coordinates to precisely the coordinates \eqref{z22} on the Poincar\'e patches of the celestial torus.  Thus, upon Wick rotation, the future (past) celestial sphere maps to the future (past) Poincar\'e patch.

\medskip

\paragraph{Where does the CCFT live?} In Minkowski space, it is natural to label scattering states as incoming or outgoing depending on whether their momenta are past or future pointing respectively. By analytic continuation, states with null momenta lying on the past Poincar\'e patch of the celestial torus may be termed incoming, and those lying on the future patch may be termed outgoing. But unlike Minkowski space where incoming momenta are never mapped to outgoing momenta by Lorentz transformations, the notion of incoming and outgoing is no longer Lorentz invariant on the celestial torus. 

This appears to contradict the philosophy that celestial CFT can be defined as living purely on either the future or the past celestial sphere. Recall that in $\R^{1,3}$, since incoming and outgoing states do not mix under Lorentz transformations, they are thought of as being dual to two independent species of local operators inserted on the same celestial sphere. But in $\R^{2,2}$, since incoming states can turn into outgoing states by the conformal group of the celestial torus, it would seem that we can define CCFT only on the full celestial torus and not on just one Poincar\'e patch. In Minkowski space, this would be the analogue of CCFT being necessarily defined on the union of the future and past celestial spheres. What follows is an attempt to ameliorate this situation by utilizing the representation theory of $\SL(2,\R)\times\SL(2,\R)$ to build operators that genuinely live within one Poincar\'e patch of the celestial torus.

%%%%%%%%%%%%%%%%%%%%%%%%%%%%%%%%%%%
%%%%%%%%%%%%%%%%%%%%%%%%%%%%%%%%%%%

\section{Conformal basis in Klein space}\label{sec:Klein_basis}

In this section, we revisit the conformal basis of on-shell scattering states constructed in \cite{Pasterski:2017kqt}. We will use them to construct certain $\Z_2$ eigenstates that transform as conformal primaries of a 2D Lorentzian CFT living on a single Poincar\'e patch of the celestial torus.

\subsection{Boost eigenstates}

Henceforth, we will solely work in $(2,2)$ signature and restrict attention to integer spin massless particles like scalars, gluons or gravitons. Let $|E,z,\bar z,J\ra$ denote the momentum eigenstate of a helicity $J\in\Z$ particle carrying null momentum 
\be\label{nullp}
p^\mu = E\,q^\mu\,,\quad q^\mu(z,\bar z) = \frac{1}{\sqrt2}\left(1+z\bar z,z+\bar z,\bar z-z,1-z\bar z\right)\,,\quad E\in\R^*\,,\;z,\bar z\in\R\,,
\ee
expressed in terms of coordinates $z,\bar z$ on a Poincar\'e patch of the celestial torus, and an overall scaling $E$ whose sign labels the two Poincar\'e patches. It is standard to decompose $E = \veps\omega$ with
\be
\veps=\sgn\,E = \sgn(p^0+p^3)\,,\qquad\omega=|E|\,.
\ee
As discussed in the previous section, $\sgn(p^0+p^3)$ labels the two Poincar\'e patches of the celestial torus. As a result, we can view the triple $(\veps,z,\bar z)\in\Z_2\times\R^{1,1}$ as ``coordinates'' on the whole torus (minus the codimension 1 locus $p^0+p^3=0$).

In Lorentzian signature, one would call this state incoming if $\veps=-1$ and outgoing if $\veps=+1$. This distinction is artificial in $(2,2)$ signature, as $\veps$ is no longer Lorentz invariant. Instead, the Lorentz group $\SL(2,\R)\times\SL(2,\R)$ acts as
\be
\begin{split}
    &z\mapsto \frac{az+b}{cz+d}\,,\qquad\bar z\mapsto \frac{\bar a\bar z+\bar b}{\bar c\bar z+\bar d}\,,\\
    &\omega\mapsto|cz+d|\,|\bar c\bar z+\bar d|\,\omega\,,\qquad\veps\mapsto\sgn\!\left[(cz+d)(\bar c\bar z+\bar d)\right]\veps\,.
\end{split}
\ee
Under these, the momentum eigenstate transforms as
\be
|\veps\omega,z,\bar z,J\ra \mapsto |cz+d|^J\,|\bar c\bar z+\bar d|^{-J}\,|\veps\omega,z,\bar z,J\ra\,.
\ee
This holds for integral $J$.\footnote{For half-integer states, depending on one's conventions, one finds an extra factor of either $(\sgn\,(cz+d))^{2J}$ or $(\sgn\,(\bar c\bar z+\bar d))^{-2J}$. We leave their exploration to future work.}

A \emph{boost eigenstate} is reached by performing a Mellin transform of the momentum eigenstates,
\be
|\veps,z,\bar z,h,\bar h\ra = \int_0^\infty\frac{\d\omega}{\omega}\,\omega^\Delta\;|\veps\omega,z,\bar z,J\ra\,,
\ee
with $\Delta\in\C$ being its conformal dimension, and
\be
h = \frac{\Delta+J}{2}\,,\qquad\bar h = \frac{\Delta-J}{2}
\ee
being its chiral and anti-chiral conformal weights. States on the principal series $\Delta\in1+\im\,\R$ form a complete basis, though one often encounters the need to work off the principal series \cite{Donnay:2020guq}. Let $\O^{\veps}_{h,\bh}(z,\bar z)$ be the operator in celestial CFT that is dual to such a boost eigenstate. These operators transform under $\SL(2,\R)\times\SL(2,\R)$ as follows
\be\label{Otrans}
{\O}^{\veps'}_{h,\bar h}\!\left(z',\bz'\right) = |cz+d|^{2h}\,|\bar c\bar z+\bar d|^{2\bar h}\,{\O}_{h,\bar h}^\veps(z,\bar z)
\ee
where
\be\label{eq:primedcoord}
z'=\frac{az+b}{cz+d},~~~\bz'=\frac{\ba\bz+\bb}{\bc\bz+\bd},~~~\veps'=\sgn\!\left[(cz+d)(\bc\bz+\bd)\right]\veps
\ee
and $ad-bc=\ba\bd-\bb\bc=1$.

Most of the previous work on celestial holography has concerned itself with computing scattering amplitudes for these Mellin transformed boost eigenstates with fixed $\veps$, as these are the states naturally reached by analytic continuation from $\R^{1,3}$ to $\R^{2,2}$~\cite{Pasterski:2017ylz}. Such objects are often termed \emph{celestial amplitudes}. We now come to a simple modification of this conformal basis that allows us to work with a celestial CFT defined on a single Poincar\'e patch of the celestial torus.

\subsection{\texorpdfstring{Boost + $\Z_2$ eigenstates}{Boost + Z2 eigenstates}}

The crucial difference between celestial operators in $(2,2)$ signature and ordinary CFT operators -- or Euclidean CCFT operators for that matter -- is that the additional label $\veps\in\{\pm1\}$ transforms non-trivially under such conformal maps. As such, $\O_{h,\bar h}^\veps$ does not transform in an irreducible representation of the conformal group.

Instead, these operators furnish a representation of the $\Z_2$ symmetry under which $\veps\mapsto-\veps$.  We can diagonalize this action by introducing the linear combinations
\begin{align}
    \O^\s_{h,\bar h}(z,\bar z) &\vcentcolon= \O^+_{h,\bar h}(z,\bar z) + \O^-_{h,\bar h}(z,\bar z)\,,\label{Os}\\
    \O^\at_{h,\bh }(z,\bar z) &\vcentcolon= \O^+_{h,\bar h}(z,\bar z) - \O^-_{h,\bar h}(z,\bar z)\label{Oa} \, ,
\end{align}
that now form genuine irreducible representations. The superscripts `s' and `a' stand for `symmetric' and `antisymmetric' combinations respectively. We will refer to these operators (or their dual states in the bulk) as symmetric or antisymmetric primaries. Under the conformal group $\SL(2,\R)\times\SL(2,\R)$, they are easily seen to transform as
\begin{align}
     \O^\s_{h,\bh }\bigg(\frac{az+b}{cz+d},\frac{\bar a\bar z+\bar b}{\bar c\bar z+\bar d}\biggr) &= |cz+d|^{2h}\,|\bar c\bar z+\bar d|^{2\bar h}\;\O^\s_{h,\bar h}(z,\bar z)\,,\\
     \O^\at_{h,\bh }\bigg(\frac{az+b}{cz+d},\frac{\bar a\bar z+\bar b}{\bar c\bar z+\bar d}\biggr) &= \sgn\!\left[(cz+d)(\bc\bz+\bd)\right]|cz+d|^{2h}\,|\bar c\bar z+\bar d|^{2\bar h}\;\O^\at_{h,\bh }(z,\bar z)\,.
\end{align}
These are two-dimensional analogues of the representations of $\SL(2,\R)$ discussed in chapter VII of \cite{Gelfand:105396}.  The latter also occur in the study of harmonic analysis on $\SL(2,\R)$ \cite{ruhl1970lorentz}. Since the antipodal map of the celestial torus acts as $\veps\mapsto-\veps$, the primaries $\O^\s_{h,\bh }$ and $\O^\at_{h,\bh }$ are physically distinguished by the property of respectively being symmetric or antisymmetric under this antipodal map.

These operators are dual to boost eigenstates that are also $\Z_2$ eigenstates in the bulk,
\begin{align}
    \O^\s_{h,\bar h}(z,\bar z) \quad&\longleftrightarrow\quad |z,\bar z,h,\bar h\ra_\text{s} \vcentcolon= |+,z,\bar z,h,\bar h\ra + |-,z,\bar z,h,\bar h\ra\,,\label{eig+}\\
    \O^\at_{h,\bh }(z,\bar z) \quad&\longleftrightarrow \quad |z,\bar z,h,\bar h\ra_\text{a} \vcentcolon= |+,z,\bar z,h,\bar h\ra - |-,z,\bar z,h,\bar h\ra\,.\label{eig-}
\end{align}
Such $\Z_2$ eigenstates span the same spectrum as the original boost eigenstates. But importantly, since they do not depend on $\veps$, the new operators $\O^\s_{h,\bar h}(z,\bar z)$ and $\O^\at_{h,\bar h}(z,\bar z)$ act as conformal primaries of a Lorentzian CFT living on the $\Z_2$ quotient $(S^1\times S^1)/\Z_2$. Hence, they can be genuinely viewed as operators of a celestial CFT living on a single Poincar\'e patch $\R^{1,1}\subset S^1\times S^1$ representing this quotient. Of course, upon Wick rotation back to Minkowski space, these operators remain well-defined as operators on a single copy of the celestial sphere, with the ${\rm s,a}$ labels replacing the $\veps$ labels.

At a practical level, the main upshot of using these $\Z_2$ eigenstates to compute celestial amplitudes will be a reorganization of the step functions that come from imposing momentum conservation. Indeed, for the symmetric wavefunctions the step functions are altogether absent. This simplifies evaluation of integral transforms like the light and shadow transforms, as already observed for correlators of the $\O^\s_{h,\bar h}$ operators in \cite{Fan:2021isc,Hu:2022syq}. They will also help us (partially) diagonalize the 2-point celestial correlators.  On the other hand, using such states naively obfuscates standard approaches for studying properties like crossing symmetry. Crossing in Minkowski space was recently studied from the viewpoint of boost eigenstates in \cite{Mizera:2022sln}. It would be interesting to see what novel lessons we can learn about crossing symmetry using amplitudes of such $\Z_2$ eigenstates on Klein space.

\subsection{Wavefunctions}

In general, we can find a direct integral transform mapping momentum eigenstates to such boost + $\Z_2$ eigenstates by combining various Mellin integrals. For instance, one can express the symmetric primary state as
\begin{align}
    |z,\bar z,h,\bar h\ra_\s &= \int_0^\infty\frac{\d\omega}{\omega}\,\omega^\Delta\left(|\omega,z,\bar z,J\ra + |-\!\omega,z,\bar z,J\ra\right)\nonumber\\
    &= \int_{-\infty}^\infty\frac{\d\omega}{|\omega|}\,|\omega|^\Delta\;|\omega,z,\bar z,J\ra\,.\label{symstate}
\end{align}
And similarly the antisymmetric primary becomes
\be\label{asymstate}
|z,\bar z,h,\bar h\ra_\at = \int_{-\infty}^\infty\frac{\d\omega}{\omega}\,|\omega|^\Delta\;|\omega,z,\bar z,J\ra\,.
\ee
These can be used to compute amplitudes of our $\Z_2$ eigenstates directly from momentum space amplitudes. The fact that the integration range of the $\omega$ integrals is now the whole real line instead of $(0,\infty)$ will rid celestial amplitudes of their step function support.

It is also instructive to compute the scattering wavefunctions associated to these $\Z_2$ eigenstates in the bulk. For the symmetric primary dual to a massless scalar, we find the wavefunction as the Mellin transform:
\begin{align}\label{phiS}
    \phi^\s_\Delta(X|z,\bar z) &= \int_0^\infty\frac{\d\omega}{\omega}\,\omega^\Delta\left(\e^{\im\omega q\cdot X}+\e^{-\im\omega q\cdot X}\right)\nonumber\\
    &= \int_0^\infty\frac{\d\omega}{\omega}\,\omega^\Delta\cdot 2\cos(\omega q\cdot X)= \frac{2}{|q\cdot X|^\Delta}\int_0^\infty\frac{\d\omega}{\omega}\,\omega^\Delta\cos\omega\nonumber\\
    &= 2\,\Gamma(\Delta)\,\cos\left(\frac{\pi\Delta}{2}\right)\frac{1}{|q\cdot X|^{\Delta}}\,.
\end{align}
Similarly, the antisymmetric primary is dual to the wavefunction
\begingroup
\allowdisplaybreaks
\begin{align}\label{phiA}
    \phi^\at_\Delta(X|z,\bar z) &= \int_0^\infty\frac{\d\omega}{\omega}\,\omega^\Delta\left(\e^{\im\omega q\cdot X}-\e^{-\im\omega q\cdot X}\right) \nonumber\\
    &= \int_0^\infty\frac{\d\omega}{\omega}\,\omega^\Delta\cdot 2\,\im\sin(\omega q\cdot X)= \frac{2\,\im\,\sgn(q\cdot X)}{|q\cdot X|^\Delta}\int_0^\infty\frac{\d\omega}{\omega}\,\omega^\Delta\sin\omega\nonumber\\
    &= 2\,\im\,\Gamma(\Delta)\,\sin\left(\frac{\pi\Delta}{2}\right)\frac{\sgn(q\cdot X)}{|q\cdot X|^{\Delta}}\,.
\end{align}
\endgroup
Notice how $\phi^\at_\Delta$ manifestly contains an extra factor of $\sgn(q\cdot X)$ on top of $\phi^\s_\Delta$, indicating its anti-symmetry under the antipodal map $q^\mu\mapsto-q^\mu$ of the celestial torus. Wavefunctions of higher spin states can now be found by dressing these scalar wavefunctions with the spin-raising tetrad vectors of \cite{Pasterski:2020pdk}.

\section{Light transforms on the celestial torus}\label{sec:LT_torus}

The 2-, 3- and 4-point celestial amplitudes of massless particles are distributional due to leftover delta functions coming from momentum conservation. For the same reason, they also come with step functions describing the channels in which a given scattering process can take place. We will later show that our $\Z_2$ eigenstates trade the step functions for signs that are much easier to handle. Similarly, to cure celestial amplitudes of their delta function supports, one turns to integral transforms like the light and shadow transforms. In this section, we provide a conformally covariant definition of the light transform that improves on some previous definitions that have been used in the literature on celestial CFT.

\subsection{Light transforms of boost eigenstates}\label{sec:LT_boost}

The light transform of an operator $\O^\veps_{h,\bar h}(z,\bar z)$ dual to the boost eigenstate $|\veps,z,\bar z,h,\bar h\ra$ is somewhat subtle to define. The sign $\veps$ changes as we move along the light cone of a point on the celestial torus. When we cross Poincar\'e patches, it jumps from $\veps$ to $-\veps$. This accounting of the $\Z_2$ covariance will modify our prescription for the light transform in contrast to the ones appearing in~\cite{Atanasov:2021cje,Sharma:2021gcz}.

\begin{figure}[t]
\centering
\vspace{-0.5em}
\begin{tikzpicture}[scale=2.1]
\definecolor{darkgreen}{rgb}{.0, 0.5, .1};
\draw[dashed, fill=magenta!20!white] (-1,-1) -- (-1,1) -- (1,1) -- (1,-1) -- (-1,-1);
\draw[fill=cyan!20!white] (0,-1)--(-1,0)--(0,1) --(1,0)--(0,-1);
\draw[->] (-1+.3+.4,0+.3-.4) -- (-1+.3+.5+.4,0+.3+.5-.4);
\draw[] (-1+.3+.5+.4,0+.3+.5-.4) -- (0+.3+.4,1+.3-.4);
\filldraw[black, thick] (-1+.3+.4,0+.3-.4)   circle (.075em);
\filldraw[black, thick] (0+.3+.4,1+.3-.4)  circle (.075em);
\node[] at (0+.3+.4,1+.3-.4-.25)  {$\sigma(z)$};
\node[] at (-1+.3+.4,0+.3-.4+.25)  {$z$};
\end{tikzpicture}
\caption{%Left: 
Contour for light transform extending to the next Poincar\'e patch.  The celestial torus is constructed by identifying along the dashed lines. $\sigma(z)$ denotes the point antipodal to $z$, having suppressed $\veps,\bar z$.
}
\label{fig:light_transform}
\end{figure}
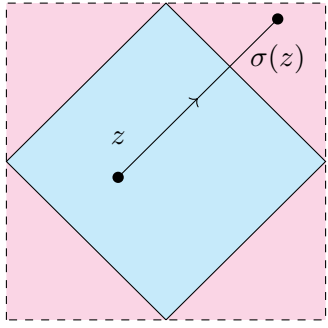

In terms of the conformal structure of $\d s^2=-\d z\,\d\bar z$ on either patch we see that lines of constant $z$ and $\bz$ are light-like.  Lifted to the celestial torus, the future light cone of a point $(\veps,z,\bar z)\in\Z_2\times\R^{1,1}$ consists of two null rays emanating from a point on the torus and re-converging at its antipodal point:
\begin{align}
    &\bullet\;\,\{(-\sgn(\al)\,\veps\,,\;z-\al^{-1}\,,\;\bar z)\in\Z_2\times\R^{1,1}\;|\;\al\in\R\}\,,\label{nullz}\\
    &\bullet\;\,\{(-\sgn(\al)\,\veps\,,\;z\,,\;\bar z-\al^{-1})\in\Z_2\times\R^{1,1}\;|\;\al\in\R\}\,.\label{nullzb}
\end{align}
The change in $\veps$ as we flow along these geodesics is found as follows. Consider the first null geodesic \eqref{nullz}. On the interval $\al\in(-\infty,0)$, we have $z<z-\al^{-1}<\infty$ and we remain in the original Poincar\'e patch. As we cross to $\al\in(0,\infty)$, the null geodesic enters the antipodal Poincar\'e patch and $\veps$ jumps to $-\veps$. On this patch, we cover the range $-\infty<z-\al^{-1}<z$. As $\al\to+\infty$, we reach the antipodal point
\be
\sigma(\veps,z,\bar z) = (-\veps,z,\bar z)\,.
\ee
This is depicted in figure \ref{fig:light_transform}. The second null geodesic \eqref{nullzb} can be dealt with analogously.

The light transform of an operator $\O^\veps_{h,\bar h}$ along the null rays \eqref{nullz} and \eqref{nullzb} are, respectively, defined by smearing these operators along these geodesics with the following choice of kernels \cite{Kravchuk:2018htv}:
\begin{align}
&\bL[\O^\veps_{h,\bar h}](z,\bar z) \vcentcolon= \int_{-\infty}^\infty\d\al\;|\al|^{-2h}\;\O^{-\sgn(\al)\veps}_{h,\bar h}(z-\al^{-1},\bar z)\,,\label{Lal}\\
&\bbL[\O^\veps_{h,\bar h}](z,\bar z) \vcentcolon= \int_{-\infty}^\infty\d\al\;|\al|^{-2\bar h}\;\O^{-\sgn(\al)\veps}_{h,\bar h}(z,\bar z-\al^{-1})\,.\label{Lbal}
\end{align}
The absolute values $|\al|^{-2h},|\al|^{-2\bar h}$ instead of just $(-\al)^{-2h},(-\al)^{-2\bar h}$ as in \cite{Kravchuk:2018htv} is consistent  with the absolute values in our definition \eqref{Otrans} of a conformal primary representation of $\SL(2,\R)\times\SL(2,\R)$.  In appendix~\ref{app:higherd} we show that the higher dimensional definition naturally reduces to this prescription and that these absolute values naturally appear due to the phase expected when we translate our operator across the Poincar\'e patches. When we light transform celestial amplitudes, having such absolute values will ensure single-valuedness of various integrands.

To study the behavior of the light transformed operators under the conformal group, we first rewrite the integrals with a more geometric parametrization. Changing the variable of integration to $w=z-\al^{-1}$ in \eqref{Lal} and $\bar w=\bar z-\al^{-1}$ in \eqref{Lbal} yields
\begin{align}
&\bL[\O^\veps_{h,\bar h}](z,\bar z) = \int_{-\infty}^\infty\frac{\d w}{|w-z|^{2-2h}}\;\O^{\sgn(w-z)\veps}_{h,\bar h}(w,\bar z)\,,\label{L}\\
&\bbL[\O^\veps_{h,\bar h}](z,\bar z) = \int_{-\infty}^\infty\frac{\d\bar w}{|\bar w-\bar z|^{2-2\bar h}}\;\O^{\sgn(\bar w-\bar z)\veps}_{h,\bar h}(z,\bar w)\,.\label{Lb}
\end{align}
Our construction ensures that the light transforms of $\O^\veps_{h,\bh}$ satisfy the following transformation laws under $\SL(2,\mathbb{R})\times\SL(2,\mathbb{R})$:
\be\label{LOtrans}
\bL[\O^{\veps'}_{h,\bar h}](z',\bz') = |cz+d|^{2(1-h)}\,|\bar c\bar z+\bar d|^{2\bar h}\,\bL[\O^\veps_{h,\bar h}](z,\bar z)\,,
\ee
and 
\be\label{bLOtrans}
\bbL[\O^{\veps'}_{h,\bar h}](z',\bz') = |cz+d|^{2h}\,|\bar c\bar z+\bar d|^{2(1-\bar h)}\,\bbL[\O^\veps_{h,\bar h}](z,\bar z)\,,
\ee
where $(z',\bz',\varepsilon')$ are given in \eqref{eq:primedcoord} and we've used that
\begin{align}\label{eq:sgnflip1}
    \sgn(w-z)\,\veps &\mapsto \sgn\!\left[(cz+d)(\bar c\bar z+\bar d)\right]\sgn\!\left[(cz+d)(cw+d)\right]\sgn(w-z)\,\veps\nonumber \\
   &= \sgn\!\left[(cw+d)(\bar c\bar z+\bar d)\right]\sgn(w-z)\,\veps\,,  \\
  \intertext{  and similarly}
    \sgn(\bar w-\bar z)\,\veps &\mapsto\sgn\!\left[(cz+d)(\bar c\bar w+\bar d)\right]\sgn(\bar w-\bar z)\,\veps\,, \label{eq:sgnflip2}
 \end{align}
in conjunction with the transformation law \eqref{Otrans} of $\O^\veps_{h,\bh}$. As discussed in \cite{Kravchuk:2018htv}, light transforms generate the action of the Weyl group of $\SL(2,\mathbb{R})\times\SL(2,\mathbb{R})$ on conformal primary representations. From \eqref{LOtrans} and \eqref{bLOtrans}, we can read off the Weyl reflections implemented by these transforms. These are summarized in table~\ref{tab:weights}.

\begin{table}[th]
\be
\begin{tabular}{c|c|c|c|c}
    $\O$ & $\Delta$ &$J$ & $h$& $\bar{h}$\\
    \hline
    $\bL \O$ & $1-J$ & $1-\Delta$& $1-h$& $\bar{h}$\\
       $\bbL \O$ & $1+J$&$-1+\Delta$ & $h$& $1-\bar{h}$\\
          $\bL\bbL \O$ & $2-\Delta$ & $-J$ & $1-h$& $1-\bar{h}$\\
\end{tabular}
\notag\ee
\caption{Weyl reflections induced by light transforms.\label{tab:weights}}
\end{table}

In what follows, we will use these adapted definitions to light transform celestial amplitudes. We will see that they prove crucial for the answers to come out conformally covariant in a manner that is consistent with the fact that Lorentz transformations mix different scattering channels in $\R^{2,2}$.

\subsection{\texorpdfstring{Light transforms of boost + $\Z_2$ eigenstates}{Light transforms of boost + Z2 eigenstates}}
\label{sec:ltZ2}

Finally, we set up the light transforms of our $\Z_2$ eigen-operators \eqref{Os} and \eqref{Oa}. Extending $\bL$ and $\bbL$ to linear combinations of $\O^\veps_{h,\bh}$ by linearity, we can apply \eqref{L} to \eqref{Os} to compute
\begin{align}
\bL[\O^\s_{h,\bh}](z,\bar z) &= \bL[\O^+_{h,\bh}](z,\bar z) + \bL[\O^-_{h,\bh}](z,\bar z)\nonumber\\
&= \int_z^\infty\frac{\d w}{|w-z|^{2-2h}}\left(\O^+_{h,\bh}(w,\bz)+\O^-_{h,\bh}(w,\bz)\right)\nonumber\\
&\hspace{3cm}+  \int_{-\infty}^z\frac{\d w}{|w-z|^{2-2h}}\left(\O^+_{h,\bh}(w,\bz)+\O^-_{h,\bh}(w,\bz)\right)\,.
\end{align}
In writing this, we broke the integration over $w\in(-\infty,\infty)$ to integrals over the ranges $w\in(-\infty,z)$ and $w\in(z,\infty)$ on which $w-z$ has a definite sign. Remembering \eqref{Os} then results in
\be\label{Ls}
\bL[\O^\s_{h,\bh}](z,\bar z) = \int_{-\infty}^\infty\frac{\d w}{|w-z|^{2-2h}}\;\O^\s_{h,\bh}(w,\bz)\,.
\ee
In the same way, we find
\be\label{Lbs}
\bbL[\O^\s_{h,\bar h}](z,\bar z) = \int_{-\infty}^\infty\frac{\d\bar w}{|\bar w-\bar z|^{2-2\bar h}}\;\O^\s_{h,\bar h}(z,\bar w)\,.
\ee
These were the light transforms considered in \cite{Hu:2022syq}, which are relatively simpler to compute than \eqref{L} and \eqref{Lb}. 

To go beyond the channel-symmetrized sector we need to take into account the $\at$-modes as well. The light transform of an antisymmetric primary reads
\begingroup
\allowdisplaybreaks
\begin{align}
\bL[\O^\at_{h,\bh}](z,\bar z) &= \bL[\O^+_{h,\bh}](z,\bar z) - \bL[\O^-_{h,\bh}](z,\bar z)\nonumber\\
&= \int_z^\infty\frac{\d w}{|w-z|^{2-2h}}\left(\O^+_{h,\bh}(w,\bz)-\O^-_{h,\bh}(w,\bz)\right)\nonumber\\
&\hspace{3cm}-  \int_{-\infty}^z\frac{\d w}{|w-z|^{2-2h}}\left(\O^+_{h,\bh}(w,\bz)-\O^-_{h,\bh}(w,\bz)\right)\,,
\end{align}
which is easily recognized to be
\be\label{La}
\bL[\O^\at_{h,\bh}](z,\bar z) = \int_{-\infty}^\infty\frac{\d w\;\sgn(w-z)}{|w-z|^{2-2h}}\;\O^\at_{h,\bh}(w,\bz)\,.
\ee
\endgroup
Similarly,
\be\label{Lba}
\bbL[\O^\at_{h,\bar h}](z,\bar z) = \int_{-\infty}^\infty\frac{\d\bar w\;\sgn(\bar w-\bar z)}{|\bar w-\bar z|^{2-2\bar h}}\;\O^\at_{h,\bar h}(z,\bar w)\,,
\ee
completing our list of light transforms. Putting these together we see that under conformal transformations \eqref{eq:primedcoord},
\begin{align}
    \bL[\O^\s_{h,\bh}](z',\bz') &= |cz+d|^{2(1-h)}\,|\bar c\bar z+\bar d|^{2\bar h}\,\bL[\O^\s_{h,\bar h}](z,\bar z)\,,\label{Lstrans}\\
    \bbL[\O^\s_{h,\bh}](z',\bz') &= |cz+d|^{2h}\,|\bar c\bar z+\bar d|^{2(1-\bar h)}\,\bbL[\O^\s_{h,\bar h}](z,\bar z)\,,\label{Lbstrans}\\
    \bL[\O^\at_{h,\bh}](z',\bz') &= \sgn\!\left[(cz+d)(\bar c\bar z+\bar d)\right]|cz+d|^{2(1-h)}\,|\bar c\bar z+\bar d|^{2\bar h}\,\bL[\O^\at_{h,\bar h}](z,\bar z)\,,\label{Latrans}\\
    \bbL[\O^\at_{h,\bh}](z',\bz') &= \sgn\!\left[(cz+d)(\bar c\bar z+\bar d)\right]|cz+d|^{2h}\,|\bar c\bar z+\bar d|^{2(1-\bar h)}\,\bbL[\O^\at_{h,\bar h}](z,\bar z)\,.\label{Lbatrans}
\end{align}
This shows that light transforms of symmetric primaries are again symmetric primaries, and similarly for the antisymmetric primaries. The set of transformations \eqref{Ls}, \eqref{Lbs}, \eqref{La} and \eqref{Lba} are two-dimensional lifts of certain intertwiners relating various $\SL(2,\R)$ representations to each other as discussed in chapter VII of \cite{Gelfand:105396}.

\subsection{Light transformed wavefunctions}

We conclude this section by calculating light transforms of the scalar wavefunctions \eqref{phiS} and \eqref{phiA} in the $\Z_2$ eigenbasis. Our result is a slight improvement on the previously discovered light transform wavefunctions of \cite{Atanasov:2021cje,Sharma:2021gcz} which were only dilatation covariant; the improved wavefunctions will be fully conformally covariant. The case of spinning particles can be dealt with analogously, though we do not touch this here for sake of brevity.

The $\bL$ transform of the symmetric primary \eqref{phiS} is given by
\be\label{Lscal}
\bL[\phi^\s_\Delta](X|z,\bar z) = 2\,\Gamma(\Delta)\,\cos\left(\frac{\pi\Delta}{2}\right)\int_{-\infty}^\infty\frac{\d w}{|w-z|^{2-\Delta}}\;\frac{1}{|q(w,\bar z)\cdot X|^{\Delta}}\,.
\ee
Naively, this integral could give a divergent result because of possible singularities at $w=z$ or $q(w,\bz)\cdot X=0$.\footnote{
In $(1,3)$ signature the singularities in the wavefunctions and the shadow transforms were avoided with an appropriate analytic continuation of the time coordinate $X^0\mapsto X^0_\pm=X^0\mp i\varepsilon$ for the in and out wavefunctions~\cite{Pasterski:2017kqt,Donnay:2020guq}.  In $(2,2)$ signature the mixing of $\varepsilon$ under global conformal transformations, and additional absolute value signs make it easier to address this problem from scratch for scattering in Klein space.  However, it is natural to expect we can merge these two methods of regulating when we consider analytically continuing between signatures which will involve complexified representations of ${\rm SL}(2,\mathbb{C})\times {\rm SL}(2,\mathbb{C})$. } It suffices to regularize the $w=z$ singularity via a small exponent $\eps$:
\be\label{Lscal1}
\lim_{\eps\to0}\int_{-\infty}^\infty\frac{\d w}{|w-z|^{2-\Delta-\eps}}\;\frac{1}{|q(w,\bar z)\cdot X|^{\Delta}}\,.
\ee
Since the parametrization $q^\mu(z,\bz)$ given in \eqref{nullp} is linear in $z$, we can Taylor expand to write
\be
q(w,\bz)\cdot X = q(z,\bz)\cdot X + (w-z)\,\p_zq(z,\bz)\cdot X\,.
\ee
Motivated from this, perform the integral substitution
\be\label{wsub}
w\mapsto z - \frac{q\cdot X}{\p_zq\cdot X}\,w\,,
\ee
where $q^\mu$ without any arguments will always be short for $q^\mu(z,\bz)$ unless stated otherwise. This reduces \eqref{Lscal1} to
\be\label{Lscal2}
\lim_{\eps\to0}\frac{|q\cdot X|^{\eps-1}}{|\p_zq\cdot X|^{\Delta-1+\eps}}\int_{-\infty}^\infty\d w\;|w|^{\Delta-2+\eps}\,|1-w|^{-\Delta}\,.
\ee
Performing the $w$ integral via a principal value prescription by breaking it over the three ranges $(-\infty,0)$, $(0,1)$ and $(1,\infty)$ yields
\be
\int_{-\infty}^\infty\d w\;|w|^{\Delta-2+\eps}\,|1-w|^{-\Delta} = \frac{\pi\eps}{\Delta-1}\,\tan\left(\frac{\pi\Delta}{2}\right) + O(\eps^2)\,.
\ee
Using the identity
\be\label{deltaid}
\lim_{\eps\to0}\eps\,|x|^{\eps-1} = 2\,\delta(x)\,,
\ee
we finally arrive at the wavefunction
\be\label{Lphis}
\bL[\phi^\s_\Delta](X|z,\bz) = 4\pi\,\Gamma(\Delta-1)\sin\left(\frac{\pi\Delta}{2}\right)\frac{\delta(q\cdot X)}{|\p_z q\cdot X|^{\Delta-1}}\,.
\ee
The $\bbL$ transform of \eqref{phiS} is found along the same lines,
\be\label{Lbphis}
\bbL[\phi^\s_\Delta](X|z,\bz) = 4\pi\,\Gamma(\Delta-1)\sin\left(\frac{\pi\Delta}{2}\right)\frac{\delta(q\cdot X)}{|\p_{\bz} q\cdot X|^{\Delta-1}}\,.
\ee
These give the light transforms of the symmetric primary.

These wavefunctions are contact terms with a singularity at $q\cdot X=0$. Fortunately, this is exactly the kind of contact term that behaves conformally covariantly~\cite{Pasterski:2020pdk}. Under the $\SL(2,\R)\times\SL(2,\R)$ transformation $z\mapsto(az+b)/(cz+d)$, $\bar z\mapsto(\bar a\bar z+\bar b)/(\bar c\bar z+\bar d)$ and the associated Lorentz transformation of $X^\mu$, we recall that $q\cdot X\mapsto q\cdot X/(cz+d)(\bar c\bar z+\bar d)$. Thus, on the support of $q\cdot X=0$, we see that $\p_zq\cdot X$ transforms as
\be
\p_zq\cdot X\mapsto (cz+d)^2\,\p_z\left(\frac{q\cdot X}{(cz+d)(\bar c\bar z+\bar d)}\right) = \frac{cz+d}{\bar c\bar z+\bar d}\;\p_zq\cdot X\,.
\ee
As a result, the wavefunction \eqref{Lphis} transforms into
\be
\bL[\phi^\s_\Delta]\!\left(X\,\biggr|\,\frac{az+b}{cz+d},\frac{\bar a\bar z+\bar b}{\bar c\bar z+\bar d}\right) = |cz+d|^{2-\Delta}\,|\bar c\bar z+\bar d|^{\Delta}\;\bL[\phi^\s_\Delta](X|z,\bz)\,,
\ee
i.e., it is a symmetric primary with weights $(1-h,\bh)=(2-\Delta,\Delta)$ as expected. It has conformal dimension $1$ and spin $1-\Delta$. Similarly, $\bbL[\phi^\s_\Delta]$ has weights $(h,1-\bh) = (\Delta,2-\Delta)$, i.e., it has conformal dimension $1$ and spin $\Delta-1$. 

The $\bL$ transform of the antisymmetric primary scalar wavefunction \eqref{phiA} has the same flavor. One starts with the regulated integral
\be
\bL[\phi^\at_\Delta](X|z,\bz) = 2\,\im\,\Gamma(\Delta)\,\sin\left(\frac{\pi\Delta}{2}\right)\lim_{\eps\to0}\int_{-\infty}^\infty\frac{\d w\;\sgn(w-z)}{|w-z|^{2-\Delta-\eps}}\;\frac{\sgn(q(w,\bz)\cdot X)}{|q(w,\bz)\cdot X|^{\Delta}}\,.
\ee
Substituting \eqref{wsub} results in
\begin{align}
    \lim_{\eps\to0}&\int_{-\infty}^\infty\frac{\d w\;\sgn(w-z)}{|w-z|^{2-\Delta-\eps}}\;\frac{\sgn(q(w,\bz)\cdot X)}{|q(w,\bz)\cdot X|^{\Delta}}\nonumber\\
    &= \sgn(\p_zq\cdot X)\,\lim_{\eps\to0}\frac{|q\cdot X|^{\eps-1}}{|\p_zq\cdot X|^{\Delta-1+\eps}}\int_{-\infty}^\infty\d w\;\sgn\!\left[w(1-w)\right]|w|^{\Delta-2+\eps}\,|1-w|^{-\Delta}\nonumber\\
    &= \sgn(\p_zq\cdot X)\,\lim_{\eps\to0}\frac{|q\cdot X|^{\eps-1}}{|\p_zq\cdot X|^{\Delta-1+\eps}}\left[\frac{\pi\eps}{\Delta-1}\,\cot\left(\frac{\pi\Delta}{2}\right)+O(\eps^2)\right]\,.
\end{align}
Taking the limit using \eqref{deltaid}, we find the light transform
\be
\bL[\phi^\at_\Delta](X|z,\bz) = 4\pi\im\,\Gamma(\Delta-1)\cos\left(\frac{\pi\Delta}{2}\right)\frac{\sgn(\p_zq\cdot X)\,\delta(q\cdot X)}{|\p_z q\cdot X|^{\Delta-1}}\,.
\ee
Its $\bbL$ counterpart is
\be
\bbL[\phi^\at_\Delta](X|z,\bz) = 4\pi\im\,\Gamma(\Delta-1)\cos\left(\frac{\pi\Delta}{2}\right)\frac{\sgn(\p_\bz q\cdot X)\,\delta(q\cdot X)}{|\p_\bz q\cdot X|^{\Delta-1}}\,.
\ee
Once again, these are easily verified to satisfy the same transformation laws \eqref{Latrans}-\eqref{Lbatrans} as their dual operators.

\subsection{Invertibility and the ambidextrous basis}

The Mellin transformed wavefunctions studied in section~\ref{sec:Klein_basis} are the natural analytic continuation of the basis of positive and negative frequency modes examined in~\cite{Pasterski:2017kqt}, modulo some extra care with the $\varepsilon$ labels to isolate irreducible representations of ${\rm SL}( 2,\mathbb{R})\times {\rm SL}(2,\mathbb{R})$. In order for us to use our light transformed states as an alternative celestial basis we need not only the correct covariance, but also need to check that our transform is invertible. 

Keeping track of the weights in table~\ref{tab:weights}, we see that the square of our light transforms $\bL^2$ and $\bbL^2$ should be intertwiners that take us back to conformal primaries with the same dimensions. Taking into account the choice of normalization, the fact that the Weyl reflections square to one would indicate that our light transforms should square to a constant function of the data labeling our irrep.  However, if one lifts the action of the global conformal group to the universal cover as in~\cite{Kravchuk:2018htv}, there can be a nontrivial dependence on the operators ${\cal T},\bar{\cal T}$ translating us from a point to its image in the next Poincar\'e patch. One can still formally renormalize the light transforms so that they square to one, but their action on operators will then involve images on each Poincar\'e patch. 

For the celestial torus, we want not the universal cover but rather an identification of it such that ${\cal T}^2=\bar{\cal T}^2=1$. Demanding that quantities are single valued on the torus results in  restrictions on the operator spectrum~\cite{Atanasov:2021oyu}. By contrast, here we have set up our light transformed operators to be consistently defined on the $\Z_2$ quotient of the celestial torus, and defined our light transformed states to be genuine irreps of ${\rm SL}(2,\mathbb{R})\times {\rm SL}(2,\mathbb{R})$ on its conformal compactification. So we are free to let our conformal weights range over $h,\bh\in\C^2$.

In practice, we can use a regularization of the light ray integrals as in section \ref{sec:ltZ2} to show that light transforms are invertible (at least for generic weights). For instance, if $\O_{h,\bh}^\s(z,\bz)$ is a symmetric primary operator of conformal weight $h$ in $z$, we can study the integral
\be
\bL^2[\O_{h,\bh}^\s](z,\bz) = \int_{\R^2}\d u\,\d w\,|u-z|^{-2h}|w-u|^{2h-2}\,\O_{h,\bh}^\s(w,\bz)\,.
\ee
The $u$ integral is performed in the same manner as the $w$ integral \eqref{Lscal2}, leading to the identity
\be
\int_{-\infty}^\infty\d u\,|u-z|^{-2h}|w-u|^{2h-2} = \cN^\s(h)\,\cN^\s(1-h)\,\delta(w-z)\,,
\ee
written in terms of a prefactor
\be
\cN^\s(h) = \frac{\Gamma(\frac{1}{2}+h)\,\Gamma(\frac{1}{2}-h)}{\Gamma(2h)} = \frac{\pi\sec(\pi h)}{\Gamma(2h)}
\ee
that can be used as a normalization factor for the light transform as long as $\frac12\pm h$ is not an integer. As a result,
\be
\bL^2[\O_{h,\bh}^\s](z,\bz) = \cN^\s(h)\,\cN^\s(1-h)\,\O_{h,\bh}^\s(z,\bz)\,,\qquad h\not\in\Z+\frac{1}{2}\,.
\ee
In this case, $\cN^\s(h)^{-1}\,\bL[\O_{h,\bh}^\s]$ becomes an involution, hence an invertible basis change.

For an antisymmetric primary $\O_{h,\bh}^\sa(z,\bz)$, we can similarly calculate
\be
\begin{split}
    \bL^2[\O_{h,\bh}^\sa](z,\bz) &= \int_{\R^2}\d u\,\d w\,\sgn[(u-z)(w-u)]\,|u-z|^{-2h}|w-u|^{2h-2}\,\O_{h,\bh}^\sa(w,\bz)\\
    &= \cN^\sa(h)\,\cN^\sa(1-h)\,\O_{h,\bh}^\sa(z,\bz)\,,\qquad h\not\in\Z\,,
\end{split}
\ee
with
\be
\cN^\sa(h) = \frac{\im\,\Gamma(h)\,\Gamma(1-h)}{\Gamma(2h)}= \frac{\pi\im\csc(\pi h)}{\Gamma(2h)}
\ee
being the analogous prefactor appearing in this case. So  $\cN^\sa(h)^{-1}\,\bL[\O_{h,\bh}^\sa]$ is an involution for the antisymmetric primaries. In the same way, $\cN^\s(\bh)^{-1}\,\bbL[\O_{h,\bh}^\s]$ and $\cN^\sa(\bh)^{-1}\,\bbL[\O_{h,\bh}^\sa]$ are also involutions and provide invertible basis changes for generic $\bh$. The resonant cases $h,\bh\in\Z+\frac12$ or $h,\bh\in\Z$ are more delicate and have been analyzed (for $\SL(2,\R)$ representation theory) in fuller detail in \cite{Gelfand:105396}. They have also been studied from a CCFT perspective in \cite{Banerjee:2022hgc}, and it is straightforward to adapt their analysis to symmetric and antisymmetric primaries if needed. 

In what follows, we will use a judicious choice of such light transforms as our scattering basis. The motivation for doing this comes from the fact that many low multiplicity celestial amplitudes of massless particles have distributional support in their kinematics (see for instance \cite{Pasterski:2017ylz}). In \cite{Sharma:2021gcz}, it was argued that in a scattering basis composed of $\bL$-transformed positive helicity boost eigenstates and $\bbL$-transformed negative helicity boost eigenstates, all gluon and graviton amplitudes become non-distributional in the celestial coordinates $z,\bz$ and take the form of standard CFT correlators. This was referred to as an \emph{ambidextrous basis}. That reference provided evidence for this proposal in the form of computing the light transforms of 2 and 3-point celestial amplitudes in the simplest crossing channels.
Using the systematic analysis of light transforms presented in sections \ref{sec:LT_boost} and \ref{sec:ltZ2}, we will now generalize this computation and confirm that they are non-distributional in all crossing channels.

\section{Light transformed celestial amplitudes}\label{sec:LT_amplitudes}

Now that we have defined our light transform prescription and checked that it is both invertible and appropriately covariant, we can begin to examine celestial correlators in this basis.  Here we present the generic form of $n\le 4$ point ambidextrous celestial correlators corresponding to massless external states. We will focus on the case where the spins are nonzero and integer valued, for simplicity. As compared to the usual Mellin-transform celestial dictionary~\cite{Pasterski:2017ylz,Arkani-Hamed:2012zlh}, we will see that this basis remedies the distributional nature of the low point correlators~\cite{Sharma:2021gcz}, while being manifestly Lorentz covariant for all channels.

We will use the following shorthand to simplify our equations.  We will denote the $n$-point momentum space amplitude (color-stripped in the case of gluons), including the momentum conserving delta function, by 
\be
A(1^{J_1}2^{J_2}\cdots n^{J_n}) \equiv A(p_1,J_1;p_2,J_2;\cdots; p_n,J_n)
\ee
for $p_i^\mu$ given by~\eqref{pi}. Similarly, we will denote the delta function stripped amplitude by $A[1^{J_1}2^{J_2}\cdots n^{J_n}]$, so that
\be
A(1^{J_1}2^{J_2}\cdots n^{J_n}) = A[1^{J_1}2^{J_2}\cdots n^{J_n}]\,\delta^4\bigg(\sum_{i=1}^np_i\bigg)\,.
\ee
The signs of the helicities $\sgn(J_i)=\pm$ determine which light transforms to perform under the ambidextrous prescription.  When the notation becomes cumbersome, we will use the following shorthand to indicate the light transformed celestial amplitudes:
\be\label{eq:shorthand}
L(1^-\cdots n^+)\equiv\langle  \bbL[\O^{\veps_1}_{h_1,\bh_1}](z_1,\bz_1)\cdots \bL[\O^{\veps_n}_{h_n,\bh_n}](z_n,\bz_n)\rangle\,,
\ee
and so forth, suppressing labels where there should be no source of confusion. Here, $\bL,\bbL$ denote the conformally covariant light transforms of boost eigenstates defined in section \ref{sec:LT_boost}. Similarly, when computing light transforms of symmetric and antisymmetric primaries as in section \ref{sec:ltZ2}, we will write
\be\label{eq:shorthandsa}
L(1_\s^-\cdots n_\at^+)\equiv\langle  \bbL[\O^{\s}_{h_1,\bh_1}](z_1,\bz_1)\cdots \bL[\O^{\at}_{h_n,\bh_n}](z_n,\bz_n)\rangle\,,
\ee
etc.

\paragraph{Spinor-helicity conventions.} For massless momenta, the momentum space amplitudes are conveniently expressed in terms of spinor-helicity variables. We use the following spinor-helicity conventions (see \cite{Elvang:2013cua} for a textbook treatment). In $\R^{2,2}$, the components of a null momentum $p^\mu$ of the form \eqref{nullp} can be arranged into a matrix of determinant $0$ as follows:
\be
p^{\al\dal} = \frac{1}{\sqrt2}\begin{pmatrix}p^0+p^3&&p^1+p^2\\p^1-p^2&&p^0-p^3\end{pmatrix} = \veps\,\omega\begin{pmatrix}1&&\bar z\\z&&z\bar z\end{pmatrix} = \veps\,\omega\begin{pmatrix}1\\z\end{pmatrix}\otimes\begin{pmatrix}1&&\bz\end{pmatrix}\,,
\ee
where $\al=0,1$, $\dal=\dot0,\dot1$ are 2-component spinor indices that transform in the fundamental of $\SL(2,\R)$ under a Lorentz transformation. They are raised or lowered using $2\times2$ Levi Civita symbols $\eps^{\al\beta}$, $\eps^{\dal\dot\beta}$, etc. Motivated by the above tensor product decomposition, spinor-helicity variables $\lambda^\al,\bar\lambda^{\dal}\in\R^2$ are defined as a pair of independent 2-component spinors satisfying
\be
p^{\al\dal} = \lambda^\al\bar\lambda^{\dal}\,.
\ee
This defines $\lambda^\al,\bar\lambda^{\dal}$ only up to little group scaling
\be
\lambda^\al\sim s\,\lambda^\al\,,\qquad\bar\lambda^{\dal}\sim s^{-1}\bar\lambda^{\dal}\,,\qquad s\in\R^*\,.
\ee
The associated spinor-helicity eigenstate is denoted $|\lambda,\bar\lambda,J\ra$ and obeys the standard transformation law
\be\label{lgsc}
|s\lambda,s^{-1}\bar\lambda,J\ra = s^{-2J}\,|\lambda,\bar\lambda,J\ra\qquad\forall\; s\in\R^*
\ee
under little group scalings.

The scattering amplitudes of states $|\lambda_i,\bar\lambda_i,J_i\ra$, $i=1,\dots,n$, are linear functionals on these states, so transform in the same way under little group scalings:
\be
\begin{split}
    A(p_i,J_i) &\equiv A(\lambda_i,\bar\lambda_i,J_i)\,,\\
    A(s_i\lambda_i,s_i^{-1}\bar\lambda_i,J_i) &= \prod_{k=1}^ns_k^{-2J_k}\,A(\lambda_i,\bar\lambda_i,J_i)\,.
\end{split}
\ee
This often helps us constrain their structure, especially at 3-points. Lorentz invariance dictates that these amplitudes are functions of the $\SL(2,\R)$ invariant contractions:
\be
\la i\,j\ra = \eps_{\beta\al}\lambda_i^\al\lambda_j^\beta\,,\qquad[i\,j] = \eps_{\dot\beta\dal}\bar\lambda_i^{\dal}\bar\lambda_j^{\dot\beta}\,.
\ee
In this section, we will usually fix little group scalings by setting
\be\label{lgfix}
\lambda_i^\al = \sqrt{\omega_i}\begin{pmatrix}1\\z_i\end{pmatrix}\,,\qquad\bar\lambda_i^{\dal} = \veps_i\sqrt{\omega_i}\begin{pmatrix}1\\\bz_i\end{pmatrix}\,.
\ee
But we will also encounter other little group fixings in the next section.

\subsection{Two-point}

The 2-point amplitude captures single particle states propagating freely through the bulk in  $\mathbb{R}^{1,3}$. In this signature, the Klein-Gordon inner product is positive definite on the subspace of on-shell wavefunctions with positive energy which, in turn, are used to define the single particle Hilbert space with inner product\footnote{ Since the single particle mode operators annihilate either the $in$ or $out$ vacuum the Celestial one point functions relevant here will vanish. 
}
\be
\langle p_1,a|p_2,b\rangle=(2\pi)^3\delta_{ab}\delta^{(3)}(\vec{p}_1-\vec{p}_2)
\ee
where $a,b$ are internal labels for the particle species and spin. In $\mathbb{R}^{2,2}$ we can still use the analogous inner products on wavefunctions but cannot make the restriction to positive frequency solutions. While this raises some questions about what to make of the Hilbert space, the extrapolate dictionary~\cite{Pasterski:2021dqe} helps point us to how to handle both signatures. Namely, for the $\mathcal{S}$-matrix in $\mathbb{R}^{1,3}$ we prepare the $in$ versus $out$ states by smearing operators on past and future components of the conformal boundary, respectively. In $\mathbb{R}^{2,2}$, there is only a single null boundary component, turning the ${\cal S}$-matrix into a ${\cal S}$-vector~\cite{Atanasov:2021oyu}.  This is no problem for our boundary-correlator interpretation.

We can always choose our operator normalization, so the only invariant content of the two point function is the particle species and data of the corresponding irreducible representation of Poincar\'e (in this case $m=0$ and helicity $\ell=\veps J$). Keeping track of the channel labels and Mellin transforming to boost eigenstates we have
\be\badat{3}\label{2point}
\scalemath{.95}{
\langle \O^{\veps_1}_{h_1,\bh_1}(z_1,\bz_1)\O^{\veps_2}_{h_2,\bh_2}(z_2,\bz_2) \rangle
=(2\pi)^3\,\Theta\Bigl(-\frac{\veps_1}{\veps_2}\Bigr)\, \boldsymbol{\delta}(\im(\Delta_1+\Delta_2-2))\,\delta_{J_1,-J_2}\,\delta^2(z_1-z_2)\,,}
\eadat
\ee
for particles of the same species, and zero otherwise. The Heaviside step function $\Theta$ is a heavy handed notation for $\delta_{\veps_1,-\veps_2}$, but we will see that it will be natural from the viewpoint of light transforms. The appearance of $\delta_{J_1,-J_2}$ arises from the fact that we label the celestial amplitudes with their outgoing helicity.  The distribution $\boldsymbol{\delta}$ was defined in~\cite{Donnay:2020guq} to allow us to analytically continue off the Principal series, which is important for understanding conformally soft modes~\cite{Donnay:2018neh}. It reduces to the ordinary Dirac delta function in $\delta(\lambda_1+\lambda_2)$ for weights on the principal series $\Delta_i=1+\im\lambda_i$.

Now if we $z$-smear the $+$ helicity and $\bz$-smear the $-$ helicity states using our ambidextrous light transform prescription we find (invoking our shorthand notation~\eqref{eq:shorthand}) 
\be\label{2point4}\badat{3}
L(1^- 2^+)
&=\Theta\left(\frac{\veps_1\bz_{12}}{\veps_2z_{12}}\right)\frac{(2\pi)^3\boldsymbol{\delta}(2\im\sh_{12})}{|z_{12}|^{2\sh_1} |\bz_{12}|^{2\bar{\sh}_1}},
\eadat\ee
where $1^-,2^+$ are particles of the same spin but opposite helicity, and the sans serif weights
\be
(\sh_i,\bar\sh_i) = \begin{cases}
(1-h_i,\bh_i)\qquad\text{if }J_i=h_i-\bh_i>0\\
(h_i,1-\bh_i)\qquad\text{if }J_i=h_i-\bh_i<0
\end{cases}
\ee
denote weights of the ambidextrously light transformed operators, i.e., Weyl reflected with respect to those of~\eqref{2point}, as in table~\ref{tab:weights}. We have also used $\Delta_1+\Delta_2-2=2(\sh_1-\sh_2)$ (valid due to $J_1+J_2=0$) and introduced the abbreviations
\be
\sh_{ij}\equiv\sh_i-\sh_j\,,\qquad z_{ij} \equiv z_{i}-z_j\,,\qquad \bar z_{ij} \equiv \bar z_{i}-\bar z_j
\ee
for the weight differences and operator separations.

We see that this transformation puts the amplitude to the standard power law form of a 2-point CFT correlator. Notice that the step function is conformally invariant under the transformations \eqref{LOtrans}-\eqref{bLOtrans}.  We can further diagonalize the two point functions by going to the symmetric and antisymmetric combinations introduced in section \ref{sec:ltZ2}, finding
\begingroup
\allowdisplaybreaks
\begin{align}
L(1^-_\s2^+_\at) &= L(1^-_\at2^-_\s)= 0\,,\\
L(1^-_\s2^+_\s)&=\frac{(2\pi)^3\,\boldsymbol{\delta}(\im\sh_{12})}{|z_{12}|^{2\sh_1} |\bz_{12}|^{2\bar{\sh}_1}}\,,\label{osos}\\
L(1^-_\at2^+_\at)&=\mathrm{sgn}(z_{12}\bar{z}_{12})\,\frac{(2\pi)^3\,\belta(\im\sh_{12})}{|z_{12}|^{2\sh_1} |\bz_{12}|^{2\bar{\sh}_1}}\, \,.\label{oaoa}
\end{align}
\endgroup
In particular, these diagonalized, non-distributional 2-point functions are enough to generate all the celestial correlators of free theories in the bulk.

\begin{figure}[t]
\centering
\vspace{-0.5em}
\begin{tikzpicture}[scale=2.1]
\filldraw[black, thick] (0,0)   circle (.075em) node[right] {$z_2,\bz_2$};
\draw[dashed] (-1,-1) --(1,1);
\draw[dashed] (-1,1) --(1,-1);
\node[] at (0,1) {$\veps_1=\veps_2$};
\node[] at (0,-1.1) {$\veps_1=\veps_2$};
\node[] at (1.1,0) {$\veps_1=-\veps_2$};
\node[] at (-1,0) {$\veps_1=-\veps_2$};
\end{tikzpicture}
\caption{
Let's keep track of the support for different channels. Here is the two point case indicating which regions defined by the $(1,1)$ lightcones emanating from operator 2
support various channels for operator 1.  
}
\label{fig:2pt_causal}
\end{figure}
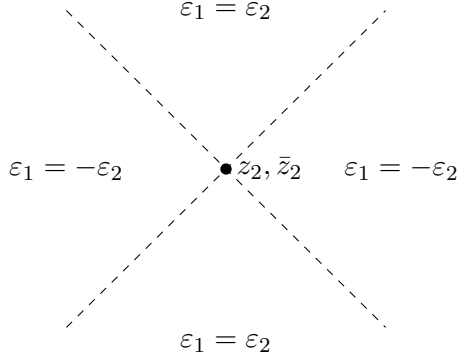

Unlike \eqref{2point}, the light transformed celestial amplitude \eqref{2point4} has a kinematics dependent support on the crossing channels: $\veps_1\bz_{12}/\veps_2z_{12}>0$. This is depicted in figure \ref{fig:2pt_causal}. In contrast, the symmetric and antisymmetric combinations \eqref{osos} and \eqref{oaoa} are supported everywhere. Thus, we learn two lessons: the light ray integrals smear away the delta function support, and passing to $\Z_2$-eigenstates smears out the step function support. The results go a long way to behaving like standard CFT$_2$ correlators.

\subsection{Three-point}

Poincar\'e invariance, little group covariance, and locality imply the momentum conservation stripped  three-point amplitude takes the form~\cite{Elvang:2013cua}
 \begin{equation}\label{3pt}
\badat{3}
    A[1^{J_1}2^{J_2}3^{J_3}] = \begin{cases} 
      c_{123}\,\langle 1\,2\rangle^{J_3-J_1-J_2}\langle 3\,2\rangle^{J_1-J_2-J_3}\langle 1\,3\rangle^{J_2-J_3-J_1},&J_1+J_2+J_3<0\\
     \tilde c_{123}\,[1\,2]^{J_1+J_2-J_3}[3\,2]^{J_2+J_3-J_1}[1\,3]^{J_3+J_1-J_2}, &J_1+J_2+J_3>0
    \end{cases}
    \eadat
\end{equation}
where $J_i$ are the outgoing helicities. In our conventions \eqref{lgfix},
\be
\la i\,j\ra = \sqrt{\omega_i\omega_j}\,z_{ij}\,,\qquad[i\,j] = \veps_i\veps_j\sqrt{\omega_i\omega_j}\,\bz_{ij}\,.
\ee
Because of this, the conditions on the spins in~\eqref{3pt} imply that these amplitudes vanish on either the $z_{ij}=0$ or $\bz_{ij}=0$ locus, and have non-trivial support on the other.

When changing from the momentum to the celestial basis, the Mellin transform splits into an overall scale integral which gives a delta function in the weights and a simplex integral over the $\sigma_i$.  For this massless three point amplitude to have non-trivial support when we restore the momentum conserving delta function, we need to work in $\mathbb{R}^{2,2}$. Upon performing this analytic continuation there are two loci of support of that will be of interest.  When $z_{ij}\neq0$ we have
\begin{align}
&\delta^{4}\Bigl(\sum\limits_i\veps_i\sigma_iq_i\Bigr)\delta\Bigl(\sum\limits_i\sigma_i-1\Bigr)\,\Bigr|_{z_{ij}\neq 0}\nonumber\\
&= \frac{\delta(\bz_{12})\delta(\bz_{13})}{4\sigma_1\sigma_2\sigma_3  D_3^2}
\,
\delta\Bigl(\sigma_1-\frac{z_{23}}{
D_3}\Bigr)\delta\Bigl(\sigma_2+\veps_1\veps_2\frac{z_{13}}{ D_3}\Bigr)
\delta\Bigl(\sigma_3-\veps_1\veps_3\frac{z_{12}}{D_3}\Bigr)\,,\nonumber\\
&\equiv\frac{\delta(\bz_{12})\delta(\bz_{13})}{4\sigma_1\sigma_2\sigma_3  D_3^2}
\,
\prod_{i=1}^3 \delta(\sigma_i -\sigma_{*i})\,,
\end{align}
where $\sigma_i= \omega_i/\sum_j\omega_j$ and the denominator is
\be\badat{3}
D_3 = (1-{\veps_1}{\veps_2})z_{13}+({\veps_1}{\veps_3}-1)z_{12} \,.
\eadat\ee
Exchanging $z_{ij}\leftrightarrow \bz_{ij}$ gives the second locus of support.  These are relevant for the MHV and anti-MHV amplitudes, respectively. 

Upon performing the simplex integrals, we find that different kinematical crossings have disjoint support. Namely, $\sigma_{*i}\in[0,1]$ when:
\be\label{ind1}
\prod_{i=1}^3
\mathbf{1}_{[0,1]}(\sigma_{*i})\neq0:~~
\begin{array}{llll}
a)~~~ 12 \ce{<-->} 3~~~&\Rightarrow z_1<z_3<z_2  ~\mathrm{or}~ z_2<z_3<z_1&\\
b)~~~ 13 \ce{<-->} 2~~~&\Rightarrow z_1<z_2<z_3  ~\mathrm{or}~ z_3<z_2<z_1&\\
c)~~~ 23 \ce{<-->} 1~~~&\Rightarrow z_3<z_1<z_2  ~\mathrm{or}~ z_2<z_1<z_3&.\\
\end{array}
\ee
Our knowledge of the general 3-point amplitude in momentum space~\eqref{3pt} lets us extend the gluon examples of~\cite{Pasterski:2017ylz}, to arbitrary spin.  Without loss of generality, we will focus on the MHV case here, for which the 3-point Mellin amplitude becomes\footnote{As we will see in the 4-point case below, tree level amplitudes have distributional support in the conformal dimensions.  However, the full celestial amplitude is expected to be analytic.  It is tempting to generalize the three point function in a manner that allows running couplings, which can similarly turn the distribution in the weights into an analytic function in the weights via $c_{123}(\bar h)=\int_0^\infty d{\mathfrak{s}} {\mathfrak{s}}^{\bar h-3} c_{123}({\mathfrak{s}})$.  This can be recast in terms of the beta function.
}
\begin{multline}\label{3ptfix}
\langle\O^{\veps_1}_{h_1,\bh_1}(z_1,\bz_1)\O^{\veps_2}_{h_2,\bh_2}(z_2,\bz_2)\O^{\veps_3}_{h_3,\bh_3}(z_3,\bz_3)\rangle\\
=\pi c_{123}\,\mathrm{sgn}(z_{12}z_{13}z_{23})^{s_1+s_2+s_3}\,\frac{\boldsymbol{\delta}(\im(\bar{h}-2))\delta(\bz_{12})\delta(\bz_{13})\prod_{i=1}^3\mathbf{1}_{[0,1]}(\sigma_{*i})}{|z_{12}|^{h_1+h_2-h_3}|z_{23}|^{h_2+h_3-h_1}|z_{31}|^{h_1+h_3-h_2}}
\end{multline}
where $\bar{h}=\sum_i \bar{h}_i$ the support of the indicator function is given in~\eqref{ind1}, and we have assumed that all of the $J_i$ are integers so that $(-1)^{\pm J_i}=(-1)^{s_i}$ for $s_i=|J_i|$.  We note that $\mathrm{sgn}(z_{12}z_{13}z_{23})$ is $\SL(2,\R)$ invariant. The theory dependence is captured by the constant $c_{123}$ determined by the three point coupling.

\begin{figure}[t]
\centering
\vspace{-0.5em}
\begin{tikzpicture}[scale=1.5]
 \draw[] (0,-2)--(-2,0)--(0,2) --(2,0)--(0,-2);
\filldraw[black, thick] (-.25,.25)   circle (.075em) node[right] {$1$};
\filldraw[black, thick] (-.25+.5,-.25-.5)   circle (.075em) node[right] {$3$};
\filldraw[black, thick] (-.25-.2,-.25-.5-.2)   circle (.075em) node[right] {$2$};
\draw[thick] (-1-.25,-1+.25) node[below left] {$\{\at,\s\}$} --(1-.25,1+.25) ;
\draw[thick] (-1+.25,-1-.25) node[below left] {$\{\at,\s\}$} --(1+.25,1-.25);
\draw[thick] (1-.25,-1-.25) node[below right] {$\{\at,\s\}$} --(-1-.25,1-.25);
\draw[->,thick] (0+.15+.3,-2-.15+.3) -- node[below right ] {$\bz$} (.25+.15+.3,-2+.25-.15+.3) ;
\draw[->,thick] (0-.15-.3,-2-.15+.3) -- node[below left] {$z$} (-.25-.15-.3,-2+.25-.15+.3) ;
\end{tikzpicture}
\caption{
The causal dependence of the 3-point correlators is controlled by the linking of the null rays over which the light transforms are performed.  Because the signs of the helicity determine which lightcone direction is integrated over, different assignments of `$\at$' and `$\s$' can lead to 3-point correlators with OPE coefficients that are not symmetric under exchanging the puncture labels.}
\label{fig:3pt_causal}
\end{figure}
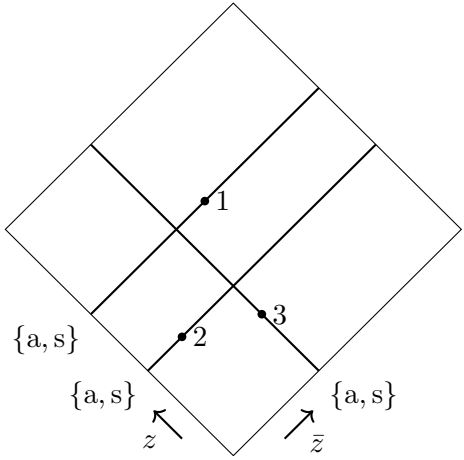

Now we can write the indicator function for the $z_i$ in~\eqref{ind1} in the more compact form
\be
\mathbf{1}_{\varepsilon_i}(z_i):=\prod_{i=1}^3
\mathbf{1}_{[0,1]}(\sigma_{*i})=\Theta\!\left(\frac{\varepsilon_3 z_{23}}{\varepsilon_1 z_{12}}\right)\Theta\!\left(\frac{\varepsilon_3 z_{31}}{\varepsilon_2 z_{12}}\right).
\ee
This presentation comes in handy when we perform the light transforms to our ambidextrous basis, where the covariance of the $\veps_i$ introduces many such step functions! Indeed, consider the MHV amplitude of~\eqref{3ptfix} in the configuration where particles $1,2$ have negative helicity and particle $3$ has positive helicity. To render it non-distributional, our ambidextrous prescription tells us to light transform in $\bar z_1$, $\bar z_2$ and $z_3$, as illustrated in figure~\ref{fig:3pt_causal}. The delta functions in $\bz_{ij}$ localize the $\bbL$ transforms of the two negative helicity operators, so that the only thing we need to take care of are these indicator functions. 

The integration kernels for the light transforms match the expected $\overline{\rm SL}(2,\mathbb{R})$-covariant form due to the following identities
\begin{align}
    &\bar\sh_2-\bar\sh_3-\bar\sh_1=1-\bar{h}_2-\bar{h}_3-1+\bar{h}_1=2\bar{h}_1-2=-2\bar{\sh}_1\,,\\
    &\bar{\sh}_1-\bar{\sh}_2-\bar{\sh}_3=1-\bar{h}_1-1+\bar{h}_2-\bar{h}_3=2\bar{h}_2-2=-2\bar{\sh}_2\,,\\
    &\bar{\sh}_3-\bar{\sh}_1-\bar{\sh}_2=\bar{h}_3-1+\bar{h}_1-1+\bar{h}_2=0
\end{align}
on the locus $\bar{h}_1+\bar{h}_2+\bar{h}_3=2$. As for the 2-point example above we are now using $\sh_i,\bar\sh_i$ to denote the light transformed weights. Meanwhile, the Mellin transformed MHV amplitude already has the expected ${\rm SL}(2,\mathbb{R})$ covariant form, and the final light transform on particle 3 will simply Weyl reflect the weight $h_3$ into $\sh_3=1-h_3$. 

Using step function identities of the type
 \be\label{eq:step_theta}
\Theta(-x)\Theta(y)+\Theta(x)\Theta(-y)=\Theta(-xy)
\ee
 we can massage our correlator into the form
 \begin{multline}\label{3pt3l}
L(1^-2^-3^+)=\pi c_{123}\,
\boldsymbol{\delta}(\im(\bar\sh_1+\bar\sh_2-\bar\sh_3))
|\bz_{12}|^{-\bar\sh_1-\bar\sh_2+\bar\sh_3}|\bz_{13}|^{-\bar\sh_1+\bar\sh_2-\bar\sh_3}|\bz_{23}|^{\bar\sh_1-\bar\sh_2-\bar\sh_3}\\
\times|z_{12}|^{-\sh_1-\sh_2+\sh_3}|z_{13}|^{-\sh_1+\sh_2-\sh_3}|z_{23}|^{\sh_1-\sh_2-\sh_3}\mathrm{sgn}(z_{12}z_{23}z_{13})^{ s_1+s_2+s_3}{\cal N}.
\end{multline}
Here we've decomposed it into a conformally covariant part times a conformally invariant factor $\cal N$ computed by the integral
\be\badat{3}\label{Nint}
{\cal N} &=\int {\d z_3'}\,\frac{|z_{12}|^{-2\sh_3+1}|z_{13}|^{\sh_1-\sh_2+\sh_3}|z_{23}|^{-\sh_1+\sh_2+\sh_3}}{|z_{3'3}|^{2\sh_3}|z_{23'}|^{\sh_2+1-\sh_3-\sh_1}|z_{3'1}|^{\sh_1+1-\sh_3-\sh_2}}\\
&\qquad\times \mathrm{sgn}\!\left(\frac{z_{23'}z_{13'}}{z_{23}z_{13}}\right)^{\sum_is_i}\Theta\!\left(\frac{\veps_3 z_{23'}\bz_{31}}{\veps_1 z_{12}z_{3'3}}\right)\Theta\!\left(\frac{\veps_3 z_{3'1}\bz_{32}}{\veps_2 z_{12}z_{3'3}}\right)\,,
\eadat\ee
where $z_{i'j} = -z_{ji'}\equiv z_i'-z_j$. The coupling constant $c_{123}$ and the normalization $\cN$ contain all the information on celestial OPE data.  To evaluate this integral, we start by treating $\{z_1,z_2,z_3,z_3'\}$ as four independent points from which we can construct the cross ratio
\be
u=\frac{z_{31}z_{23'}}{z_{12}z_{3'3}},
\ee
as well as the conformally invariant signs
\be\label{eta}
\eta_1=\mathrm{sgn}\!\left(\frac{\veps_3\bz_{13}}{\veps_1 z_{13}}\right),~~~\eta_2=\mathrm{sgn}\!\left(\frac{\veps_3\bz_{23}}{\veps_2 z_{23}}\right).
\ee
 The integral~\eqref{Nint} then reduces to
\be\label{Nint2}\scalemath{.98}{\badat{3}
{\cal N}&=(-\eta_1\eta_2)^{\sum_is_i}\int_{-\infty}^\infty\d u\, |u|^{\sh_3+\sh_1-\sh_2-1}|1-u|^{\sh_2+\sh_3-\sh_1-1}\, \Theta\!\left(\eta_1 u\right)\Theta\!\left(\eta_2(1-u)\right).
\eadat}\ee
where the $u$ integral is over the full real line.\footnote{
 Note that $z_3'\mapsto\pm \infty$ as $u\mapsto (\frac{z_{13}}{z_{12}})^\mp$, while  $u\mapsto \pm\infty$ as $z_3'\mapsto z_3^\pm$. These integration ranges glue together to restore the standard $u\in(-\infty,\infty)$ contour.
} Also, the factor of $\sgn(z_{23'}z_{13'})$ has been simplified by noting that it equals $-\eta_1\eta_2\,\sgn(z_{23}z_{13})$ on the support of the step functions.

The integral over $u$ can be easily done by breaking it into domains $(-\infty,0)$, $(0,1)$ and $(1,\infty)$ on which both $u$, $1-u$ have fixed signs.
Indeed of the four possible assignments of the signs $\eta_i$, only three are allowed and these partition the range of $u$ into the following subsets:
\be
\begin{array}{ccc}
\eta_1 &~~\eta_2 &~~ u\\
+ &~~ + &~~ (0,1)\\
+ &~~ - &~~ (1,\infty)\\
- &~~ + &~~ (-\infty,0)\\
- &~~ - &~~ \times 
\end{array}
\ee
Putting everything together we have 
\be\scalemath{.95}{\badat{3}
&{\cal N}(\sh_i,\eta_i)= \Big[(-1)^{s_1+s_2+s_3}\Theta(\eta_1)\Theta(\eta_2) B(\sh_3 + \sh_1- \sh_2, \sh_2 + \sh_3 - \sh_1) \\
&+\Theta(\eta_1)\Theta(-\eta_2)B(1 - 2 \sh_3, \sh_2 + \sh_3-\sh_1)+\Theta(-\eta_1)\Theta(\eta_2)B(1 - 2 \sh_3, \sh_3 + \sh_1 - \sh_2)\Big]
\eadat}\ee
where $B(x,y) = \Gamma(x)\,\Gamma(y)/\Gamma(x+y)$ is the Euler beta function.  We thus find an answer that is fully conformally covariant as well as non-distributional in all channels. It moreover contains all the beta function OPE coefficients that are known to occur in the celestial OPE of two like-helicity boost eigenstates \cite{Pate:2019lpp}.

Let us close our discussion of the ambidextrous three point function by commenting on some symmetries exhibited by our correlators. Assuming that this three point data indeed determines the higher point correlators, it is meaningful to ask such questions at this stage.  First, we see that our amplitude obeys a conservation law for the $\bar{\sh}$ weights at an MHV vertex. The same is true for the ${\sh}$ weights at an anti-MHV vertex.  Meanwhile, the two point function preserves both $\sh$ and $\bar{\sh}$, so we can use this to glue together such vertices in a CSW-like expansion. 

Besides this continuous symmetry, we also observe a discrete parity symmetry under $\vec{\varepsilon}\mapsto-\vec{\varepsilon}$, where $\vec\veps=(\veps_1,\veps_2,\veps_3)$. This is because the conformally invariant signs \eqref{eta} only depend on $\vec\veps$ through the entries of the \emph{sign vector}
\be\label{svec}
\vec{s}=(\varepsilon_1\varepsilon_3,\varepsilon_2\varepsilon_3)\,.
\ee
It is invariant under $\vec{\varepsilon}\mapsto-\vec{\varepsilon}$ so we can fix $\varepsilon_3=1$ if we want to pick a representative of $\vec{\veps}$ modulo this $\mathbb{Z}_2$ action.  There is then a one to one map between crossing channels and entries of $\vec{s}$. 

This also reflects in the light transformed amplitudes of the symmetric and antisymmetric primaries. When we sum over channels we include all chambers of the step functions. To be precise, let 
\be
\mathfrak{p}_i\in\{\s,\at\}
\ee
denote the symmetric/antisymmetric label for the $i^\text{th}$ particle. For every symmetric operator we weight the two channels equally, while for every antisymmetric operator we weight the channels with a relative phase. In total, the light transformed celestial correlator of symmetric/antisymmetric primaries reads
\be
L(1^{J_1}_{\mathfrak{p}_1}2^{J_2}_{\mathfrak{p}_2}\cdots n^{J_n}_{\mathfrak{p}_n}) = \sum_{\veps_i=\pm1}\prod_{\mathfrak{p}_j=\at}\veps_j\,L(1^{J_1}2^{J_2}\cdots n^{J_n})\,,
\ee
where $L(1^{J_1}2^{J_2}\cdots n^{J_n})$ implicitly depends on the $\veps_j$.

At three points, we can again write this as a covariant prefactor times an invariant normalization:
\begin{multline}
L(1^{-}_{\mathfrak{p}_1}2^{-}_{\mathfrak{p}_2}3^+_{\mathfrak{p}_3}) = \mathcal{N}^{\mathfrak{p}_1\mathfrak{p}_2\mathfrak{p}_3}\pi c_{123}\,
\boldsymbol{\delta}(\im(\bar\sh_1+\bar\sh_2-\bar\sh_3))
|\bz_{12}|^{-\bar\sh_1-\bar\sh_2+\bar\sh_3}|\bz_{13}|^{-\bar\sh_1+\bar\sh_2-\bar\sh_3}|\bz_{23}|^{\bar\sh_1-\bar\sh_2-\bar\sh_3}\\
\times|z_{12}|^{-\sh_1-\sh_2+\sh_3}|z_{13}|^{-\sh_1+\sh_2-\sh_3}|z_{23}|^{\sh_1-\sh_2-\sh_3}\mathrm{sgn}(z_{12}z_{23}z_{13})^{ s_1+s_2+s_3}.
\end{multline}
Because of the $\mathbb{Z}_2$ symmetry, the sum over all channels gives a factor of two in each $\mathcal{N}^{\mathfrak{p}_1\mathfrak{p}_2\mathfrak{p}_3}$. For example,
\be\badat{3}
\mathcal N^{\bar \s\bar \s\s} &= 2 \Big[(-1)^{s_1+s_2+s_3} B(\sh_3 + \sh_1- \sh_2, \sh_2 + \sh_3 - \sh_1) \\
&\qquad+B(1 - 2 \sh_3, \sh_2 + \sh_3-\sh_1)+B(1 - 2 \sh_3, \sh_3 + \sh_1 - \sh_2)\Bigr].
\eadat\ee
All step functions have been absorbed by the sum over channels, as in~\cite{Hu:2022syq}. One can further check that
\be
\mathcal{N}^{\bar \s\bar \s\at}=\mathcal{N}^{\bar \s\bar \at\s}=\mathcal{N}^{\bar \at\bar \s\s}=\mathcal{N}^{\bar \at\bar \at\at}=0,
\ee
corresponding to the fact that the discrete symmetry forces correlators with an odd number of $A$ operators to vanish.  For completeness we note that
\be\badat{3}
\mathcal{N}^{\bar \s\bar \at\at}&=2\,\sgn(z_{23}\bz_{23}) \Big[(-1)^{s_1+s_2+s_3} B(\sh_3 + \sh_1- \sh_2, \sh_2 + \sh_3 - \sh_1) \\
&\qquad+B(1 - 2 \sh_3, \sh_2 + \sh_3-\sh_1)-B(1 - 2 \sh_3, \sh_3 + \sh_1 - \sh_2)\Bigr]\,,
\eadat\ee
while
\be\badat{3}
\mathcal{N}^{\bar \at\bar \s\at}&=2\,\sgn(z_{13}\bar z_{13}) \Big[(-1)^{s_1+s_2+s_3} B(\sh_3 + \sh_1- \sh_2, \sh_2 + \sh_3 - \sh_1) \\
&\qquad-B(1 - 2 \sh_3, \sh_2 + \sh_3-\sh_1)+B(1 - 2 \sh_3, \sh_3 + \sh_1 - \sh_2)\Bigr]\,,
\eadat\ee
and
\be\badat{3}
\mathcal{N}^{\bar \at\bar \at\s}&=2\,\sgn(z_{13}z_{23}\bz_{13}\bz_{23}) \Big[(-1)^{s_1+s_2+s_3} B(\sh_3 + \sh_1- \sh_2, \sh_2 + \sh_3 - \sh_1) \\
&\qquad-B(1 - 2 \sh_3, \sh_2 + \sh_3-\sh_1)-B(1 - 2 \sh_3, \sh_3 + \sh_1 - \sh_2)\Bigr]\,.
\eadat\ee
Here we draw attention to the sign flips relative to ${\cal N}^{\bar \s\bar \s\s}$ in addition to the modified loci of support.  Figure~\ref{fig:3pt_causal} illustrates why the `s' and `a' label assignments aren't expected to give permutation symmetric normalization/OPE coefficients $\cN^{\mathfrak{p}_1\mathfrak{p}_2\mathfrak{p}_3}$ for the MHV amplitudes; to make this more obvious we have added bars on the `s' and `a' labels here to indicate the different smearings (equivalently the different helicities).  We indeed see that the channel-averaged amplitudes studied in~\cite{Fan:2021isc,Hu:2022syq} are simplest, even in this ambidextrous basis. Still, the other combinations allowed by our $\mathbb{Z}_2$ selection rule have a clean dependence on the causal separation of the operators.  

%%%%%%%%%%%%%%%%%%%%%%%%

\subsection{Four-point}

Let us now proceed to the four-point amplitude. The momentum conservation stripped amplitude has a more complicated form than at three points~\eqref{3pt}. However, we can still factor out a part that captures the appropriate little group covariance for the momentum space amplitude (see for instance appendix A in \cite{Arkani-Hamed:2020gyp})
\be\label{A4}
A[1^{J_1}2^{J_2}3^{J_3}4^{J_4}]=H(\lambda_i,\bar\lambda_i)M(s,t),~~~H(\lambda_i,\bar\lambda_i)=\prod_{i<j}\left(\frac{\langle i\,j\rangle}{[i\,j]}\right)^{\frac{1}{2}(\frac{1}{3}\sum_kJ_k-J_i-J_j)}
\ee
where $s=(p_1+p_2)^2$ and $t=(p_1+p_3)^2$ are the usual Mandelstam invariants. Meanwhile, the locus of momentum conservation now takes the following form~\cite{Pasterski:2017ylz}
\begin{align}\label{eq:d2}
&\delta^{4}\bigg(\sum_{i=1}^4\veps_i\sigma_iq_i\bigg)\delta\bigg(\sum_{i=1}^4\sigma_{i}-1\bigg)
= \frac{1}{4}\,\delta(|z_{12}z_{34}\bz_{13}\bz_{24}-z_{13}z_{24}\bz_{12}\bz_{34}|)
\notag\\
&\times\delta\!\left(\sigma_1+
\frac{\varepsilon_1\varepsilon_4}{ D_4 }
\frac{z_{24}\bz_{34}}{z_{12}\bz_{13}}\right)
\delta\!\left(\sigma_2-
\frac{\varepsilon_2\varepsilon_4}{ D_4}
\frac{z_{34}\bz_{14}}{z_{23}\bz_{12}}
\right)\delta\!\left(\sigma_3+\frac{\varepsilon_3\varepsilon_4}{ D_4}
\frac{z_{24}\bz_{14}}{z_{23}\bz_{13}}\right)
\delta\!\left(\sigma_4-\frac{1}{D_4}\right)  \notag\\
&= \frac{1}{4}\,\delta(|z_{12}z_{34}\bz_{13}\bz_{24}-z_{13}z_{24}\bz_{12}\bz_{34}|)
\sum_{i=1}^4 \delta(\sigma_i  - \sigma_{*i})\,,
\end{align}
where the denominator $D_4$ is given by
\begin{align}\label{D4}
D_4 = \left( 1-{\varepsilon_1\varepsilon_4}\right)
\frac{z_{24} \bz_{34}}{ z_{12}\bz_{13}}
+\left( {\varepsilon_2\varepsilon_4}-1\right)
\frac{z_{34} \bz_{14} }{ z_{23} \bz_{12}}
+\left(1-{\varepsilon_3\varepsilon_4}\right)
\frac{z_{24}\bz_{14}}{ z_{23}\bz_{13}} \,
\end{align}
and, as for the three-point case above, we have used the integration variables $\sigma_i= \omega_i/\sum_j\omega_j$.
Introducing the conformally invariant cross ratios
\begin{align}\label{zbz}
    z &= \frac{z_{12}z_{34}}{z_{13}z_{24}}\,,\quad\bz = \frac{\bar z_{12}\bar z_{34}}{\bar z_{13}\bar z_{24}}\,,
 \end{align}
the support of~\eqref{eq:d2} restricts us to $z=\bz$, i.e., to real cross ratios.    

Upon performing the Mellin transforms we localize the simplex integrals $\sigma_i$ and again find that different kinematical crossings have distinct regions of support in the cross ratios.  In $(1,3)$ signature the different $2\leftrightarrow2$ processes partition $z\in\mathbb{R}$ as follows~\cite{Pasterski:2017ylz}
\be\label{ind2}
\prod_{i=1}^4\mathbf{1}_{[0,1]}(\sigma_{*i})\neq0:~~\begin{array}{lll}
a)~~~ 12 \ce{<-->} 34~~~&\Rightarrow~~~ 1<z \\ 
b)~~~ 13 \ce{<-->} 24~~~&\Rightarrow~~~ 0<z<1 \\
c)~~~ 14 \ce{<-->} 23~~~&\Rightarrow~~~  z<0 \, .
\end{array}
\ee
Namely, $\sigma_{*i}\in[0,1]$ in these ranges. This follows from the fact that $z_{ij}\bz_{ij}>0$ in this signature. In $(2,2)$ signature one needs to be on the lookout for additional channels with non-trivial support~(see also \cite{Hu:2022syq}). To handle either signature, let us introduce the following set of lightcone indicator variables 
\be\label{etaij}
\eta_{ij}=\sgn\left(\varepsilon_i\varepsilon_jz_{ij}\bz_{ij}\right),
\ee
generalizing~\eqref{eta}.   We see that these reduce to the sign vectors~\eqref{svec} labeling channels $\vec{\varepsilon}/\mathbb{Z}_2$ for $z_{ij}\bz_{ij}>0$. From the transformation laws~\eqref{eq:sgnflip1}-\eqref{eq:sgnflip2} 
\be\badat{3}
\eta_{ij}&\mapsto {\rm sgn}[(cz_i+d)(\bc\bz_i+\bd)]\,\veps_i\,{\rm sgn}\,[(cz_j+d)(\bc\bz_j+\bd)]\,\veps_j \\ &~~~\times{\rm sgn}
\left[\frac{z_{ij}}{(cz_i+d)(cz_j+d)}\frac{\bz_{ij}}{(\bc\bz_i+\bd)(\bc\bz_j+\bd)}\right]=\eta_{ij}
\eadat\ee
we that the $\eta_{ij}$ are Lorentz invariant in any signature and are thus more natural than the $\vec{\varepsilon}$'s once we want to analytically continue our amplitudes.

More generally, they are determined by the channel labels for any fixed puncture configuration. Now there are six possible choices of pairs $(ij)$ for $i\neq j$. However, for a given configuration of punctures the $\eta_{ij}=\eta_{ji}$ are not all independent because they encode the causal relations between the 4 points pairwise, and not only do the four punctures live in $\mathbb{R}^{1,1}$, they are also restricted to the locus $z=\bz$.  In particular
\be
\frac{\eta_{13}\eta_{24}}{\eta_{14}\eta_{23}}={\rm sgn}(z\bar z) = \sgn\,z^2 = +1\,.
\ee
The same is true for the other permutations of the puncture labels and we can use this to eliminate
\be
\eta_{23}=\eta_{13}\eta_{14}\eta_{24},~~~\eta_{34}=\eta_{12}\eta_{13}\eta_{24}
\ee
leaving $\eta_{12},\eta_{13},\eta_{14},\eta_{24}$ as the independent conformally invariant signs. For a fixed set of punctures, we can freely assign $\varepsilon_i$ such that $\eta_{13}$, $\eta_{14}$ and $\eta_{23}$ take a sign of our choosing via
\be
\varepsilon_1=\eta_{13}\varepsilon_3\sgn (z_{13}\bz_{13}),~~\varepsilon_4=\eta_{14}\varepsilon_1\sgn (z_{14}\bz_{14}),~~\varepsilon_2=\eta_{23}\varepsilon_3 \sgn (z_{23}\bz_{23})\,.
\ee
We further see that flipping $\varepsilon_3\mapsto-\varepsilon_3$ with fixed $\eta_{13},\eta_{14},\eta_{23}$ flips the signs of the other $\veps_i$'s. 
So, for a fixed puncture configuration, the vector $\{\eta_{13},\eta_{14},\eta_{23}\}$ parameterizes $\vec{\varepsilon}/\mathbb{Z}_2$.

For our manipulations in $(2,2)$ signature, it will turn out to be simpler to start from the more explicit expression for the indicator functions,
\be\label{thetastep}
\prod_{i=1}^4\mathbf{1}_{[0,1]}(\sigma_{*i})=\Theta\!\left(-
\frac{\varepsilon_4}{\varepsilon_1}
\frac{z_{24}\bz_{34}}{z_{12}\bz_{13}}\right)
\Theta\!\left(\frac{\varepsilon_4}{\varepsilon_2}
\frac{z_{34}\bz_{14}}{z_{23}\bz_{12}}
\right)\Theta\!\left(-\frac{\varepsilon_4}{\varepsilon_3}
\frac{z_{24}\bz_{14}}{ z_{23}\bz_{13}}
\right)\,,
\ee
since our improved light transforms \eqref{L}, \eqref{Lb} are specified in terms of flipping the signs of the $\veps_i$ on either side of their insertion points. \eqref{thetastep} is easily obtained by demanding positivity of the $\sigma_i$'s obtained by integrating out the delta functions in the second line of \eqref{eq:d2}. We can also recast these step functions 
in terms of the Lorentz invariant $\eta_{ij}$ and cross ratio $z$.
Comparing with the notation of~\cite{Hu:2022syq},  we see that
\be\badat{3}\label{sigmaeta} \prod_{i=1}^4\mathbf{1}_{[0,1]}(\sigma_{*i})&=\Theta\left(-\frac{\eta_{24}}{\eta_{12}}\,z\right)
\Theta\left(\frac{\eta_{14}}{\eta_{12}}\frac{z}{1-z}\right)
\Theta\left(-\frac{\eta_{13}}{\eta_{14}}\,(1-z)\right)\,.
\eadat\ee
In analogy with \eqref{ind2}, this provides a demarcation of scattering channels in $(2,2)$ signature.

Turning to the Mellin transformed amplitude, we will want to strip off a kinematical factor that takes into account the external weights. Certain choices of these `leg factors' are more convenient for different contexts, and  they differ by Lorentz invariant factors.  In~\eqref{A4} we stripped off a particular choice of spinor-helicity variables / powers of $z_{ij}$ and $\bz_{ij}$, to match~\cite{Pasterski:2017ylz,Arkani-Hamed:2020gyp}.   It was maximally symmetric under external label exchanges. In what follows, it will be helpful to use a slightly modified decomposition of the correlator described in~\cite{Law:2019glh,Law:2020xcf} 
\begin{multline}\scalemath{.97}{\label{celamp4}
\bigg\langle\prod_{i=1}^4\O^{\veps_i}_{h_i,\bh_i}(z_i,\bz_i)\bigg\rangle= \delta(z-\bz)\,Y(z_j,\bz_j)\, G(z)
\prod_{i=1}^4\mathbf{1}_{[0,1]}(\sigma_{*i})\prod_{j<k}(\eta_{jk})^{\frac{1}{2}(\frac13\sum_\ell J_\ell-J_j-J_k)}\,, }
\end{multline}
where the support of the indicator functions is given in \eqref{thetastep} and the leg factor is taken to be
\begingroup
\allowdisplaybreaks
\begin{align}
    Y(z_i,\bz_i) &=\left|\frac{z_{24}}{z_{14}}\right|^{h_{12}}\left|\frac{z_{14}}{z_{13}}\right|^{h_{34}}\left|\frac{\bar{z}_{24}}{\bar{z}_{14}}\right|^{\bar{h}_{12}}\left|\frac{\bar{z}_{14}}{\bar{z}_{13}}\right|^{\bar{h}_{34}}\frac{|1-z|^{\frac{1}{2}(h_{12}-h_{34})}}{|z_{12}|^{h_1+h_2}| z_{34}|^{h_3+h_4}}  \frac{|1-\bz|^{\frac{1}{2}(\bh_{12}-\bh_{34})}}{|\bar{z}_{12}|^{\bar{h}_1+\bar{h}_2}|\bar{z}_{34}|^{\bar{h}_3+\bar{h}_4}}\nonumber\\
    &= \left|\frac{1-z}{z^2}\right|^{\frac{1}{6}\sum_kh_k}\left|\frac{1-\bz}{\bz^2}\right|^{\frac{1}{6}\sum_k\bh_k}\prod_{i<j}\left|z_{ij}\right|^{\frac13\sum_kh_k-h_i-h_j}\left|\bz_{ij}\right|^{\frac13\sum_k\bh_k-\bh_i-\bh_j} 
\end{align}
\endgroup
with $h_{ij}=h_i-h_j$, $\bh_{ij}=\bh_i-\bh_j$. The conformally invariant data within the 4-point amplitude takes the general form 
\be
G(z)= \frac{z^2}{2^{\beta/2}}\int_0^\infty\frac{\d\omega}{\omega}\, \omega^{\beta}\,M\bigl(\eta_{12}\omega^2,-\eta_{12}\omega^2z^{-1}\bigr)\,.
\ee
Here, $\beta=\sum_{i=1}^4(\Delta_i-1)$ is the net boost weight, and $M(s,t)$ is the little group invariant part of the momentum space amplitude as described in \eqref{A4}. This has been simplified using the support of $z=\bz$ as well as the indicator functions. Using the same, we can also simplify the explicit sign factor in \eqref{celamp4}
\be\label{simplesign}
\prod_{j<k}(\eta_{jk})^{\frac{1}{2}(\frac13\sum_\ell J_\ell-J_j-J_k)} = \eta_{12}^{\frac12(J_3+J_4-J_1-J_2)}\eta_{13}^{\frac12(J_2+J_4-J_1-J_3)}\eta_{14}^{\frac12(J_2+J_3-J_1-J_4)}\,.
\ee
In all these calculations, we nicely see that the invariant data of the CFT encodes the invariant data of the $(2,2)$ signature amplitude.

\medskip

Now our only remaining task is to work out its ambidextrous light transform!  Without loss of generality we will consider the helicity configuration $1^-2^-3^+4^+$. Let us denote the corresponding ambidextrous light transform as 
\be\label{L4}
\begin{split}
    L(1^-2^-3^+4^+) &= \int_{\R^4}\frac{\d\bar z_1'}{|\bar z_{1'1}|^{2-2\bsh_1}}\,\frac{\d\bar z_2'}{|\bar z_{2'2}|^{2-2\bsh_2}}\,\frac{\d z_3'}{|z_{3'3}|^{2-2\sh_3}}\,\frac{\d z_4'}{|z_{4'4}|^{2-2\sh_4}} \\
    &\qquad\times\left\la\prod_{\ell=1}^2\O^{\veps_\ell'}_{h_\ell,\bh_\ell}(z_\ell,\bz_\ell')\prod_{r=3}^4\O^{\veps_r'}_{h_r,\bh_r}(z_r',\bz_r)\right\ra
\end{split}
\ee
where $z_{i'j}=z_i'-z_j$, $\bz_{i'j} = \bz_i'-\bz_j$, the helicities are $h_\ell-\bh_\ell<0$ for $\ell=1,2$ and $h_r-\bh_r>0$ for $r=3,4$, and the light transformed signs $\veps'_i$ are
\be\label{epsprime}
(\veps_1',\veps_2',\veps_3',\veps_4') = (\sgn(\bz_{1'1})\varepsilon_1,\sgn(\bz_{2'2})\varepsilon_2,\sgn(z_{3'3})\varepsilon_3,\sgn(z_{4'4})\varepsilon_4)\,.
\ee
Again, we would like to decompose this into a conformally covariant leg factor and the conformally invariant dynamical data.

To compute this, we introduce the following change of integration variables (see also parallel work in \cite{De:2022gjn}):
\be\label{ui}
x = \frac{z_{12}z_{3'4'}}{z_{13'}z_{24'}}\,,\qquad \bar x = \frac{\bz_{1'2'}\bz_{34}}{\bz_{1'3}\bz_{2'4}}\,,\qquad t = \frac{z_{12}z_{34'}}{z_{13}z_{24'}}\,,\qquad \bar t = \frac{\bz_{12'}\bz_{34}}{\bz_{13}\bz_{2'4}}\,.
\ee
which are each cross ratios constructed from the integrated and external puncture locations. The (absolute value of the) Jacobian of this transformation is found to be
\be\badat{3}
\d x\,\d\bar x\,\d t\,\d\bar t
&= |1-t||1-\bar t|\,\frac{z_{12}^2}{z_{13'}^2z_{24'}^2}\frac{\bz_{34}^2}{\bz_{1'3}^2\bz_{2'4}^2}\;\d\bz_{1}'\, \d\bz_{2}'\,\d z_{3}'\,\d z_{4}'\,.
\eadat\ee
In these new variables, the delta function in \eqref{celamp4} maps to
\be
\delta(z-\bz)\mapsto\delta\bigg(\frac{z_{12}z_{3'4'}}{z_{13'}z_{24'}} - \frac{\bz_{1'2'}\bz_{34}}{\bz_{1'3}\bz_{2'4}}\bigg)=\delta(x-\bar x)
\ee
under the light transform. Thus, we will be easily able to integrate it out by setting $\bar x=x$. 

We will again use the conformally invariant cross ratios and signs defined in~\eqref{zbz} and~\eqref{etaij} above, but now they will be built from the $z_i,\bz_i$ insertions of the light transformed operators. For the conformally invariant signs, it will be slightly cleaner to use $\eta_{ij}$ for $i=1,2$ and $j=3,4$ as a `basis' to express our result. While the single particle transforms were defined in terms of the individual signs $\varepsilon_i$, the multi-particle correlators are more naturally recast as a function of the Lorentz invariant $\eta_{ij}$. 

Not only do we have the step functions in~\eqref{sigmaeta} at the primed coordinates, our light transform now mixes different crossing channels depending on the ranges of $\bz_1',\bz_2',z_3',z_4'$.  The spinor-helicity factor has no explicit epsilon dependence, so the only $\varepsilon$ dependence is implicit in the $\eta_{i'j}$, etc. Starting from~\eqref{thetastep} and substituting $\veps_i\mapsto\veps'_i$, $z_i\mapsto z'_i$, etc., our indicator function transforms into
\be\badat{3}\label{theta4}
\Theta_4(\eta_{ij},z,\bz,x,t,\bar t) \equiv \Theta\bigg(\eta_{13}\,\frac{x-t}{x-\bar t}\bigg)\Theta\bigg(\eta_{14}\,\frac{(z-1)(x-\bar t)}{(z-t)(1-\bar t)}\bigg)\Theta\bigg(\eta_{24}\,\frac{(1-t)(\bz - \bar t)}{(z-t)(1-\bar t)}\bigg)\,.\\
\eadat\ee
This is now expressed in terms of the final puncture positions and has been simplified using the support of $\bar x=x$. 

On the support of this, the sign-dependent factor \eqref{simplesign} also transforms to (working with integer external spins as before)
\begin{multline}\label{fraks}
\mathfrak{s}_4(\eta_{ij},z,\bz,x,t,\bar t) \equiv \Bigl(\eta_{12}\,\sgn[\bz(1-\bar t)(x-\bar t)(\bz-\bar t)]\Bigr)^{\frac12\sum_iJ_i}\\
\times\left(\sgn\,x\right)^{\frac12(J_3+J_4-J_1-J_2)}\left(\sgn\,(1-x)\right)^{\frac12(J_2+J_3-J_4-J_1)}\,.
\end{multline}
The $x,t,\bar t$ dependent factor in this expression will appropriately dress the light ray integrals. Although it looks like this factor needs us to define a choice of branch for the square roots, the combinations of spins appearing in the exponents in this expression will generally turn out to be integral in practice. For example, for a gluon MHV configuration, one will have $J_3=J_4=-J_1=-J_2=1$ so that $\frac12(J_3+J_4-J_1-J_2)=2$, $\frac12\sum_iJ_i=0$, etc. When this is not the case, one is free to accordingly alter the factor $H(z_i,\bz_i)$ stripped out from the amplitude in \eqref{A4}.

After performing the light transform, we will strip off a slightly simpler leg factor that is better suited to conformal block expansions as well as alpha space transforms:
\be\scalemath{.90}{\badat{3}
K(z_i,\bz_i)&=\left|\frac{z_{24}}{z_{14}}\right|^{\sh_{12}}\left|\frac{z_{14}}{z_{13}}\right|^{\sh_{34}}\left|\frac{\bar{z}_{24}}{\bar{z}_{14}}\right|^{\bsh_{12}}\left|\frac{\bar{z}_{14}}{\bar{z}_{13}}\right|^{\bsh_{34}}\frac{1}{|z_{12}|^{\sh_1+\sh_2}| z_{34}|^{\sh_3+\sh_4}}  \frac{1}{|\bar{z}_{12}|^{\bsh_1+\bsh_2}|\bar{z}_{34}|^{\bsh_3+\bsh_4}}
  \eadat}
\ee
where $\sh_i,\bsh_i$ are the light transformed weights as before, and $\sh_{ij}=\sh_i-\sh_j,\bsh_{ij}=\bsh_i-\bsh_j$. The complete ambidextrously light transformed 4-point amplitude then takes the form
\be
L(1^-2^-3^+4^+) = K(z_i,\bz_i)\,\sL(z,\bz)\,.
\ee
Having performed the $\bar x$ integral to set $\bar x=x$, the resulting conformally invariant data is encapsulated by the following set of leftover integrals
\begin{align}\label{eq:G_hi}
    \sL(z,\bz) &= 2^{-\frac\beta2}\,\frac{|z|^{\sh_3+\sh_4}|\bz|^{\bsh_1+\bsh_2}}{|1-z|^{\sh_{34}-\sh_{12}}}\int_{-\infty}^\infty\d x\;|x|^{\bsh_1+\bsh_2+\sh_3+\sh_4-2}|1-x|^{\frac{1}{2}(J_{12}-J_{34})}\,\nonumber\\
    &\times\int_{-\infty}^\infty\d t\;|1-t|^{\sh_3+\sh_4-\sh_{12}-1}|x-t|^{-2\sh_3}|z-t|^{-2\sh_4}\nonumber\\
    &\times\int_{-\infty}^\infty\d\bar t\;|1-\bar t|^{\bsh_1+\bsh_2-\bsh_{34}-1}|x-\bar t|^{-2\bsh_1}|\bz-\bar t|^{-2\bsh_2}\nonumber\\
    &\times\Theta_4(\eta_{ij},z,\bz,x,t,\bar t)\,\mathfrak{s}_4(\eta_{ij},z,\bz,x,t,\bar t)\int_0^\infty\frac{\d\omega}{\omega}\,\omega^{\beta}\,M\!\left(\eta_{1'2'}\omega^2,\eta_{1'2'}\omega^2x^{-1}\right)
\end{align}
where $J_{ij}=J_i-J_j$ are differences of the original spins $J_i=h_i-\bh_i$, the transformed indicator function $\Theta_4$ is given by \eqref{theta4}, the transformed sign $\mathfrak{s}_4$ by \eqref{fraks}, and the sign $\eta_{12}$ has been replaced by
\be
\begin{split}
    \eta_{1'2'} &= \eta_{12}\,\sgn[x\bz(1-\bar t)(x-\bar t)(\bz-\bar t)] \\
    &= \eta_{34}\,\sgn[xz(1-t)(x-t)(z-t)]
\end{split}
\ee
post light transforms.

The step functions in the integrand of~\eqref{eq:G_hi} depend on the signs of the following 10 variables
\be
\{\{\eta_{13}, \eta_{14}, \eta_{24},z-1\},\{1-t,x-t,z-t\},
\{ 1-\bar t,  x-\bar t, \bar z-\bar t\}\}
\ee
which we've grouped suggestively. Using our observation above that $\{\eta_{13},\eta_{14},\eta_{23}\}$ parameterizes $\vec{\varepsilon}/\mathbb{Z}_2$ for a fixed puncture configuration and channel, we can repeatedly use identities of the form~\eqref{eq:step_theta} to split up the step functions in~\eqref{eq:G_hi} for each channel.  The $t$ and $\bar t$ integrals are of Gauss hypergeometric type, with the step functions carving up different integration ranges. Namely, the fundamental building blocks are the integrals
\be\label{fint0}
\int_D \d t |1-t|^{a-2}|x-t|^{b-1}|w-t|^{c-1} \Theta(\pm(1- t))\Theta(\pm(x-t))\Theta(\pm(w-t))
\ee
where the integration domain $D$ is taken to be any of the four intervals cut out by $-\infty,0,1,w,\infty$ for $w=z$ or $w=\bz$. Each window of support can be written in terms of (a sum of) $_2 F_1$ hypergeometric functions times step functions in $w$.

Using \eqref{eq:G_hi}, it is cumbersome but completely straightforward to build linear combinations that compute light transformed amplitudes of the symmetric and antisymmetric primaries. Instead of listing them in full generality, we will evaluate these integrals explicitly in some simple gluon examples like factorization limits (optical theorem) and MHV amplitudes in the following sections. 

Another somewhat painful point of this derivation was the step involving trying to come up with an appropriate substitution like \eqref{ui}. It is natural to ask why such a clever substitution exists in the first place, and whether there is some general mathematical principle for choosing judicious integration variables while computing light transforms. As we discuss below, it turns out that this question is crucially related to the twistorial approach to scattering amplitudes \cite{ArkaniHamed:2009si,Mason:2009sa} and the associated Grassmannian formulae in the case of gauge theory \cite{ArkaniHamed:2009dn}.

%%%%%%%%%%%%%%%%%%%%%%%%%%%%%%%%%%%
%%%%%%%%%%%%%%%%%%%%%%%%%%%%%%%%%%%

\section{Deconstructing gluon amplitudes}\label{sec:gluon_decomp}

Having studied the generalities of low-multiplicity light transformed celestial amplitudes, we now come to a case study of tree-level gluon amplitudes. We begin by providing efficient expressions for light transform integrals in terms of integrals over spinor-helicity variables. These allow one to directly and painlessly jump from momentum amplitudes to light transformed celestial amplitudes. Together with these, we use Grassmannian formulae for tree-level gluon amplitudes to efficiently derive expressions for the light transformed amplitude of four gluon symmetric primaries. We study its expansion in alpha space and extract data on the spectrum of celestial CFT dual to gauge theory in $\R^{2,2}$. 

For the reader's benefit, a slightly simpler example of such calculations is also included in appendix \ref{app:fac}. There, we work in the factorization limit and are able to perform much more explicit conformal block expansions. For the $s$-channel unitarity cut of a tree-level gluon MHV amplitude, we analytically find a four term conformal block expansion given by equation \eqref{facdec}, showing that only a \emph{finite} number of operators are exchanged as we approach the factorization limit. Taken together with the sections below, it provides a useful starting point to delve into calculations involving the ambidextrous basis.

\subsection{Light transforms from momentum eigenstates}

Continuing with the philosophy of using light transformed operators as a basis for celestial CFT, we would like to find a way to map momentum space amplitudes directly to those of light transformed symmetric and antisymmetric primaries. This is best done using spinor-helicity variables.

As before, we will focus on massless states of integral helicity $J\in\Z$. Recall that spinor components of null momenta $p^\mu$ decompose as $p^{\al\dal}=\lambda^\al\bar\lambda^{\dal}$ in spinor-helicity variables (see section \ref{sec:LT_amplitudes} for our spinor-helicity conventions). Define the scale fixed spinors
\be
\zeta^\al = (1,z)\,,\qquad\bar\zeta^{\dal} = (1,\bz)\,.
\ee
One conventionally fixes little group scaling by setting
\be
\lambda^\al = \sqrt{\omega}\,\zeta^\al\,,\qquad\bar\lambda^{\dal} = \veps\sqrt{\omega}\,\bar\zeta^{\dal}\,.
\ee
The variables $\veps,\omega,z,\bar z$ are then extracted via
\be\label{varrel}
\veps = \sgn(\lambda^0\bar\lambda^{\dot0})\,,\quad\omega = |\lambda^0\bar\lambda^{\dot0}|\,,\quad z = \frac{\lambda^1}{\lambda^0}\,,\quad\bar z = \frac{\bar\lambda^{\dot1}}{\bar\lambda^{\dot0}}\,.
\ee
The associated momentum eigenstate of a helicity $J$ particle was denoted $|\veps\omega,z,\bz, J\ra$ in section \ref{sec:Klein_basis}. Conversely, taking \eqref{varrel} to be the definition of $\veps,\omega,z,\bz$, one can define a general spinor-helicity eigenstate
\be\label{sheig}
|\lambda,\bar\lambda,J\ra = \sgn(\lambda^0)^{-2J}\,\bigg|\frac{\bar\lambda^{\dot0}}{\lambda^0}\bigg|^J\;|\veps\omega,z,\bz, J\ra
\ee
labeled by spinor-helicity variables without a choice of little group fixing. It is easily checked that they transform in accordance with \eqref{lgsc} under little group scalings.

Using the states \eqref{sheig} and their little group transformations, we can recast the symmetric primary state \eqref{symstate} as
\begin{align}
|z,\bz,h,\bh\ra_\s &= \int_{-\infty}^\infty\frac{\d\omega}{|\omega|}\,|\omega|^{\Delta}\,\left|\,|\omega|^{\frac12}\zeta\,,\,\sgn(\omega)|\omega|^{\frac12}\bar\zeta\,,\,J\right\ra\nonumber\\
&= \int_{-\infty}^\infty\frac{\d\omega}{|\omega|}\,|\omega|^{2\bh}\,|\zeta,\omega\bar\zeta,J\ra\\
&= \int_{-\infty}^\infty\frac{\d\omega}{|\omega|}\,|\omega|^{2h}\,|\omega\zeta,\bar\zeta,J\ra\,,
\end{align}
having used $J\in\Z$ to set a factor of $\sgn(\omega)^{2J}$ to $1$ while getting the last line. Similarly, for the antisymmetric primary \eqref{asymstate} one gets the expressions
\begin{align}
|z,\bz,h,\bh\ra_\at &= \int_{-\infty}^\infty\frac{\d\omega}{\omega}\,|\omega|^{2 h}\,|\omega\zeta,\bar\zeta,J\ra \\
&= \int_{-\infty}^\infty\frac{\d\omega}{\omega}\,|\omega|^{2\bar h}\,|\zeta,\omega\bar\zeta,J\ra\,.
\end{align}
Let us plug these into the definitions of the corresponding light transforms.

Just as for the CCFT operator \eqref{Ls}, the $\bL$ transform of a symmetric primary state in the bulk now reads
\begin{align}
    \bL|z,\bz,h,\bh\ra_\s &= \int_{-\infty}^\infty\frac{\d w}{|w-z|^{2-2h}}\;|w,\bz,h,\bh\ra_\s\nonumber\\
    &= \int_{-\infty}^\infty\d w\int_{-\infty}^\infty|\omega|\,\d\omega\;|\omega(w-z)|^{2h-2}\,|\omega\zeta,\bar\zeta,J\ra\,.
\end{align}
Introducing a combined 2-component integration variable
\be
\kappa^\al = \omega\,(1,w)\,,
\ee
this can be written in a much more compact fashion:
\be\label{Lssh}
\bL|z,\bz,h,\bh\ra_\s = \int_{\R^2}\d^2\kappa\;|\la\kappa\,\zeta\ra|^{2h-2}\,|\kappa,\bar\zeta,J\ra\,.
\ee
Here, we have used the notation $\la\kappa\,\zeta\ra\equiv \eps_{\beta\al}\kappa^\al\zeta^\beta = \omega(w-z)$ for spinor contractions.

One can similarly derive the following integral expressions for the remaining kinds of light transforms:
\begingroup
\allowdisplaybreaks
\begin{align}
    \bbL|z,\bz,h,\bh\ra_\s &= \int_{\R^2}\d^2\bar\kappa\;|[\bar\kappa\,\bar\zeta]|^{2\bh-2}\,|\zeta,\bar\kappa,J\ra\,,\label{Lbssh}\\
    \bL|z,\bz,h,\bh\ra_\at &= \int_{\R^2}\d^2\kappa\;|\la\kappa\,\zeta\ra|^{2h-2}\,\sgn(\la\kappa\,\zeta\ra)\,|\kappa,\bar\zeta,J\ra\,,\label{Lash}\\
    \bbL|z,\bz,h,\bh\ra_\at &= \int_{\R^2}\d^2\bar\kappa\;|[\bar\kappa\,\bar\zeta]|^{2\bh-2}\,\sgn([\bar\kappa\,\bar\zeta])\,|\zeta,\bar\kappa,J\ra\,,\label{Lbash}
\end{align}
\endgroup
where $[\bar\kappa\,\bar\zeta] \equiv \eps_{\dot\beta\dal}\bar\kappa^{\dal}\bar\zeta^{\dot\beta}$. Written in this manner, the light transforms are reminiscent of Witten's half-Fourier transforms to twistor/dual twistor space \cite{Witten:2003nn} that were key in the discovery of many mathematical structures hidden in scattering amplitudes. A preliminary study of the connections between the two subjects was performed in \cite{Sharma:2021gcz}. In what follows, we will instead focus on putting equations \eqref{Lssh}-\eqref{Lbash} to computational use.

\subsection{4-gluon amplitude}

Color-stripped tree-level gluon amplitudes find many elegant descriptions in the literature. The description based on the modern subject of Grassmannian formulae \cite{Arkani-Hamed:2012zlh} is best suited for computing light transformed celestial correlators.

The momentum space 4-gluon amplitude is given by the Parke-Taylor formula,
\be\label{4ptPT}
A(1^-\,2^-\,3^+\,4^+) = \frac{\la1\,2\ra^4}{\la1\,2\ra\la2\,3\ra\la3\,4\ra\la4\,1\ra}\;\delta^4\biggl(\sum_{i=1}^4\lambda_i\bar\lambda_i\biggr)\,.
\ee
This admits an integral representation as an integral over a Grassmannian $\Gr(2,4)$ \cite{ArkaniHamed:2009si,ArkaniHamed:2009dn}
\begin{multline}\label{4ptgr}
A(1^-\,2^-\,3^+\,4^+) = \int\displaylimits_{\R^4}\frac{\d c_{13}\,\d c_{14}\,\d c_{23}\,\d c_{24}}{c_{13}c_{24}(c_{13}c_{24}-c_{23}c_{14})}\;\delta^2(\bar\lambda_1+c_{13}\bar\lambda_3+c_{14}\bar\lambda_4)\,\delta^2(\bar\lambda_2+c_{23}\bar\lambda_3+c_{24}\bar\lambda_4)\\
\times\delta^2(\lambda_3-c_{13}\lambda_1-c_{23}\lambda_2)\,\delta^2(\lambda_4-c_{14}\lambda_1-c_{24}\lambda_2)\,.
\end{multline}
The integration variables $c_{13},c_{14}$, etc.\ are known as ``link variables'' and act as affine coordinates on $\Gr(2,4)$ when arranged into the $2\times4$ matrix
\be
C = \begin{pmatrix}1&&0&&c_{13}&&c_{14}\\
0&&1&&c_{23}&&c_{24}\end{pmatrix}\,.
\ee
Integrating out the last four delta functions of \eqref{4ptgr} fixes
\be
c_{13} = \frac{\la3\,2\ra}{\la1\,2\ra}\,,\quad c_{14} = \frac{\la4\,2\ra}{\la1\,2\ra}\,,\quad c_{23} = \frac{\la1\,3\ra}{\la1\,2\ra}\,,\quad c_{23} = \frac{\la1\,4\ra}{\la1\,2\ra}
\ee
and results in a Jacobian of $\la1\,2\ra^{-2}$. For these values of the link variables, the remaining integrand of \eqref{4ptgr} readily reduces to \eqref{4ptPT}.

We can use the Grassmannian formula \eqref{4ptgr} and the integrals \eqref{Lssh}-\eqref{Lbash} to compute the corresponding ambidextrously light transformed celestial amplitudes. For simplicity, take all four gluon states to be symmetric primaries with null momenta decomposed as
\be
\begin{split}
    p_i^{\al\dal} &= \zeta_i^{\al}\bar\kappa_i^{\dal}\,,\qquad\zeta_i^\al=(1,z_i)\,,\qquad i=1,2\,,\\
    p_i^{\al\dal} &= \kappa_i^{\al}\bar\zeta_i^{\dal}\,,\qquad\bar\zeta_i^{\dal}=(1,\bar z_i)\,,\qquad i=3,4\,.
\end{split}
\ee
We then need to compute
\begin{multline}
    L(1^-_\s\,2^-_\s\,3^+_\s\,4^+_\s) = \prod_{i=1}^2\int_{\R^2}\d^2\bar\kappa_i\,|[\bar\kappa_i\,\bar\zeta_i]|^{\Delta_i-1}\prod_{j=3}^4\int_{\R^2}\d^2\kappa_j\,|\la\kappa_j\,\zeta_j\ra|^{\Delta_j-1}\\
    \times\int_{\R^4}\frac{\d c_{13}\,\d c_{14}\,\d c_{23}\,\d c_{24}}{c_{13}c_{24}(c_{13}c_{24}-c_{23}c_{14})}\,\prod_{i=1}^2\delta^2\biggl(\bar\kappa_i+\sum_{k=3}^4c_{ik}\bar\zeta_k\biggr)\,\prod_{j=3}^4\delta^2\biggl(\kappa_j-\sum_{l=1}^2c_{lj}\zeta_l\biggr)\,.
\end{multline}
Integrating out $\bar\kappa_1,\bar\kappa_2$ and $\kappa_3,\kappa_4$ instead of the $c_{ij}$, we immediately land on the integral formula
\begin{multline}\label{l4gr}
    L(1^-_\s\,2^-_\s\,3^+_\s\,4^+_\s) = \int\displaylimits_{\R^4}\frac{\d c_{13}\,\d c_{14}\,\d c_{23}\,\d c_{24}}{c_{13}c_{24}(c_{13}c_{24}-c_{23}c_{14})}\;|c_{13}\bz_{13}+c_{14}\bz_{14}|^{\Delta_1-1}\,|c_{23}\bz_{23}+c_{24}\bz_{24}|^{\Delta_2-1}\\
    \times|c_{13}z_{13}+c_{23}z_{23}|^{\Delta_3-1}\,|c_{14}z_{14}+c_{24}z_{24}|^{\Delta_4-1}\,.
\end{multline}
Similar calculations yield amplitudes containing anti-symmetric primaries without extra effort. We will focus on \eqref{l4gr} in the rest of this section.

To further study \eqref{l4gr}, we start by substituting
\be\label{csubs}
\begin{split}
    &c_{13}\mapsto\frac{1}{z_{14}}\frac{1}{\bar z_{23}}\,c_{13}\,,\qquad c_{14}\mapsto\frac{1}{z_{14}}\frac{\bar z_{13}}{\bar z_{14}\bar z_{23}}\,c_{14}\,,\\
    &c_{23}\mapsto\frac{z_{13}}{z_{14}z_{23}}\frac{1}{\bar z_{23}}\,c_{23}\,,\qquad c_{24}\mapsto \frac{z_{13}}{z_{14}z_{23}}\frac{\bar z_{13}}{\bar z_{14}\bar z_{23}}\,c_{24}\,.
\end{split}
\ee
This turns \eqref{l4gr} into
\begin{multline}\label{l4gr2}
    L(1^-_\s\,2^-_\s\,3^+_\s\,4^+_\s) = \left|\frac{z_{13}}{z_{14}}\right|^{\Delta_2+\Delta_3-2}\left|\frac{\bar z_{13}}{\bar z_{23}}\right|^{\Delta_1+\Delta_4-2}\frac{|z_{14}|^{1-\Delta_1}\,|z_{23}|^{1-\Delta_2}}{|\bar z_{23}|^{\Delta_3-1}\,|\bar z_{14}|^{\Delta_4-1}}\int\frac{\d c_{13}\,\d c_{14}\,\d c_{23}\,\d c_{24}}{c_{13}c_{24}(c_{13}c_{24}-c_{23}c_{14})}\\
    \times |c_{13}+c_{14}|^{\Delta_1-1}\left|c_{23}+\frac{c_{24}}{1-\bz}\right|^{\Delta_2-1}\,|c_{13}+c_{23}|^{\Delta_3-1}\left|c_{14}+\frac{c_{24}}{1-z}\right|^{\Delta_4-1}\,,
\end{multline}
where we have defined the cross ratios
\be
z = \frac{z_{12}z_{34}}{z_{13}z_{24}}\,,\qquad\bz = \frac{\bz_{12}\bz_{34}}{\bz_{13}\bz_{24}}\,.
\ee
To perform the integrals, we introduce a judicious choice of integration variables $t_1,t_2,x$ defined by
\be
c_{14} = \frac{1-\bz}{t_1-1}\,c_{13}\,,\qquad c_{23} = \frac{t_2-1}{1-x}\,c_{13}\,,\qquad c_{24} = \frac{(1-\bz)(t_2-1)}{t_1-1}\,c_{13}\,.
\ee
The $c_{13}$ integral then separates out to give
\be
\int_{-\infty}^\infty\frac{\d c_{13}}{|c_{13}|}\,|c_{13}|^{\sum_{i=1}^4(\Delta_i-1)} = 4\pi\,\delta(\im\beta)
\ee
where $\beta = \Delta_1+\Delta_2+\Delta_3+\Delta_4-4$ is the net boost weight.

The resulting correlator decomposes into a conformally covariant prefactor times a function of the cross ratios:
\be\label{Lssss}
L(1^-_\s\,2^-_\s\,3^+_\s\,4^+_\s) = 4\pi\,\delta(\im\beta)\,K(z_i,\bar z_i)\,z^{\sh_{12}}\bz^{\bar\sh_{12}}\,\sL(z,\bz)\,,
\ee
where the covariant kinematic prefactor $K$ is chosen to be symmetric in $1,2$ and $3,4$ separately,
\be\label{Ksc}
K(z_i,\bar z_i) = \biggl|\frac{z_{24}\bar z_{14}}{z_{14}\bar z_{24}}\biggr|^{\sh_{12}}\biggl|\frac{z_{14}\bar z_{13}}{z_{13}\bar z_{14}}\biggr|^{\sh_{34}}\biggl|\frac{\bar z_{12}}{z_{12}}\biggr|^{\sh_1+\sh_2}\biggl|\frac{\bar z_{34}}{z_{34}}\biggr|^{\sh_3+\sh_4}\,,
\ee
the $\sh_i$ are the left-moving conformal weights of the light-transformed primaries:
\be
\begin{split}
    \sh_1 &= \frac{\Delta_1-1}{2}\,,\quad \sh_2 = \frac{\Delta_2-1}{2}\,,\quad\sh_3 = \frac{1-\Delta_3}{2}\,,\quad \sh_4 = \frac{1-\Delta_4}{2}\,,
\end{split}
\ee
and $\sh_{ij}=\sh_i-\sh_j$. The right-moving weights will be denoted $\bar\sh_i$, so that $\bar\sh_{ij}=\bar\sh_i-\bar\sh_j$, though we have not made them explicit in \eqref{Ksc} since they are related to the left-moving weights in the spin 1 case: $\bar\sh_i=-\sh_i$. Most importantly, the dynamical data $\mathscr{L}(z,\bz)$ is found to be
\begin{multline}\label{Fgrfin}
\sL(z,\bz) = \left|\frac{z}{\bz}\right|^{\Delta_2-1}\left|\frac{1-z}{1-\bz}\right|^{\Delta_1+\Delta_3-2}\\
\times\int_{-\infty}^\infty\frac{|1-x|^{\Delta_1+\Delta_4-2}}{x\,(x-1)}\;\sJ(\Delta_2,\Delta_3,\Delta_4\,|\,x,z)\,\sJ(\Delta_3,\Delta_2,\Delta_1\,|\,x,\bz)\;\d x\,,
\end{multline}
where $\sJ$ denotes the integral
\be\label{masterint}
\sJ(a,b,c\,|\,u,v) \vcentcolon= \int_{-\infty}^\infty\d t\,|1-t|^{a-2}\,|u-t|^{b-1}\,|v-t|^{c-1}\,.
\ee
The result \eqref{Fgrfin} displays a beautiful factorization property into chiral and anti-chiral halves, convolved together through the $x$ integral. As our derivation shows, this simplicity originates from the underlying Grassmannian geometry of the 4-gluon amplitude \eqref{4ptgr}.

\eqref{Fgrfin} can be shown to match the result obtained in the parallel work of \cite{De:2022gjn} by applying to $\sJ(\Delta_2,\Delta_3,\Delta_4\,|\,x,z)$ the identity
\be
\sJ(a,b,c\,|\,u,v) = |1-u|^{a+b-2}|1-v|^{a+c-2}\,\sJ(4-a-b-c,c,b\,|\,u,v)\,.
\ee
This identity can be derived by substituting $t\mapsto1-(1-u)(1-v)/(1-t)$ in \eqref{masterint}. The integral \eqref{masterint} was referred to as a \emph{four marked point integral} in \cite{Hu:2022syq}, where it was evaluated by breaking the integration domain into regions on which $1-t,u-t,v-t$ take definite signs. When the dust settles, one ends up with
\be
    \sJ(a,b,c\,|\,u,v) = \Theta\!\left(\frac{u-1}{1-v}\right)\,\sJ_+(a,b,c\,|\,u,v)+\Theta\!\left(\frac{1-u}{1-v}\right)\,\sJ_-(a,b,c\,|\,u,v)\,.
\ee
The two pieces $\sJ_\pm$ admit expressions in terms of linear combinations of Gauss hypergeometric functions:
\begin{align}
    \sJ_+(a,b,c\,|\,u,v) &= \frac{\sC(a-1,\,b)}{|1-u|^{2-a-b}\,|1-v|^{1-c}}\;{}_2F_1\biggl(a-1,\,1-c\,;\,a+b-1\,;\,\frac{1-u}{1-v}\biggr)\nonumber\\
    &\quad+ \frac{\sC(a+b-2,\,c)}{|1-v|^{3-a-b-c}}\;{}_2F_1\biggl(3-a-b-c,\,1-b\,;\,3-a-b\,;\,\frac{1-u}{1-v}\biggr)\,,\label{Jplusdef}\\
    \sJ_-(a,b,c\,|\,u,v) &= \frac{\sC(b,\,c)}{|u-v|^{1-b-c}\,|1-u|^{2-a}}\;{}_2F_1\biggl(2-a,\,b\,;\,b+c\,;\,\frac{u-v}{u-1}\biggr)\nonumber\\
    &\hspace{-0.3cm}+ \frac{\sC(b+c-1,\,a-1)}{|1-u|^{3-a-b-c}}\;{}_2F_1\biggl(3-a-b-c,\,1-c\,;\,2-b-c\,;\,\frac{u-v}{u-1}\biggr)\,,\label{Jminusdef}
\end{align}
written in terms of coefficient functions built out of Euler Beta functions,
\be
\sC(a,b) = B(a,b) + B(a, 1-a-b) + B(b, 1-a-b)\,.
\ee
As probed in more detail in \cite{Hu:2022syq}, the $\sC(a,b)$ contain information about celestial OPE coefficients.

Performing the $x$ integral directly proves to be unilluminating. Instead, in the next section, we will transform \eqref{Fgrfin} to alpha space, just as was done for the four-scalar celestial amplitude in \cite{Atanasov:2021cje}. This will help us make a preliminary study of the spectrum of operators exchanged in the correlator of four light transformed symmetric primaries.

%%%%%%%%%%%%%%%%%%%%%

\subsection{Alpha space decomposition}

Conformal covariance of correlators in a CFT implies that we can decompose any 4-point function into a so-called \emph{conformal partial wave expansion}: the conformally invariant part of the correlator can be expressed as an expansion over the orthonormal basis of solutions to the two-particle conformal Casimir equation. In the context of this work, the CCFT operators we have defined transform in irreducible representations of $\SL(2,\R) \times \SL(2,\R)$, therefore we want to decompose the four-point function in $\SL(2,\R)$ conformal blocks, and thus we will want to work in terms of the $SL(2,\R)$ Casimir equation (or rather, study its Sturm-Liouville problem). This technology was formalised in \cite{Hogervorst:2017sfd, Rutter:2020vpw} and it is generally known as the \emph{alpha space decomposition}.\footnote{In this case, the conformal partial waves have support for real $z,\bar z\in [0,1]\times[0,1]$ making them natural for correlators on the celestial torus. By contrast the traditional Euclidean~\cite{PhysRevD.13.887,Dobrev1977} and Lorentzian inversion~\cite{Caron-Huot:2017vep,Simmons-Duffin2018} formulae conformal waves are supported on the complex plane or its analytic continuation, rather than starting with operators intrinsically defined in $\mathbb{R}^{1,1}$.  Furthermore, the Lorentzian inversion formula gives us OPE coefficients at a given dimension and spin, but doesn't tell us which primaries will appear in the OPE expansion.  Here we are interested in better understanding the operator content of celestial CFT, rather than assuming it matches what follows from the single particle spectrum of the free theory~\cite{Pasterski:2017kqt}.
}

\subsubsection{Alpha Space: a crash course }

Alpha space decomposition was introduced for one dimensional theories in \cite{Hogervorst:2017sfd}. We will briefly introduce it as it is a stepping stone for the two-dimensional case. Consider a four-point correlation function in the $s$-channel\footnote{We will be using as an example the $s$ channel case. The $t$ channel case can be found in \cite{Hogervorst:2017sfd}.} in a one-dimensional CFT. From first principles in the theory of CFTs, we know that this four-point function can be split into two parts: a conformally covariant part, and a conformally invariant part:

\begin{equation}
    F_{1234}^s (z_1,\dots,z_4)= K^s(z_1,z_2,z_3,z_4) \cF^s(z)\,, \,\quad z= \frac{z_{12}z_{34}}{z_{13}z_{24}}\,.
\end{equation}

The conformally invariant part, $\cF^s$, can be decomposed into an infinite sum of $\SL(2,\R)$ blocks, $k_h^s(z)$, times the appropriate OPE coefficients,
\be\label{confblock}
\cF^s(z) = \sum_\O C_{12\O}C_{34\O} k^s_{h_\O}(z), \quad k^s_h=z^{h+a} \,_2F_1(h+a,h+b;2h;z),%\quad a=-h_{12},~b=h_{34}.
\ee
where $a=-h_{12},~b=h_{34}$. To guarantee OPE convergence in the crossing equation, we will restrict $0<z<1$. 

Although the conformal blocks themselves are eigenfunctions of the quadratic Casimir of $\SL(2,\R)$, they do not form a complete basis for the space of solutions: we need to instead use the so-called \emph{conformal partial waves}.

Following \cite{Hogervorst:2017sfd}, the inner product we are working with is defined as
\begin{equation}\label{innerprodalpha}
    \langle f,g \rangle_s = \int_0^1 \d z\, w_s(z) \overline{f(z)} g(z)\,,\qquad w_s(z) = \frac{(1-z)^{a+b}}{z^{2+2a}}\,,
\end{equation}
and one arrives at the following eigenfunctions 
\begin{equation}
\begin{aligned}
    \Psi^s_\alpha(z) &= \frac12 \left( Q(\al)k^s_{1/2+\al}(z) + Q(-\al)k^s_{1/2-\al}(z)\right)\\ &=\,_2F_1\!\left(\frac12+a+\al, \frac12+a-\al;1+a+b; \frac{z-1}{z}\right)\,,
    \end{aligned}
    \label{confwaves}
\end{equation}
where 
\be
Q(\al)=\frac{2\Gamma(-2\al)\Gamma(1+a+b)}{\Gamma(1/2+a-\al)\Gamma(1/2+b-\al)} \, .
\ee

Utilising these, the alpha space transform and inverse transform of $\cF^s(z)$ are defined as follows:
\be \label{alphatransformofF}
\hat{\cF}^s(\al)= \int_0^1 \d z\, w_s(z)\, \cF^s(z)\,\Psi_\al^s(z)\,,\qquad \cF^s(z)=\int\frac{[\d \al]}{N_s(\al)}\,\hat{\cF}^s(\al)\,\Psi_\al^s(z)\,,
\ee
with 
\be 
N_s(\al) = \frac{2\Gamma(\pm2\al)\Gamma^2(1+a+b)}{\Gamma(\frac12+a\pm\al)\Gamma(\frac12+b\pm\al)}\,,\,\quad \Gamma(p\pm q)\vcentcolon= \Gamma(p+q)\Gamma(p-q).
\ee

The authors of \cite{Rutter:2020vpw} argued that for the two dimensional functions $f(z,\bz)$ defined on the Lorentzian square we can define the alpha space transform as
\begin{align}\label{2dalphaspace}
\hat{f}(\al,\bal)&= \int_0^1 \d z\, w_s(z) \Psi_\al(z) \int_0^1\d\bar z\, \bw_s(\bz) \Psi_\bal(\bz) f(z,\bz)\,,\\
f(z,\bz) & = \int[d\al]\int[d\bal]\, \frac{4\Psi_\al(z)\Psi_\bal(\bz)}{N_s(\al)\bar N_s(\bal)}\hat{f}(\al,\bal)\,,\label{inverse2dalphaspace}
\end{align}
where $\bar w_s$ and $\bar N_s$ are obtained from $w_s$ and $N_s$ by replacing $h_i\mapsto\bar h_i$. From here, following the argument of \cite{Rutter:2020vpw}, we can find the contributions to the conformal block expansion through twin poles in the alpha space density:
\begin{equation}\label{alphapoles}
    \hat{f}(\al, \bal)\supset \frac{R}{(\al-h+1/2)(\bal-\bh+1/2)}+ (\al \leftrightarrow \bal)\,
\end{equation}
where $R$ denotes the residues of these twin poles.
Note that these functions shall be invariant under Weyl reflections, so we actually have a set of eight twin poles on alpha space for every conformal block (exchanging $\al \leftrightarrow -\al$ and $\bal \leftrightarrow - \bal$ independently).

Due to how these give rise to a  corresponding conformal block in position space, we can read off the OPE coefficients as
\begin{equation}
    C_{12\O}C_{34\O} = \frac{4R}{Q(-h+1/2) \bar Q (-\bh+1/2)},
    \label{eq: ope from residue}
\end{equation}
taking into account that we are only considering one pair of the eight Weyl reflections, as they all give rise to a single operator. 

\subsubsection{4-gluon amplitude in alpha space}\label{sec:alpha}

We would now like to decompose the amplitude \eqref{Lssss} to examine features of the 2D CFT it encodes. 

To begin, we have to make an important note: what we are looking to transform is the dynamical data $\sL(z,\bz)$ from \eqref{Fgrfin}, which includes an integral over the variable $x$. This integral has singularities at $x=0,1$. Because we are integrating over $z,\bz\in(0,1)$, it is natural to define it by dividing it in two sectors:
\begin{itemize}
    \item A first sector where $x\in (-\infty, 1)$ and the factor $\sJ(z,\bz)$ from \eqref{masterint} is simply $\sJ_-(z,\bz)$ given by \eqref{Jminusdef}. The corresponding contribution to $\sL(z,\bz)$ will be denoted $\sL_-(z,\bz)$.
    \item A second sector where $x\in (1,\infty)$ and the factor $\sJ(z,\bz)$ from \eqref{masterint} is simply $\sJ_+(z,\bz)$ given by \eqref{Jplusdef}. The corresponding contribution to $\sL(z,\bz)$ will be denoted $\sL_+(z,\bz)$.
\end{itemize}
The sector $(-\infty,1)$ will be further divided into the sectors $(-\infty,0)$ and $(0,1)$. In total, this amounts to defining the $x$ integral via principal value prescription.

Nevertheless, the $x$ integral does not seem to have an analytic solution that we can use to then perform the alpha space transform of $\sL$ and obtain the polology we want. Therefore, we are going to assume convergence of the $x$ integral (after appropriately adding the three sectors if needed) and commute it with the alpha-space transform. We also encounter issues with doing the latter analytically. However, we find that we can solve it via a mix of Mellin-Barnes representations and series expansions of the hypergeometric functions involved. Let's see this with an example.

\paragraph{An example: a term of $\sJ_+$.} We have chosen to show the computation for one of the terms of $\sJ_+$. All others are performed in a similar fashion, and so is the sector of $x\in(-\infty,0)$ for $\sJ_-$. The sector where $x\in (0,1)$ is especially complicated to alpha transform, but we still can find its pole structure after careful treatment of the integrals.

The total integral we want to do is as follows:
\begin{equation}\label{inttot}
    \begin{aligned}
        &\int_0^1 \d z\, w_s(z)\int_0^1\d\bz \, \bw_s(\bz)\; \sL_+(z,\bz)\, \Psi_\al(z)\,\Psi_\bal(\bz) \\
        &= \int_1^\infty \d x\, \frac{(x-1)^{\Delta_1+\Delta_4-2}}{x(x-1)} \int_0^1 \d z\,w_s(z)\,\Psi_\al(z)\,\frac{\sJ_+(\Delta_2,\Delta_3,\Delta_4\,|\, x,z)}{z^{1-\Delta_2}(1-z)^{2-\Delta_1-\Delta_3}}\\
        &\hspace{4cm}\times \int_0^1 \d\bz \,\bw_s(\bz)\,\Psi_\bal(\bz)\,
        \frac{\sJ_+(\Delta_3,\Delta_2,\Delta_1\,|\,x,\bz)}{\bz^{\Delta_2-1}(1-\bz)^{\Delta_1+\Delta_3-2}}\,.
    \end{aligned}
\end{equation}
Here, the kernels $w_s(z)$, $\bar w_s(\bz)$ are evaluated at the light transformed weights $\sh_i,\bsh_i$. The factor of $(1-z)^{-\sh_{12}+\sh_{34}}(1-\bz)^{-\bsh_{12}+\bsh_{34}}$ in $w_s(z)\bar w_s(\bz)$ cancels out completely against the prefactor $(\frac{1-z}{1-\bz})^{\Delta_1+\Delta_3-2}$ on the support of $\sh_1+\sh_2=\sh_3+\sh_4$.

Let us for now exclude the $x$ integral and leave it for last, and focus only on the $z$ integral (under the assumption our integrals are indeed convergent) 
\be
\int_0^1 \d z\,w_s(z)\,\Psi_\al(z)\,\frac{\sJ_+(\Delta_2,\Delta_3,\Delta_4\,|\, x,z)}{z^{1-\Delta_2}(1-z)^{2-\Delta_1-\Delta_3}}.
\ee
The $\bz$ integral is analogous. To illustrate, let us focus on one term in the $\sJ_+$ integral, containing the following sub-integral: 
\begin{equation}
\begin{aligned}
    \hat{\sJ}_+(\al,x)=\int_0^1 &\d z\; (1-z)^{-2 \sh_4} z^{ 2\sh_1-2} (x-1)^{2\sh_2 -2\sh_3} \\
    &\times\,{}_2F_1\left(2 \sh_2, 2 \sh_4;1-2 \sh_{23};\frac{1-x}{1-z}\right)\\
    &\times{}_2F_1\left(\frac12-\alpha -\sh_{12},\frac12+\alpha
   -\sh_{12};1-\sh_{12}+\sh_{34};\frac{z-1}{z}\right).
\end{aligned}
\label{term3-jplus}
\end{equation}
Since the results we need are only related to its polar structure in $z$, we are going to try and separate the $x$-dependence. We expand the ${}_2F_1$ coming from the amplitude (first line) as a power series on its argument, and then we express the partial wave (second line) in its Mellin-Barnes form
\begingroup
\allowdisplaybreaks
\addtolength{\jot}{1em}
\begin{align}
        &\hat{\sJ}_+ (\al,x)= \sum_{q=0}^\infty (-1)^{q}\int_{-\im\infty}^{\im\infty} \frac{\d t}{2\pi\im} \int_0^1 \d z \, z^{ 2\sh_1-2} (x-1)^{2\sh_{23}+q}\nonumber \\
        & \times(1-z)^{-2 \sh_4-q}\frac{ \Gamma \left(1+2 \sh_{23}\right) \Gamma \left(2 \sh_2+q\right) \Gamma \left(2 \sh_4+q\right) }{\Gamma \left(2 \sh_2\right) \Gamma \left(2 \sh_4\right) \Gamma (1+q) \Gamma \left(1+2 \sh_{23}+q\right)}\nonumber \\
        & \times \left(\frac{1-z}{z}\right)^t\frac{ \Gamma \left(1-\sh_{12}+\sh_{34}\right)  \Gamma (-t) \Gamma \left(\frac12 -\alpha -\sh_{12}+t\right) \Gamma \left(\frac12 +\alpha -\sh_{12}+t\right)}{ \Gamma \left(\frac12-\alpha -\sh_{12}\right) \Gamma \left(\frac12 +\alpha -\sh_{12}\right) \Gamma \left(1-\sh_{12}+\sh_{34}+t\right)}\nonumber\\
        =& \sum_{q=0}^\infty(-1)^{q}\int_{-\im\infty}^{\im\infty}\frac{\d t}{2\pi\im}  \,(x-1)^{2 \sh_{23}+q}\\
        &\times \frac{  \Gamma \left(1+2 \sh_{23}\right) \Gamma \left(1-\sh_{12}+\sh_{34}\right) \Gamma \left(2 \sh_2+q\right) \Gamma \left(2 \sh_4+q\right)    }{  \Gamma \left(2 \sh_2\right) \Gamma \left(2 \sh_4\right) \Gamma (q+1) \Gamma \left(\frac12-\alpha -\sh_{12}\right) \Gamma \left(\frac12+\alpha -\sh_{12}\right) \Gamma \left(1+2 \sh_{23}+q\right) \Gamma \left(2 \sh_{14}-1\right) }\nonumber\\
        &\times \frac{\Gamma (-t) \Gamma \left(2 \sh_1-1-t\right)\Gamma \left(1-2\sh_4-q+t\right)\Gamma \left(\frac{1}{2}-\alpha -\sh_{12}+t\right) \Gamma \left(\frac12+\alpha -\sh_{12}+t\right)}{\Gamma\left(1-\sh_{12}+\sh_{34}+t\right)}.\nonumber
\end{align}
\endgroup
At this point, we can identify the Mellin-Barnes integral as a Meijer-G function:
\begin{multline}
     \int_{-\im\infty}^{\im\infty}\frac{\d t}{2\pi\im}\, \frac{\Gamma (-t) \Gamma \left(2 \sh_1-1-t\right)\Gamma \left(1-2\sh_4-q+t\right)\Gamma \left(\frac{1}{2}-\alpha -\sh_{12}+t\right) \Gamma \left(\frac12+\alpha -\sh_{12}+t\right)}{\Gamma\left(1-\sh_{12}+\sh_{34}+t\right)}\\
= G_{3,3}^{2,3}\left(1\left|
\begin{array}{c}
 \alpha +\sh_{12}+\frac{1}{2},-\alpha +\sh_{12}+\frac{1}{2},2 \sh_4+q \\
 2 \sh_1-1,0,\sh_{12}-h_{34} \\
\end{array}
\right.\right)\,.
\end{multline}
This Meijer G can be expressed as a sum over two terms, which leaves us with the following expression overall:
\begingroup
\addtolength{\jot}{1em}
\begin{align}
         &\hat{\sJ}_+(\al,x) = \sum_{q=0}^\infty (x-1)^{2\sh_{23}+q} (-1)^{q}\nonumber\\
        &\times  \frac{  \Gamma \left(1+2 \sh_{23}\right) \Gamma \left(1-\sh_{12}+\sh_{34}\right) \Gamma \left(2 \sh_2+q\right) \Gamma \left(2 \sh_4+q\right)    }{  \Gamma \left(2 \sh_2\right) \Gamma \left(2 \sh_4\right) \Gamma (q+1) \Gamma \left(1+2 \sh_{23}+q\right) \Gamma \left(2 \sh_{14}-1\right) }\nonumber \\
         & \times \bigg[\;\frac{\Gamma \left(1-2 \sh_1\right)\highlight{ \Gamma \left(-\frac12-\alpha +\sh_1+\sh_2\right) \Gamma \left(-\frac12\alpha +\sh_1+\sh_2\right)} \Gamma \left(2 \sh_{14}-q\right) }{\Gamma \left(2\sh_3\right)\Gamma \left(\frac12-\alpha -\sh_{12}\right) \Gamma \left(\frac12+\alpha -\sh_{12}\right) }\nonumber\\
         & \quad \quad \times \, _3F_2\left(-\frac12-\alpha +\sh_1+\sh_2,-\frac12+\alpha +\sh_1+\sh_2,2 \sh_{14}-q;2 \sh_1,2\sh_3;1\right)\nonumber\\
        &\quad + \frac{\Gamma \left(2 \sh_1-1\right)  \Gamma \left(-q-2 \sh_4+1\right) }{\Gamma \left(-\sh_{12}+\sh_{34}+1\right) }\nonumber\\
        &\quad\times \, _3F_2\left(\frac12-\alpha -\sh_{12},\frac12+\alpha -\sh_{12},1-q-2 \sh_4;2-2 \sh_1,1-\sh_{12}+\sh_{34};1\right)\bigg].
       \label{jplus-term1-finalform}
\end{align}
\endgroup
We have highlighted in a \highlight{\text{light gray}} the only terms in this expression that yield poles, which are two gamma functions. In our case, to determine the spectrum of operators present in the conformal block expansion, we are only going to look at values of $\al$ yielding poles in first of the two highlighted gamma functions, as the other will correspond to Weyl reflections of these operators. The poles in $\alpha$ we choose are $\alpha= -1/2+\sh_1+\sh_2+\N$.  These poles indicate the presence of an infinite tower of states with left-moving dimension $h = \sh_1+\sh_2+\N$. 

At this stage, one might wonder whether it is beneficial to also do the $x$ integral explicitly. In principle, this is necessary to determine the OPE coefficients. However, we end up with relatively obscure expressions, of which we showcase an example in Appendix \ref{app:ope}, where we compute the contribution to the OPE coefficients of the term \eqref{jplus-term1-finalform} and one of its antiholomorphic counterparts. The resulting function of a generic term of the $\sL$ integral has a shape compatible with a generalized hypergeometric function of four variables (the likes of a Meijer-G function of four variables), from which one can read the same polar structure, with more difficulty. We will leave further explorations in this direction to future adventurers.

\subsubsection{Spectrum}

We obtain the following sets of poles in our integrals, noting that these come from the alpha-transform of $\sL(z,\bz)$ defined in \eqref{Fgrfin} in all regions of $x$:
\begin{itemize}
    \item Poles at $\alpha=-\frac{1}{2} +\sh_1+\sh_2+\mathbb{N} $
    \item Poles at $\bal = -\frac12 +\bsh_1+\bsh_2+ \mathbb{N}$
    \item Poles at $\al = \frac{\Z^+}{2}$
    \item Poles at $\bal = \frac{\Z^+}{2}$
\end{itemize}
Meanwhile, there are poles at the Weyl reflections of these values, but these represent the same operators.

Note that in the example shown before we can only find the first set of poles, and these are the only ones that appear in the two main sectors of the $x$ integral, $x\in(-\infty, 0) \cup (1,\infty)$, as the functions $\sJ(x,z),\sJ(x,\bz)$ in those regimes have an analytic structure that leads to the same polar structure in $\al$. The poles at half-integer $\al$ values (and the corresponding $\bal$ ones) are only found in the $x\in(0,1)$ regime, and they are related to a unique feature in this regime: the presence of $| x-z|$, which in this region partitions the $z$-integral in two subsectors $z<x$ and $z>x$ (in the other sectors, $|x-z|$ always attains a fixed sign because $z\in(0,1)$). This change causes a critical change in the analytic structure of the $\al$-transform which leads to an appearance of the new poles. They are simply gamma functions, just as the ones shown in the example, but the overall formulas are more complicated. 

The poles in $\bar\al$ are obtained by performing the $\bz$ integral in \eqref{inttot}, in the same way but independently of the poles in $\al$. 

As shown in \eqref{alphapoles}, the poles in $\al,\bar\al$ are related to dimensions of the exchanged operators by a shift of a $\frac12$. Hence, the polar structure listed above indicates the presence of the following two towers of operators
\begin{enumerate}
    \item $h = \sh_1+\sh_2+\N =  h_1+h_2+\N$,\\ $\bh =  \bsh_1+\bsh_2+\N = 2 - \bh_1 -\bh_2+\N$,
    \item $h = (1+\Z^+)/2$, $\bh = (1+\Z^+)/2$.
\end{enumerate}
These correspond to operator exchanges with dimensions and spins parametrized by a pair of integers $K,L$:
\begin{align}
&\bullet\;(\Delta,J) = 
(K,\Delta_1+\Delta_2-2+L)\,,\qquad K\geq1\,,\;\;1-K\leq L\leq K-1\,,\\
&\bullet\;(\Delta,J) = \bigg(\frac{3+K}{2},\frac{L}{2}\bigg)\,,\qquad\hspace{1.45cm} K\geq1\,,\;\;1-K\leq L\leq K-1\,.
\end{align}
The $K=L=0$ operator from the first set of poles corresponds to the $\bbL$ light transform of a negative helicity gluon of weight $\Delta_1+\Delta_2-1$. This is consistent with expectations from celestial OPE \cite{Pate:2019lpp}. The other exchanges in the first set agree (to varying degrees) with the exchanges observed in the conformal block decompositions studied in \cite{Fan:2021isc,Fan:2021pbp,Hu:2022syq}. 

Among the second set of exchanges, the ones for which $L=K-1$ are similar in character to the discrete set of operator exchanges observed in the conformal block expansion of scalar celestial amplitudes in $t$-channel blocks \cite{Atanasov:2021cje}. It will be very interesting to understand the precise relation between the two, as well as identify the bulk states dual to these operators. It is plausible that they are related to soft gluon exchanges in the $t$-channel, as the wedge condition $1-K\leq L\leq K-1$ is reminiscent of generators of the $S$-algebra \cite{Strominger:2021mtt}.

\section{Conclusions} \label{sec:conclusions}
Now that we've constructed an improved ambidextrous basis, written the generic form of low-point massless amplitudes in this basis, and examined the gluon example in detail, we will close by discussing two natural questions.

\paragraph{What is the natural basis of local operators?}
The question of which scattering basis should map to local operators in the dual theory has perplexed the early celestial literature~\cite{Pasterski:2017kqt}.  The manner in which the saddle point approximation identifies the position and momentum space celestial spheres for massless particles~\cite{He:2014laa} seems to prefer the usual Mellin transform dictionary, however the symmetry generators are constructed from what we now recognize as the shadow transforms~\cite{Cheung:2016iub,Kapec:2016jld}. Different basis choices change which symmetries are manifest~\cite{Banerjee:2022wht}.

However, since we are only looking at the global conformal symmetries here there is a sense in which we don't encounter one being preferred over the other. The correlation functions in the Mellin basis take non-standard distributional form~\cite{Pasterski:2017ylz}. The ambidextrous basis~\cite{Sharma:2021gcz} remedies the distributional nature of the low point correlators. Our improvement herein guarantees the  expected conformal covariance. With this in mind we want to emphasize that in exploring this ambidextrous basis we want to stay open to the possibility that these are the {\it local} operators of the 2D theory (rather than treating them as light transformed operators in 2D).  Mechanically this affects how one uses certain results from the 2D Lorentzian CFT literature. Using the extrapolate dictionary~\cite{Pasterski:2021dqe} we have a better chance of understanding the expected time ordering etc. of the dimensionally reduced theory living on the celestial torus.

\paragraph{How do our correlators differ from those of a standard 2D CFT?}

It is with this in mind that we want to interpret our results from section~\ref{sec:gluon_decomp}. The correlation functions themselves have a more standard form, however the Weyl reflections enacted by the light transforms yield an operator spectrum with continuous spins and discrete dimensions that we would not expect to get by Wick rotating an ordinary Euclidean CFT to Lorentzian signature.  When it comes to the structure of the correlators we see that the $x$ integral glues together holomorphic and antiholomorphic sectors.  This non standard form should arise from the fact that our correlators are intrinsically higher dimensional and have been projected down to 2D~\cite{Hu:2022txx}.  While the integrals we encountered in section~\ref{sec:alpha} are difficult to do analytically, in principle one could evaluate these numerically for an amplitude of interest.  Understanding which configurations of the external weights to probe is something we're building a better appreciation of from more general investigations of celestial amplitudes and their symmetries~\cite{Arkani-Hamed:2020gyp,Guevara:2021abz,Donnay:2022sdg}.   

In each case, we know that mechanically what we are doing is introducing additional intertwiners between representations of the Lorentz group. Once the underlying physics is extracted in a convenient basis, the appropriate representations of the Poincar\'e group should tell us how to convert this to the data here. Even if the ambidextrous basis turns out not to be preferred from the point of view of dimensionally reducing the conformal boundary, or introduces unnecessary additional intertwiners onto simpler representations of the partial waves, it seems very likely to have a deep connection to twistorial presentations of the same physics~\cite{Sharma:2021gcz,Adamo:2021lrv,Adamo:2021zpw,Sharma:2022arl,Costello:2022wso,Costello:2022upu,Bu:2022dis,Bittleston:2022jeq,Adamo:2022wjo}. Both of these approaches deserve additional exploration, and will be the focus of future work.

\section*{Acknowledgements}

We are grateful to Scott Collier for collaboration at early stages of this work.  We thank Murat Kologlu, Lionel Mason, Joseph McGovern,  Prahar Mitra, and Herman Verlinde for useful conversations. CJD is supported by a Schreder Music Award. SP is supported by the Celestial Holography Initiative at the Perimeter Institute for Theoretical Physics and has been supported by the Sam B. Treiman Fellowship at the Princeton Center for Theoretical Science. Research at the Perimeter Institute is supported by the Government of Canada through the Department of Innovation, Science and Industry Canada and by the Province of Ontario through the Ministry of Colleges and Universities.  AS is supported by a Black Hole Initiative fellowship, which is funded by the Gordon and Betty Moore Foundation and the John Templeton Foundation. He has also received support from the ERC grant GALOP ID: 724638 during earlier stages of this work.

\appendix

\section{From higher dimensions to the ambidextrous prescription}\label{app:higherd}

Certain aspect of celestial CFT, or amplitudes more generally, are either more interesting or easier when the bulk dimension is four. For example, the spinor-helicity variables and twistor constructions are engineered for this case.   Indeed, besides taming the distributional nature of the low point celestial amplitudes in the standard Mellin dictionary~\eqref{eq:mellinmap}, the ambidextrous light transforms were shown~\cite{Sharma:2021gcz} to have a close connection to the eponymous twistorial half-Fourier transform prescription of~\cite{Witten:2003nn}. Because one would hope our construction of a flat space hologram exists for generic dimensions, it is useful to keep track of what generalizes and what features are ${\cal M}_4/{\rm CCFT}_2$ specific.   The light transform is one case where the $d=2$ limit of the generic $d$ definition makes a difficult choice for us, landing us on the ambidextrous prescription, whereby the choice of which light ray to smear along is correlated with the operator's spin.

If we start in $d$ dimensions, the light transform of a spin $|J|$, dimension $\Delta$ operator takes the form~\cite{Kravchuk:2018htv}
\be\label{light}
\int_{-\infty}^{\infty}\d\alpha\, (-\alpha)^{-\Delta-|J|} \O\Bigl(x-\frac{z}{\alpha},z\Bigr)
\ee
where we've introduced the embedding space coordinates
\be
x^\mu = \frac{1}{\sqrt2}\,\left(1+w\bar w,w+\bar w,\bar w-w,1-w\bar w\right)
\ee
and contracted the $d$-dimensional vector indices with a null vector $z^\mu$ 
\be
\O(x,z)=\O^{\mu_1...\mu_J}z_{\mu_1}\cdots z_{\mu_J}\,.
\ee
These expressions simplify in 2D where
\be
\O_{\Delta,|J|}(x)=\O_{\Delta,w\cdots w}(x)\,,~~~\O_{\Delta,-|J|}(x)=\O_{\Delta,\bw\cdots\bw}(x)
\ee
in terms of the standard basis of polarization vectors
\be\label{polarization}
z_w^\mu=\p_wx^\mu,~~~z_\bw^\mu=\p_{\bar w}x^\mu
\ee
used for positive/negative helicity states. 

In terms of the lightcone coordinates we see that for positive helicity operators we have  $|J|=J$ and
\be
x^\mu\mapsto x^\mu-\alpha^{-1}z_w^\mu~~~ \Leftrightarrow~~~ w\mapsto w-\alpha^{-1},~~~\bw\mapsto \bw
\ee
so that, writing $\O_{h,\bh}(w,\bw) \equiv \O_{\Delta,J}(x)$,
\be\label{lightint}
\begin{split}
    \bL[\O_{h,\bh}](w,\bw)&=\int_{-\infty}^\infty \d\alpha\, (-\alpha)^{-2h}\,\O_{h,\bh}(w-\alpha^{-1},\bw)=\int_{w}^{\mathcal{T}w} \frac{\d z}{(z-w)^{2h-2}}\,\O_{h,\bh}(z,\bw)
\end{split}
\ee
where we've used  the change of variables 
\be\label{zchange}
z=w-\alpha^{-1},~~~\d z=\frac{\d\alpha}{\alpha^2}
\ee
as in section~\ref{sec:LT_boost} above. Here, $\mathcal{T}w$ denotes the point $w$ in the next Poincar\'e patch, just as our antipodal map $\sigma$ did on the celestial torus. Meanwhile for negative helicity operators we have $|J|=-J$
\be
x^\mu\mapsto x^\mu-\alpha^{-1}z_\bw^\mu~~~ \Leftrightarrow~~~ w\mapsto w,~~~\bw\mapsto \bw-\alpha^{-1}
\ee
so that
\be\label{lightintbar}
\begin{split}
    \bbL[\O_{h,\bh}](w,\bw)&=\int_{-\infty}^\infty\d\alpha\, (-\alpha)^{-2\bh}\,\O_{h,\bh}(w,\bw-\alpha^{-1})\\
    &=\int_{\bw}^{\mathcal{\bar T}\bw} \d\bz\, (\bz-\bw)^{2\bh-2}\,\O_{h,\bh}(w,\bz)\,.
\end{split}
\ee
We have suppressed the additional $\varepsilon$ labels on our celestial operators here. 

Now from our discussion above we know how to restore the $\varepsilon$ labels when these integration contours cross to the second Poincar\'e patch, as well as how to map the integral over the torus interval to the $\mathbb{Z}_2$ quotient. If we use the fact that $\cal T$ and $\cal \bar T$ should introduce the following phases~\cite{Kravchuk:2018htv,Guevara:2021tvr} 
\be\label{Tphase}
\mathcal{T}\O_{h,\bar{h}}(w,\bw)|0\rangle=e^{2\pi\im h}\O_{h,\bar{h}}(w,\bw)|0\rangle,~~~\mathcal{\bar T}\O_{h,\bar{h}}(w,\bw)|0\rangle=e^{2\pi\im \bh}\O_{h,\bar{h}}(w,\bw)|0\rangle
\ee
when we go to the second Poincar\'e patch, the powers of $(\bz-\bw)$ turn into the absolute values we used in~\eqref{Lal}-\eqref{Lbal}. Putting everything together we have
\begin{align}
    &\bL[\O^\veps_{h,\bar{h}}](w,\bar w) = \int_{\R}\frac{\d z}{|z-w|^{2-2h}}\;\O^{\sgn(z-w)\veps}_{h,\bar{h}}(z,\bar w)\,\qquad h-\bh = +|J|\,,\\
    &\bbL[O^\veps_{h,\bar{h}}](w,\bar w) = \int_{\R}\frac{\d\bar z}{|\bar z-\bar w|^{2-2\bar h}}\;\O^{\sgn(\bz-\bw)\veps}_{h,\bar{h}}(w,\bar z)\qquad h-\bh = -|J|\,.
\end{align}
This is our modified ambidextrous prescription. Note that in 2D these integral transformations can be applied to operators with either spin. When we want to distinguish the $z$ from the $\bz$ transform  it is still natural we will denote them by $\bL$ and $\bbL$, respectively, as we have done in the main text above. 

%%%%%%%%%%%%%%%%%%%%%%%%%%%%%%%
%%%%%%%%%%%%%%%%%%%%%%%%%%%%%%%

\section{Conformal block expansions of unitarity cuts}
\label{app:fac}

Conformal block and partial wave expansions of celestial amplitudes have been studied from various perspectives over the course of many works \cite{Nandan:2019jas, Atanasov:2021cje, Law:2020xcf, Fan:2021isc, Fan:2021pbp, Hu:2022syq, De:2022gjn,Garcia-Sepulveda:2022lga}. A precursor to these studies was the conformal partial wave expansion of the optical theorem for scalar celestial amplitudes in arbitrary dimensions performed in \cite{Lam:2017ofc}. In this section, we perform an analogous calculation for unitarity cuts of light transforms of 4D gluon celestial amplitudes. The highlight of our method is that -- in contrast to the partial wave expansion of \cite{Lam:2017ofc} -- the output of our light transforms will automatically emerge in the form of a conformal \emph{block} expansion. Furthermore, it will only involve a \emph{finite} number of operator exchanges, as one would expect in the factorization limit. This provides a self-contained toy example of the more involved analysis carried out in section \ref{sec:gluon_decomp}. It also further highlights the power of Grassmannian formulae for efficiently computing light transforms.

By unitarity, the residue of a momentum space tree amplitude $A(1^-2^-3^+4^+)$ at an $s$-channel factorization pole is a product of two 3-point tree amplitudes. This defines for us the $s$-channel unitarity cut
\be\label{facs}
\begin{split}
F_s(1^-2^-3^+4^+) &= \int\frac{\d^2\lambda\,\d^2\bar\lambda}{\vol\,\R^*}\,A(1^-2^-p^+)\,A(-p^-3^+4^+)\\
&= \frac{4\,\la1\,2\ra^2\,[3\,4]^2}{u}\,\delta(s)\,\delta^4\biggl(\sum_{i=1}^4p_i\bigg)\,,
\end{split}
\ee
where $s=2\,\la1\,2\ra[1\,2]$ and $u=2\,\la1\,4\ra[1\,4]$, and the integral is over on-shell phase space parametrized by spinor-helicity variables $\lambda_\al,\bar\lambda_{\dal}$, modulo the $\R^*$ little group. This is obtained by cutting the exchanged particle's $s$-channel propagator $1/s$ using standard Cutkosky rules. The extra delta function $\delta(s)$ is what will make it possible to evaluate all the light transform integrals analytically.

Just like the full gluon amplitude $A(1^-2^-3^+4^+)$, this can be expressed as an integral over the Grassmannian $\Gr(2,4)$:
\begin{multline}
    F_s(1^-2^-3^+4^+) = \int\frac{1}{\vol\,\R^*}\prod_{i=1}^4\frac{\d\al_i}{\al_i}\,\delta^2(\bar\lambda_1+\al_1\al_3\bar\lambda_3+\al_1\al_4\bar\lambda_4)\,\delta^2(\bar\lambda_2+\al_2\al_3\bar\lambda_3+\al_2\al_4\bar\lambda_4)\\
    \times\delta^2(\lambda_3-\al_1\al_3\lambda_1-\al_2\al_3\lambda_2)\,\delta^2(\lambda_4-\al_1\al_4\lambda_1-\al_2\al_4\lambda_2)\,\sgn(\la1\,2\ra[3\,4])\,.
\end{multline}
The Grassmannian elements $c_{ij}$ present in \eqref{4ptgr} have factorized into $c_{ij} = \al_i\al_j$. Also present are a pair of explicit sign factors $\sgn(\la1\,2\ra[3\,4])$ that are required to get the correct little group scaling in split signature. Most importantly, there is now a factor of $1/\vol\,\R^*$ that acts on these new integration variables as $(\al_1,\al_2,\al_3,\al_4)\mapsto (t\al_1,t\al_2,t^{-1}\al_3,t^{-1}\al_4)$.

Computing its light transform by taking all four external gluons to be symmetric primaries and applying the formulae \eqref{Lssh}, \eqref{Lbssh} yields
\begin{multline}
    L_s(1^-2^-3^+4^+) = \int\frac{1}{\vol\,\R^*}\prod_{i=1}^4\frac{\d\al_i}{\al_i}\,|\al_1\al_3\bz_{13}+\al_1\al_4\bz_{14}|^{\Delta_1-1}\,|\al_2\al_3\bz_{23}+\al_2\al_4\bz_{24}|^{\Delta_2-1}\\
    \times|\al_1\al_3z_{13}+\al_2\al_3z_{23}|^{\Delta_3-1}\,|\al_1\al_4z_{14}+\al_2\al_4z_{24}|^{\Delta_4-1}\,\sgn(z_{12}\bar z_{34})\,.
\end{multline}
We can use the $\R^*$ quotient to set $\al_2=1$ (this does not generate any Faddeev-Popov determinant). Then we can rescale $\al_4\mapsto\al_3\al_4$ to extract out the $\al_3$ integral, generating the prefactor
\be
\int_{-\infty}^\infty\frac{\d\al_3}{\al_3}\,\sgn(\al_3)\,|\al_3|^{\Delta_1+\Delta_2+\Delta_3+\Delta_4-4} = 4\pi\,\delta(\im\beta)\,,
\ee
where the factor of $\sgn(\al_3)$ comes from the transformation $\d\al_4/\al_4\mapsto|\al_3|\d\al_4/(\al_3\al_4)$. Here, $\beta=\sum_{i=1}^4(\Delta_i-1)$ is again the net boost weight. This leaves us with just two nontrivial hypergeometric integrals,
\begin{multline}
    L_s(1^-2^-3^+4^+) = 4\pi\,\delta(\im\beta)\,\sgn(z_{12}\bar z_{34})\int\frac{\d\al_1}{\al_1}\,|\al_1|^{\Delta_1-1}\,|\al_1z_{13}+z_{23}|^{\Delta_3-1}\,|\al_1z_{14}+z_{24}|^{\Delta_4-1}\\
    \times\int\frac{\d\al_4}{\al_4}\,|\al_4|^{\Delta_4-1} |\al_4\bz_{14}+\bz_{13}|^{\Delta_1-1}\,|\al_4\bz_{24}+\bz_{23}|^{\Delta_2-1}\,.
\end{multline}
The $\al_1$ and $\al_4$ integrals now completely decouple, as expected from the factorization limit.

Performing the substitutions
\be
\al_1\mapsto \frac{z_{24}}{z_{14}}\,(\al_1-1)\,,\qquad\al_4\mapsto\frac{\bz_{13}}{\bz_{14}}\,(\al_4-1)\,,
\ee
we simplify this to
\be
L_s(1^-2^-3^+4^+) = 4\pi\,\delta(\im\beta)\,S(z_i,\bz_i)\,z^{\sh_{12}}\bar z^{\bar\sh_{12}}\mathscr{L}_s(z,\bz)\,.
\ee
Here $z=z_{12}z_{34}/z_{13}z_{24}$, $\bz = \bz_{12}\bz_{34}/\bz_{13}\bz_{24}$, and $\sh_{ij}=\sh_i-\sh_j$, $\bar\sh_{ij}=\bar\sh_i-\bar\sh_j=-\sh_{ij}$ are differences of the light transformed conformal weights. $S(z_i,\bz_i)$ is the conformally covariant prefactor
\be
S(z_i,\bar z_i) = \sgn(z_{12}z_{24}z_{41}\bz_{13}\bz_{34}\bz_{41})\,\biggl|\frac{z_{24}\bar z_{14}}{z_{14}\bar z_{24}}\biggr|^{\sh_{12}}\biggl|\frac{z_{14}\bar z_{13}}{z_{13}\bar z_{14}}\biggr|^{\sh_{34}}\biggl|\frac{\bar z_{12}}{z_{12}}\biggr|^{\sh_1+\sh_2}\biggl|\frac{\bar z_{34}}{z_{34}}\biggr|^{\sh_3+\sh_4}\,,
\ee
which involves some conformally invariant signs but otherwise is the same as the prefactor $K(z_i,\bz_i)$ given in \eqref{Ksc}. The conformally invariant data $\mathscr{L}_s(z,\bz)$ is found to factorize:
\be\label{curL}
\mathscr{L}_s(z,\bz) = \left|\frac{z}{\bz}\right|^{2\sh_2}\,\mathscr{I}(\Delta_1,\Delta_4,\Delta_3\,|\,z)\,\mathscr{I}(\Delta_4,\Delta_1,\Delta_2\,|\,\bz)\,,
\ee
where
\be\label{facJ}
\mathscr{I}(a,b,c\,|\,z) = \int_{-\infty}^\infty\d\al\;\sgn(1-\al)\,|1-\al|^{a-2}\,|\al|^{b-1}\,|z-\al|^{c-1}\,.
\ee
But despite this factorization, the result is not trivially a product of two 3-point light transformed amplitudes. It will have a nontrivial conformal block decomposition.

$\mathscr{I}(a,b,c\,|\,z)$ is very close to being the four-marked-point integral of \cite{Hu:2022syq} (the marked points being $1,0,z,\infty$), the only new ingredient being the factor of $\sgn(1-\al)$. So its computation isn't too hard either. We will content ourselves with computing it in the region $0<z<1$. It divides into four parts:
\be
\mathscr{I}_{a,b,c}(z) = I_{-\infty,0}+I_{0,z}+I_{z,1}-I_{1,\infty}\qquad(0<z<1)\,.
\ee
Using standard M\"obius transformations, the various pieces in this expression yield
\begingroup
\allowdisplaybreaks
\begin{align}
    I_{-\infty,0} &= \int_{-\infty}^0\d\al\;(1-\al)^{a-2}\,(-\al)^{b-1}\,(z-\al)^{c-1}\nonumber\\
    &= B(b,3-a-b-c)\,{}_2F_1(1-c,3-a-b-c\,;3-a-c\,;1-z)\,,\\
    I_{0,z} &= \int_0^z\d\al\;(1-\al)^{a-2}\,\al^{b-1}\,(z-\al)^{c-1}\nonumber\\
    &= z^{b+c-1}\,B(b,c)\,{}_2F_1(2-a,b\,;b+c\,;z)\,,\\
    I_{z,1} &= \int_z^1\d\al\;(1-\al)^{a-2}\,\al^{b-1}\,(\al-z)^{c-1}\nonumber\\
    &= (1-z)^{a+c-2}\,B(a-1,c)\,{}_2F_1(a-1,1-b\,;a+c-1\,;1-z)\,,\\
    I_{1,\infty} &= \int_1^\infty\d\al\;(\al-1)^{a-2}\,\al^{b-1}\,(\al-z)^{c-1}\nonumber\\
    &= B(a-1,3-a-b-c)\,{}_2F_1(1-c,3-a-b-c\,;2-b-c\,;z)\,.
\end{align}
\endgroup
Putting together \eqref{facJ} by applying an Euler transformation on $I_{z,1}$ and the identity
\begin{multline}
{}_2F_1(a,b;c;1-z) = \frac{\Gamma(a+b-c)\Gamma(c)}{\Gamma(a)\Gamma(b)}\,z^{c-a-b}{}_2F_1(c-a,c-b;c-a-b+1;z) \\
+ \frac{\Gamma(c-a-b)\Gamma(c)}{\Gamma(c-a)\Gamma(c-b)}\,{}_2F_1(a,b;a+b-c+1;z)
\end{multline}
on $I_{-\infty,0}$, $I_{z,1}$, we land on
\begin{multline}\label{Iabc}
    \mathscr{I}(a,b,c\,|\,z) = z^{b+c-1}\,\mathscr{C}(b,c)\,{}_2F_1(2-a,b\,;b+c\,;z) \\
    + \mathscr{D}(a-1,3-a-b-c)\,{}_2F_1(1-c,3-a-b-c\,;2-b-c\,;z)\,,
\end{multline}
in the region $0<z<1$. This has been written in terms of the coefficients
\begin{align}
    \mathscr{C}(a,b) &= B(a,1-a-b)+B(b,1-a-b)+B(a,b)\,,\\
    \mathscr{D}(a,b) &= B(a,1-a-b)+B(b,1-a-b)-B(a,b)\,,
\end{align}
that often appear as OPE coefficients of light transformed 3-point amplitudes.

Following \cite{Hogervorst:2017sfd}, introduce the $s$-channel chiral conformal block
\be
k^s_{\sh}(z) = z^{\sh-\sh_{12}}\,{}_2F_1(\sh-\sh_{12},\sh+\sh_{34};2\sh;z)\,.
\ee
Then notice that, since $\beta/2 = \sh_1+\sh_2-\sh_3-\sh_4=0$,
\begin{align}
    k^s_{\sh_1+\sh_2}(z) &= z^{2\sh_2}\,{}_2F_1(2\sh_2,2\sh_3\,;2\sh_1+2\sh_2\,;z)\,,\\
    k^s_{1-\sh_1-\sh_2}(z) &= z^{1-2\sh_1}\,{}_2F_1(1-2\sh_1,1-2\sh_4\,;2-2\sh_3-2\sh_4\,;z)\,.
\end{align}
Using these identities and the result \eqref{Iabc} for our integrals, we find the beautifully simple decomposition
\be
    \mathscr{I}(\Delta_1,\Delta_4,\Delta_3\,|\,z) = z^{-2\sh_2}\Bigl\{\mathscr{D}(2\sh_1,2\sh_2)\,k^s_{\sh_1+\sh_2}(z)\\
    + \mathscr{C}(1-2\sh_3,1-2\sh_4)\,k^s_{1-\sh_1-\sh_2}(z)\Bigr\}\,.
\ee
Similarly, defining the antichiral conformal block
\be
k^s_{\bar\sh}(\bz) = \bz^{\bar\sh-\bar\sh_{12}}\,{}_2F_1(\bar\sh-\bar\sh_{12},\bar\sh+\bar\sh_{34};2\bar\sh;\bz)\,,
\ee
we find
\be
\mathscr{I}(\Delta_4,\Delta_1,\Delta_2\,|\,\bz) =  \bz^{-2\bar\sh_2}\Bigl\{\mathscr{D}(2\bar\sh_3,2\bar\sh_4)\, k^s_{\bar\sh_1+\bar\sh_2}(\bz) + \mathscr{C}(1-2\bar\sh_1,1-2\bar\sh_2)\, k^s_{1-\bar\sh_1-\bar\sh_2}(\bz)\Bigr\}\,.
\ee
Plugging these back into the conformally invariant data of the factorization limit given in \eqref{curL}, we are led to an extremely simple four term conformal block decomposition
\be\label{facdec}
\begin{split}
    \mathscr{L}_s(z,\bz) &= \mathscr{C}(1-2\sh_3,1-2\sh_4)\mathscr{C}(1-2\bar\sh_1,1-2\bar\sh_2)\,k^s_{1-\sh_1-\sh_2}(z)k^s_{1-\bar\sh_1-\bar\sh_2}(\bz)\\
    &\qquad+ \mathscr{C}(1-2\sh_3,1-2\sh_4)\mathscr{D}(2\bar\sh_3,2\bar\sh_4)\,k^s_{1-\sh_1-\sh_2}(z)k^s_{\bar\sh_1+\bar\sh_2}(\bz)\\
    &\qquad+ \mathscr{C}(1-2\bar\sh_1,1-2\bar\sh_2)\mathscr{D}(2\sh_1,2\sh_2)\,k^s_{\sh_1+\sh_2}(z)k^s_{1-\bar\sh_1-\bar\sh_2}(\bz)\\
    &\qquad+ \mathscr{D}(2\sh_1,2\sh_2)\mathscr{D}(2\bar\sh_3,2\bar\sh_4)\,k^s_{\sh_1+\sh_2}(z) k^s_{\bar\sh_1+\bar\sh_2}(\bz)
\end{split}
\ee
in the region $0<z,\bz<1$.

Let $\O_{h_1+h_2,\bh_1+\bh_2-1}^\s$ be a symmetric primary, negative helicity gluon of weight $\Delta_1+\Delta_2-1$. Relating the pre- and post-light-transform weights, it is easily confirmed that the various terms in \eqref{facdec} correspond to the exchanges:
\begin{itemize}
    \item Term 1: $\bL[\O_{h_1+h_2,\bh_1+\bh_2-1}^\s]$,
    \item Term 2: $\mathbf{S}[\O_{h_1+h_2,\bh_1+\bh_2-1}^\s]$,
    \item Term 3: $\O_{h_1+h_2,\bh_1+\bh_2-1}^\s$,
    \item Term 4: $\overline\bL[\O_{h_1+h_2,\bh_1+\bh_2-1}^\s]$,
\end{itemize}
where $\mathbf{S}=\bL\circ\bbL$ is the shadow transform. Since these operators are related by Weyl reflection, it is natural to conjecture that only $\overline\bL[\O_{h_1+h_2,\bh_1+\bh_2-1}^\s]$ (the element of our ambidextrous basis) is an independent physical exchange and the rest appear purely due to representation theory. It would be interesting to understand this phenomenon in greater generality, as it would help in proving or disproving whether our ambidextrous basis of states is dual to a complete basis of the spectrum of celestial CFT. Lastly, though we will avoid listing them explicitly, we have checked that all four terms contain the right product of OPE coefficients of the associated 3-point celestial amplitudes (some of these are listed in appendix B of \cite{Hu:2022syq} for instance). 

This toy example really brings out the advantages of working in the ambidextrous light transform basis: the physics of celestial operator exchanges takes center stage from the get go.

\section{Contributions to the OPE coefficients}
\label{app:ope}

In this appendix, we are going to compute the contribution to the OPE coefficients of the $(h,\bh)= (\N + \sh_1+\sh_2,\N + \bsh_1+\bsh_2)$ operators exchanged in the $\alpha$ space decomposition coming from a product of one holomorphic and one antiholomorphic term present in the $\sJ_+$ portion of \eqref{Fgrfin}. As an example, we are going to look at the polar contributions from the term
\begin{align}
    &\sL_+^{(1,1)} (\al,\bal) =\int_0^1 \d z\,w_s(z)\, \Psi_\al^s(z) \int_0^1 \d\bz\,\bar w_s(\bz)\, \Psi_\bal^s(\bz)~\left(\frac{z}{\bz}\right)^{\Delta_2-1}\left(\frac{1-z}{1-\bz}\right)^{\Delta_1+\Delta_3-2}\nonumber\\
&\times\int_{1}^\infty\d x\frac{(x-1)^{\Delta_1+\Delta_4-2}}{x\,(x-1)}(x-1)^{2 \sh_2-2 \sh_3} (1-z)^{-2 \sh_4} \, _2F_1\left(2 \sh_2,2 \sh_4;1+2 \sh_{23};\frac{1-x}{1-z}\right)\nonumber \\
& \times (x-1)^{2 \bsh_3-2 \bsh_2} (1-\bz)^{-2 \bsh_1} \, _2F_1\left(2 \bsh_1,2 \bsh_3;1-2 \bsh_{23};\frac{1-x}{1-\bz}\right).
\label{eq:exampleopestartingpoint}
\end{align}

After following the same steps for the $\bal$ decomposition as we highlighted in section \ref{sec:alpha}, we arrive at the following expressions for the alpha space transform -- we are only including terms with poles in $\alpha,\bal$, and we will only be highlighting the poles that give rise to the aforementioned towers, and not their Weyl reflections,\footnote{Here, we have chosen a slightly different convention to the one used in \cite{Hogervorst:2017sfd,Rutter:2020vpw}, as per the particulars of our case it makes sense to partition the $\al$ space decomposition in this manner, as we are only counting one family of Weyl reflections, and constructing the tower of operators accordingly.} as these will be the ones used to compute the contribution of this term to conformal block decompositions,
\begingroup
\allowdisplaybreaks
\addtolength{\jot}{1em}
\begin{align}
    &\sL_+^{(1,1)} (\al,\bal) =\sum_{q,\bar{q}=0}^\infty\int_{1}^\infty\d x\frac{(x-1)^{2\sh_1-2\sh_4}}{x\,(x-1)}(x-1)^{2\sh_{23}}(1-x)^q(x-1)^{-2 \bsh_{23}}(1-x)^{{\bar{q}}}\nonumber\\
&\times\bigg[\frac{ \Gamma \left(1-2 \sh_1\right) \Gamma \left(1+2 \sh_{23}\right) \Gamma \left(1-\sh_{12}+\sh_{34}\right)   \Gamma \left(2 \sh_2+q\right) \Gamma \left(2 \sh_{14}-q\right) \Gamma \left(2 \sh_4+q\right) }{\Gamma \left(2 \sh_2\right) \Gamma \left(2 \sh_3\right) \Gamma \left(2 \sh_4\right) \Gamma (1+q) \Gamma \left(\frac12-\alpha -\sh_{12}\right) \Gamma \left(\frac12+\alpha -\sh_{12}\right) \Gamma \left(1+2 \sh_{23}+q\right) \Gamma \left(2 \sh_{14}-q\right)}\nonumber \\
& \quad \quad\times \highlight{\Gamma \left(-\frac12-\alpha +\sh_1+\sh_2\right) }\Gamma \left(-\frac12+\alpha +\sh_1+\sh_2\right)\nonumber \\
& \quad \quad \times  \, _3F_2\left(-\frac12-\alpha +\sh_1+\sh_2,-\frac12\alpha +\sh_1+\sh_2,2 \sh_{14}-q;2 \sh_1,2 \sh_3;1\right)\bigg]\nonumber \\
& \times \bigg[\frac{ \Gamma \left(1-2 \bsh_1\right) \Gamma \left(1-2 \bsh_{23}\right) \Gamma \left(1-\bsh_{12}+\bsh_{34}\right)    \Gamma \left(2 \bsh_1+{\bar{q}}\right) \Gamma \left(2 \bsh_3+{\bar{q}}\right) }{\Gamma \left(2 \bsh_1\right) \Gamma \left(2 \bsh_3\right) \Gamma \left(2 \bsh_3\right) \Gamma (1+{\bar{q}}) \Gamma \left(\frac12-\bal-\bsh_{12}\right) \Gamma \left(\frac12+\bal-\bsh_{12}\right) \Gamma \left(1-2 \bsh_{23}+{\bar{q}}\right)}\nonumber \\
& \quad \quad \times \highlight{\Gamma \left(-\frac12-\bal+\bsh_1+\bsh_2\right) }\Gamma \left(-\frac12+\bal+\bsh_1+\bsh_2\right)\nonumber \\
&\quad \quad \times \, _3F_2\left(-\frac12-\bal+\bsh_1+\bsh_2,-\frac12+\bal+\bsh_1+\bsh_2,-{\bar{q}};2 \bsh_1,2 \bsh_3;1\right) \bigg].
\label{eq:exampleope -step 2}
\end{align}
\endgroup
We now compute the $x$ integral, yielding the following result:
\begingroup
\allowdisplaybreaks
\addtolength{\jot}{1em}
\begin{align}
    &\sL_+^{(1,1)} (\al,\bal) =\highlight{\Gamma \left(-\frac12-\alpha +\sh_1+\sh_2\right) \Gamma \left(-\frac12-\bal+\bsh_1+\bsh_2\right) } \\
    &\times \frac{\Gamma \left(-\frac12+\alpha +\sh_1+\sh_2\right)\Gamma \left(-\frac12+\bal+\bsh_1+\bsh_2\right)}{\Gamma \left(\frac12-\alpha -\sh_{12}\right) \Gamma \left(\frac12+\alpha -\sh_{12}\right)\Gamma \left(\frac12-\bal-\bsh_{12}\right) \Gamma \left(\frac12 +\bal-\bsh_{12}\right)}\nonumber \\
    & \times \frac{\Gamma \left(1-2 \sh_1\right) \Gamma \left(1+2 \sh_{23}\right) \Gamma \left(1-\sh_{12}+\sh_{34}\right) \Gamma \left(1-2 \bsh_1\right) \Gamma \left(1-2 \bsh_{23}\right) \Gamma \left(1-\bsh_{12}+\bsh_{34}\right) }{\Gamma \left(2 \sh_2\right) \Gamma \left(2 \sh_3\right) \Gamma \left(2 \sh_4\right) \Gamma \left(2 \bsh_1\right) \Gamma \left(2 \bsh_3\right){}^2} \nonumber\\
   & \times \sum_{q,\bar q =0}^\infty
\frac{  (-1)^{  (q+\bar{q})} \Gamma \left(2 \sh_2+q\right) \Gamma \left(2 \sh_4+q\right)  \Gamma \left(2 \bsh_1+\bar{q}\right) \Gamma \left(2 \bsh_3+\bar{q}\right) \pi \csc \left(\pi  \left(q+\bar{q}-2 \bsh_{23}\right)\right) }{ \Gamma (1+q) \Gamma (1+\bar{q})  \Gamma \left(1+2 \sh_{23}+q\right)  \Gamma \left(1-2 \bsh_{23}+q\right)}\nonumber \\
& \times  \, _3F_2\left(-\frac12-\alpha +\sh_1+\sh_2,-\frac12\alpha +\sh_1+\sh_2,2 \sh_{14}-q;2 \sh_1,2 \sh_3;1\right) \nonumber \\
&\times \, _3F_2\left(-\frac12-\bal+\bsh_1+\bsh_2,-\frac12+\bal+\bsh_1+\bsh_2,-{\bar{q}};2 \bsh_1,2 \bsh_3;1\right) \, .
\label{eq:exampleope -step 3- xintegral}
\end{align}
\endgroup

Now, using \eqref{eq: ope from residue}, we see that we need to compute the residue of \eqref{eq:exampleope -step 3- xintegral} in order to find the contribution to the OPE coefficient of the poles $(h_n,\bh_m)= (1/2+n + \sh_1+\sh_2,1/2+m + \bsh_1+\bsh_2)$. This residue is found to be
\begingroup
\addtolength{\jot}{1em}
\begin{align}
    &R_+^{(1,1)} (n,m) = \frac{ (-1)^{m+n} \Gamma \left(n+2 \sh_1+2 \sh_2-1\right)  \Gamma \left(m+2 \bsh_1+2 \bsh_2-1\right)  }{m! \,n!\, \Gamma \left(-n-2 \sh_1+1\right) \Gamma \left(n+2 \sh_2\right) \Gamma \left(-m-2 \bsh_1+1\right) \Gamma \left(m+2 \bsh_2\right)} \nonumber \\
   & \times \sum_{q,\bar q =0}^\infty
\frac{  (-1)^{  (q+\bar{q})} \Gamma \left(2 \sh_2+q\right) \Gamma \left(2 \sh_4+q\right)  \Gamma \left(2 \bsh_1+\bar{q}\right) \Gamma \left(2 \bsh_3+\bar{q}\right) \pi \csc \left(\pi  \left(q+\bar{q}-2 \bsh_{23}\right)\right) }{ \Gamma (1+q) \Gamma (1+\bar{q})  \Gamma \left(1+2 \sh_{23}+q\right)  \Gamma \left(1-2 \bsh_{23}+q\right)}\nonumber \\
& \, _3F_2\left(-n,n+2 \sh_1+2 \sh_2-1,-q+2 \sh_1-2 \sh_4;2 \sh_1,2 \sh_3;1\right)  \nonumber \\
&\, _3F_2\left(-m,m+2 \bsh_1+2 \bsh_2-1,-\bar{q};2 \bsh_1,2 \bsh_3;1\right)\, .
\label{eq:exampleope -step 4- residue}
\end{align}
\endgroup
Here we should note that, remarkably, for each pair $n,m$, the hypergeometric series terminates, and we are left with a polynomial. This behaviour is in agreement with results appearing in \cite{Hogervorst:2017sfd}.

With this residue, we can write the contribution of this term to the overall conformal block decomposition as
\begin{equation}
\sL_{+}^{(1,1)} (z,\bz) = \sum_{n,m=0}^\infty \frac{4 R_{+}^{(1,1)}(n,m)\,k_{n+\sh_1+\sh_2}(z) k_{m+\sh_1+\sh_2}(z)}{Q(1/2-n-\sh_1-\sh_2)Q(1/2-m-\bsh_1-\bsh_2)}.
\end{equation}
To obtain the full OPE coefficient, we have to identify the residues for all terms in $\sJ_+$ and in $\sJ_-$ in all the regions of $x$, perform the $x$ integral, and then sum them (i.e. in the previous formula we would substitute $R_+^{(1,1)}$ by $R_+^{(\text{all terms } \sJ_+)} + R_-^{(\text{all terms } \sJ_-)}$. As mentioned above, there are technical complications arising from the structure of $\sJ_-$. These lead to a quadruple sum for each term of the residues and require a regularisation prescription when we partition the $x$ integral into three regions. 

The full conformal block decomposition should include both these poles and a similar term with the additional poles found in the $(0,1)$ region: $(h,\bh)= (\Z^+/2,\Z^+/2)$. The overall procedure has technical complications due to the expressions arising when doing the $\al$-space integral.  Nevertheless, it would be extremely interesting to verify whether these residues factorize into coefficients of 3-point amplitudes, and we hope to return to this in the future.

\bibliographystyle{JHEP}
\bibliography{combined}

\providecommand{\href}[2]{#2}\begingroup\raggedright\begin{thebibliography}{10}

\bibitem{Pasterski:2021raf}
S.~Pasterski, M.~Pate, and A.-M. Raclariu, {\it {Celestial Holography}},  in
  {\em {2022 Snowmass Summer Study}}, 11, 2021.
\newblock \href{http://arxiv.org/abs/2111.11392}{{\tt arXiv:2111.11392}}.

\bibitem{Costello:2022jpg}
K.~Costello, N.~M. Paquette, and A.~Sharma, {\it {Top-down holography in an
  asymptotically flat spacetime}},  \href{http://arxiv.org/abs/2208.14233}{{\tt
  arXiv:2208.14233}}.

\bibitem{Strominger:2017zoo}
A.~Strominger, {\em {Lectures on the Infrared Structure of Gravity and Gauge
  Theory}}.
\newblock {Princeton University Press}, 2018.

\bibitem{Pasterski:2016qvg}
S.~Pasterski, S.-H. Shao, and A.~Strominger, {\it {Flat Space Amplitudes and
  Conformal Symmetry of the Celestial Sphere}},  {\em Phys. Rev. D} {\bf 96}
  (2017), no.~6 065026, [\href{http://arxiv.org/abs/1701.00049}{{\tt
  arXiv:1701.00049}}].

\bibitem{Pasterski:2017kqt}
S.~Pasterski and S.-H. Shao, {\it {Conformal basis for flat space amplitudes}},
   {\em Phys. Rev. D} {\bf 96} (2017), no.~6 065022,
  [\href{http://arxiv.org/abs/1705.01027}{{\tt arXiv:1705.01027}}].

\bibitem{Pasterski:2017ylz}
S.~Pasterski, S.-H. Shao, and A.~Strominger, {\it {Gluon Amplitudes as 2d
  Conformal Correlators}},  {\em Phys. Rev. D} {\bf 96} (2017), no.~8 085006,
  [\href{http://arxiv.org/abs/1706.03917}{{\tt arXiv:1706.03917}}].

\bibitem{deBoer:2003vf}
J.~de~Boer and S.~N. Solodukhin, {\it {A Holographic reduction of Minkowski
  space-time}},  {\em Nucl. Phys.} {\bf B665} (2003) 545--593,
  [\href{http://arxiv.org/abs/hep-th/0303006}{{\tt hep-th/0303006}}].

\bibitem{Cheung:2016iub}
C.~Cheung, A.~de~la Fuente, and R.~Sundrum, {\it {4D scattering amplitudes and
  asymptotic symmetries from 2D CFT}},  {\em JHEP} {\bf 01} (2017) 112,
  [\href{http://arxiv.org/abs/1609.00732}{{\tt arXiv:1609.00732}}].

\bibitem{Mizera:2022sln}
S.~Mizera and S.~Pasterski, {\it {Celestial geometry}},  {\em JHEP} {\bf 09}
  (2022) 045, [\href{http://arxiv.org/abs/2204.02505}{{\tt arXiv:2204.02505}}].

\bibitem{Fotopoulos:2019tpe}
A.~Fotopoulos and T.~R. Taylor, {\it {Primary Fields in Celestial CFT}},  {\em
  JHEP} {\bf 10} (2019) 167, [\href{http://arxiv.org/abs/1906.10149}{{\tt
  arXiv:1906.10149}}].

\bibitem{Pate:2019lpp}
M.~Pate, A.-M. Raclariu, A.~Strominger, and E.~Y. Yuan, {\it {Celestial
  operator products of gluons and gravitons}},  {\em Rev. Math. Phys.} {\bf 33}
  (2021), no.~09 2140003, [\href{http://arxiv.org/abs/1910.07424}{{\tt
  arXiv:1910.07424}}].

\bibitem{Fotopoulos:2019vac}
A.~Fotopoulos, S.~Stieberger, T.~R. Taylor, and B.~Zhu, {\it {Extended BMS
  Algebra of Celestial CFT}},  {\em JHEP} {\bf 03} (2020) 130,
  [\href{http://arxiv.org/abs/1912.10973}{{\tt arXiv:1912.10973}}].

\bibitem{Donnay:2020guq}
L.~Donnay, S.~Pasterski, and A.~Puhm, {\it {Asymptotic Symmetries and Celestial
  CFT}},  {\em JHEP} {\bf 09} (2020) 176,
  [\href{http://arxiv.org/abs/2005.08990}{{\tt arXiv:2005.08990}}].

\bibitem{Sharma:2021gcz}
A.~Sharma, {\it {Ambidextrous light transforms for celestial amplitudes}},
  {\em JHEP} {\bf 01} (2022) 031, [\href{http://arxiv.org/abs/2107.06250}{{\tt
  arXiv:2107.06250}}].

\bibitem{Lam:2017ofc}
H.~T. Lam and S.-H. Shao, {\it {Conformal Basis, Optical Theorem, and the Bulk
  Point Singularity}},  {\em Phys. Rev. D} {\bf 98} (2018), no.~2 025020,
  [\href{http://arxiv.org/abs/1711.06138}{{\tt arXiv:1711.06138}}].

\bibitem{Nandan:2019jas}
D.~Nandan, A.~Schreiber, A.~Volovich, and M.~Zlotnikov, {\it {Celestial
  Amplitudes: Conformal Partial Waves and Soft Limits}},  {\em JHEP} {\bf 10}
  (2019) 018, [\href{http://arxiv.org/abs/1904.10940}{{\tt arXiv:1904.10940}}].

\bibitem{Atanasov:2021cje}
A.~Atanasov, W.~Melton, A.-M. Raclariu, and A.~Strominger, {\it {Conformal
  block expansion in celestial CFT}},  {\em Phys. Rev. D} {\bf 104} (2021),
  no.~12 126033, [\href{http://arxiv.org/abs/2104.13432}{{\tt
  arXiv:2104.13432}}].

\bibitem{Guevara:2021tvr}
A.~Guevara, {\it {Celestial OPE blocks}},
  \href{http://arxiv.org/abs/2108.12706}{{\tt arXiv:2108.12706}}.

\bibitem{Hu:2022syq}
Y.~Hu, L.~Lippstreu, M.~Spradlin, A.~Y. Srikant, and A.~Volovich, {\it
  {Four-point correlators of light-ray operators in CCFT}},  {\em JHEP} {\bf
  07} (2022) 104, [\href{http://arxiv.org/abs/2203.04255}{{\tt
  arXiv:2203.04255}}].

\bibitem{Banerjee:2022hgc}
S.~Banerjee, R.~Basu, and S.~A. Bhatkar, {\it {Light transformed gluon
  correlators in CCFT}},  \href{http://arxiv.org/abs/2203.06657}{{\tt
  arXiv:2203.06657}}.

\bibitem{Fan:2021isc}
W.~Fan, A.~Fotopoulos, S.~Stieberger, T.~R. Taylor, and B.~Zhu, {\it {Conformal
  blocks from celestial gluon amplitudes}},  {\em JHEP} {\bf 05} (2021) 170,
  [\href{http://arxiv.org/abs/2103.04420}{{\tt arXiv:2103.04420}}].

\bibitem{Fan:2021pbp}
W.~Fan, A.~Fotopoulos, S.~Stieberger, T.~R. Taylor, and B.~Zhu, {\it {Conformal
  blocks from celestial gluon amplitudes. Part II. Single-valued correlators}},
   {\em JHEP} {\bf 11} (2021) 179, [\href{http://arxiv.org/abs/2108.10337}{{\tt
  arXiv:2108.10337}}].

\bibitem{Fan:2022vbz}
W.~Fan, A.~Fotopoulos, S.~Stieberger, T.~R. Taylor, and B.~Zhu, {\it {Elements
  of celestial conformal field theory}},  {\em JHEP} {\bf 08} (2022) 213,
  [\href{http://arxiv.org/abs/2202.08288}{{\tt arXiv:2202.08288}}].

\bibitem{Crawley:2021ivb}
E.~Crawley, N.~Miller, S.~A. Narayanan, and A.~Strominger, {\it {State-operator
  correspondence in celestial conformal field theory}},  {\em JHEP} {\bf 09}
  (2021) 132, [\href{http://arxiv.org/abs/2105.00331}{{\tt arXiv:2105.00331}}].

\bibitem{ss}
S.~Pasterski, {\it Soft shadows},  {\em
  \href{https://physicsgirl.com/ss.pdf}{978-0-9863685-4-7}} (2017).

\bibitem{Hogervorst:2017sfd}
M.~Hogervorst and B.~C. van Rees, {\it {Crossing symmetry in alpha space}},
  {\em JHEP} {\bf 11} (2017) 193, [\href{http://arxiv.org/abs/1702.08471}{{\tt
  arXiv:1702.08471}}].

\bibitem{Atanasov:2021oyu}
A.~Atanasov, A.~Ball, W.~Melton, A.-M. Raclariu, and A.~Strominger, {\it {(2,
  2) Scattering and the celestial torus}},  {\em JHEP} {\bf 07} (2021) 083,
  [\href{http://arxiv.org/abs/2101.09591}{{\tt arXiv:2101.09591}}].

\bibitem{Mason:2005qu}
L.~J. Mason, {\it {Global anti-self-dual Yang-Mills fields in split signature
  and their scattering}},  \href{http://arxiv.org/abs/math-ph/0505039}{{\tt
  math-ph/0505039}}.

\bibitem{Gelfand:105396}
I.~M. Gelfand, G.~E. Shilov, M.~I. Graev, N.~Y. Vilenkin, and I.~I.
  Pyatetskii-Shapiro, {\em {Generalized functions, Vol. 5}}.
\newblock AMS Chelsea Publishing. Academic Press, New York, NY, 1964.

\bibitem{ruhl1970lorentz}
W.~Ruhl, {\em {The Lorentz Group and Harmonic Analysis}}.
\newblock Mathematical physics monograph series. W. A. Benjamin, 1970.

\bibitem{Pasterski:2020pdk}
S.~Pasterski and A.~Puhm, {\it {Shifting spin on the celestial sphere}},  {\em
  Phys. Rev. D} {\bf 104} (2021), no.~8 086020,
  [\href{http://arxiv.org/abs/2012.15694}{{\tt arXiv:2012.15694}}].

\bibitem{Kravchuk:2018htv}
P.~Kravchuk and D.~Simmons-Duffin, {\it {Light-ray operators in conformal field
  theory}},  {\em JHEP} {\bf 11} (2018) 102,
  [\href{http://arxiv.org/abs/1805.00098}{{\tt arXiv:1805.00098}}].

\bibitem{Arkani-Hamed:2012zlh}
N.~Arkani-Hamed, J.~L. Bourjaily, F.~Cachazo, A.~B. Goncharov, A.~Postnikov,
  and J.~Trnka, {\em {Grassmannian Geometry of Scattering Amplitudes}}.
\newblock Cambridge University Press, 4, 2016.

\bibitem{Elvang:2013cua}
H.~Elvang and Y.-t. Huang, {\it {Scattering Amplitudes}},
  \href{http://arxiv.org/abs/1308.1697}{{\tt arXiv:1308.1697}}.

\bibitem{Pasterski:2021dqe}
S.~Pasterski, A.~Puhm, and E.~Trevisani, {\it {Revisiting the conformally soft
  sector with celestial diamonds}},  {\em JHEP} {\bf 11} (2021) 143,
  [\href{http://arxiv.org/abs/2105.09792}{{\tt arXiv:2105.09792}}].

\bibitem{Donnay:2018neh}
L.~Donnay, A.~Puhm, and A.~Strominger, {\it {Conformally Soft Photons and
  Gravitons}},  {\em JHEP} {\bf 01} (2019) 184,
  [\href{http://arxiv.org/abs/1810.05219}{{\tt arXiv:1810.05219}}].

\bibitem{Arkani-Hamed:2020gyp}
N.~Arkani-Hamed, M.~Pate, A.-M. Raclariu, and A.~Strominger, {\it {Celestial
  amplitudes from UV to IR}},  {\em JHEP} {\bf 08} (2021) 062,
  [\href{http://arxiv.org/abs/2012.04208}{{\tt arXiv:2012.04208}}].

\bibitem{Law:2019glh}
Y.~T.~A. Law and M.~Zlotnikov, {\it {Poincar\'e constraints on celestial
  amplitudes}},  {\em JHEP} {\bf 03} (2020) 085,
  [\href{http://arxiv.org/abs/1910.04356}{{\tt arXiv:1910.04356}}].

\bibitem{Law:2020xcf}
Y.~A. Law and M.~Zlotnikov, {\it {Relativistic partial waves for celestial
  amplitudes}},  {\em JHEP} {\bf 11} (2020) 149,
  [\href{http://arxiv.org/abs/2008.02331}{{\tt arXiv:2008.02331}}].

\bibitem{De:2022gjn}
S.~De, Y.~Hu, A.~Yelleshpur~Srikant, and A.~Volovich, {\it {Correlators of four
  light-ray operators in CCFT}},  {\em JHEP} {\bf 10} (2022) 170,
  [\href{http://arxiv.org/abs/2206.08875}{{\tt arXiv:2206.08875}}].

\bibitem{ArkaniHamed:2009si}
N.~Arkani-Hamed, F.~Cachazo, C.~Cheung, and J.~Kaplan, {\it {The S-Matrix in
  Twistor Space}},  {\em JHEP} {\bf 03} (2010) 110,
  [\href{http://arxiv.org/abs/0903.2110}{{\tt arXiv:0903.2110}}].

\bibitem{Mason:2009sa}
L.~J. Mason and D.~Skinner, {\it {Scattering Amplitudes and BCFW Recursion in
  Twistor Space}},  {\em JHEP} {\bf 01} (2010) 064,
  [\href{http://arxiv.org/abs/0903.2083}{{\tt arXiv:0903.2083}}].

\bibitem{ArkaniHamed:2009dn}
N.~Arkani-Hamed, F.~Cachazo, C.~Cheung, and J.~Kaplan, {\it {A Duality For The
  S Matrix}},  {\em JHEP} {\bf 03} (2010) 020,
  [\href{http://arxiv.org/abs/0907.5418}{{\tt arXiv:0907.5418}}].

\bibitem{Witten:2003nn}
E.~Witten, {\it {Perturbative gauge theory as a string theory in twistor
  space}},  {\em Commun. Math. Phys.} {\bf 252} (2004) 189--258,
  [\href{http://arxiv.org/abs/hep-th/0312171}{{\tt hep-th/0312171}}].

\bibitem{Rutter:2020vpw}
D.~Rutter and B.~C. Van~Rees, {\it {Applications of Alpha Space}},  {\em JHEP}
  {\bf 12} (2020) 048, [\href{http://arxiv.org/abs/2003.07964}{{\tt
  arXiv:2003.07964}}].

\bibitem{PhysRevD.13.887}
V.~K. Dobrev, V.~B. Petkova, S.~G. Petrova, and I.~T. Todorov, {\it Dynamical
  derivation of vacuum operator-product expansion in euclidean conformal
  quantum field theory},  {\em Phys. Rev. D} {\bf 13} (Feb, 1976) 887--912.

\bibitem{Dobrev1977}
V.~K. Dobrev, G.~Mack, V.~B. Petkova, S.~G. Petrova, and I.~T. Todorov, {\it
  {Harmonic Analysis on the n-Dimensional Lorentz Group and Its Application to
  Conformal Quantum Field Theory}},  {\em Lect. Notes Phys.} {\bf 63} (1977)
  1--280.

\bibitem{Caron-Huot:2017vep}
S.~Caron-Huot, {\it {Analyticity in Spin in Conformal Theories}},  {\em JHEP}
  {\bf 09} (2017) 078, [\href{http://arxiv.org/abs/1703.00278}{{\tt
  arXiv:1703.00278}}].

\bibitem{Simmons-Duffin2018}
D.~Simmons-Duffin, D.~Stanford, and E.~Witten, {\it {A spacetime derivation of
  the Lorentzian OPE inversion formula}},  {\em JHEP} {\bf 07} (2018) 085,
  [\href{http://arxiv.org/abs/1711.03816}{{\tt arXiv:1711.03816}}].

\bibitem{Strominger:2021mtt}
A.~Strominger, {\it {$w_{1+\infty}$ Algebra and the Celestial Sphere: Infinite
  Towers of Soft Graviton, Photon, and Gluon Symmetries}},  {\em Phys. Rev.
  Lett.} {\bf 127} (2021), no.~22 221601.

\bibitem{He:2014laa}
T.~He, V.~Lysov, P.~Mitra, and A.~Strominger, {\it {BMS supertranslations and
  Weinberg's soft graviton theorem}},  {\em JHEP} {\bf 05} (2015) 151,
  [\href{http://arxiv.org/abs/1401.7026}{{\tt arXiv:1401.7026}}].

\bibitem{Kapec:2016jld}
D.~Kapec, P.~Mitra, A.-M. Raclariu, and A.~Strominger, {\it {2D Stress Tensor
  for 4D Gravity}},  {\em Phys. Rev. Lett.} {\bf 119} (2017), no.~12 121601,
  [\href{http://arxiv.org/abs/1609.00282}{{\tt arXiv:1609.00282}}].

\bibitem{Banerjee:2022wht}
S.~Banerjee and S.~Pasterski, {\it {Revisiting the Shadow Stress Tensor in
  Celestial CFT}},  \href{http://arxiv.org/abs/2212.00257}{{\tt
  arXiv:2212.00257}}.

\bibitem{Hu:2022txx}
Y.~Hu and S.~Pasterski, {\it {Celestial Conformal Colliders}},
  \href{http://arxiv.org/abs/2211.14287}{{\tt arXiv:2211.14287}}.

\bibitem{Guevara:2021abz}
A.~Guevara, E.~Himwich, M.~Pate, and A.~Strominger, {\it {Holographic symmetry
  algebras for gauge theory and gravity}},  {\em JHEP} {\bf 11} (2021) 152,
  [\href{http://arxiv.org/abs/2103.03961}{{\tt arXiv:2103.03961}}].

\bibitem{Donnay:2022sdg}
L.~Donnay, S.~Pasterski, and A.~Puhm, {\it {Goldilocks modes and the three
  scattering bases}},  {\em JHEP} {\bf 06} (2022) 124,
  [\href{http://arxiv.org/abs/2202.11127}{{\tt arXiv:2202.11127}}].

\bibitem{Adamo:2021lrv}
T.~Adamo, L.~Mason, and A.~Sharma, {\it {Celestial $w_{1+\infty}$ Symmetries
  from Twistor Space}},  {\em SIGMA} {\bf 18} (2022) 016,
  [\href{http://arxiv.org/abs/2110.06066}{{\tt arXiv:2110.06066}}].

\bibitem{Adamo:2021zpw}
T.~Adamo, W.~Bu, E.~Casali, and A.~Sharma, {\it {Celestial operator products
  from the worldsheet}},  {\em JHEP} {\bf 06} (2022) 052,
  [\href{http://arxiv.org/abs/2111.02279}{{\tt arXiv:2111.02279}}].

\bibitem{Sharma:2022arl}
A.~Sharma, {\em {Twistor sigma models}}.
\newblock PhD thesis, Oxford U., 2022.

\bibitem{Costello:2022wso}
K.~Costello and N.~M. Paquette, {\it {Celestial holography meets twisted
  holography: 4d amplitudes from chiral correlators}},  {\em JHEP} {\bf 10}
  (2022) 193, [\href{http://arxiv.org/abs/2201.02595}{{\tt arXiv:2201.02595}}].

\bibitem{Costello:2022upu}
K.~Costello and N.~M. Paquette, {\it {On the associativity of one-loop
  corrections to the celestial OPE}},
  \href{http://arxiv.org/abs/2204.05301}{{\tt arXiv:2204.05301}}.

\bibitem{Bu:2022dis}
W.~Bu and E.~Casali, {\it {The 4d/2d correspondence in twistor space and
  holomorphic Wilson lines}},  {\em JHEP} {\bf 11} (2022) 076,
  [\href{http://arxiv.org/abs/2208.06334}{{\tt arXiv:2208.06334}}].

\bibitem{Bittleston:2022jeq}
R.~Bittleston, {\it {On the associativity of 1-loop corrections to the
  celestial operator product in gravity}},
  \href{http://arxiv.org/abs/2211.06417}{{\tt arXiv:2211.06417}}.

\bibitem{Adamo:2022wjo}
T.~Adamo, W.~Bu, E.~Casali, and A.~Sharma, {\it {All-order celestial OPE in the
  MHV sector}},  \href{http://arxiv.org/abs/2211.17124}{{\tt
  arXiv:2211.17124}}.

\bibitem{Garcia-Sepulveda:2022lga}
D.~Garc\'\i{}a-Sep\'ulveda, A.~Guevara, J.~Kulp, and J.~Wu, {\it {Notes on
  resonances and unitarity from celestial amplitudes}},  {\em JHEP} {\bf 09}
  (2022) 245, [\href{http://arxiv.org/abs/2205.14633}{{\tt arXiv:2205.14633}}].

\end{thebibliography}\endgroup

\end{document}